\documentclass[12pt,preprint]{aastex}
\usepackage{epsf}

\def\linebreak{\hfil\break}

\def\
{\hfil\linebreak}

\def\akari{{\it AKARI}}
\def\iso{{\it ISO}}
\def\iras{{\it IRAS}}

\def\spitz{{\it Spitzer}}
\def\wise{{\it WISE}}
\def\orch{{\it Orchestra}}
\def\nbody{{$n$-body}}

\def\deg{\ifmmode {^\circ}\else {$^\circ$}\fi}
\def\degree{\ifmmode {^\circ}\else {$^\circ$}\fi}
\def\mum{\ifmmode {\rm \mu {\rm m}}\else $\rm \mu {\rm m}$\fi}
\def\arcsec{\ifmmode ^{\prime \prime}\else $^{\prime \prime}$\fi}

\def\inch{\ifmmode ^{\prime \prime}\else $^{\prime \prime}$\fi}
\def\arcmin{\ifmmode ^{\prime}\else $^{\prime}$\fi}
\def\qprime{\ifmmode q^{\prime}\else $q^{\prime}$\fi}

\def\degree{\ifmmode {^\circ}\else {$^\circ$}\fi}
\def\arcsec{\ifmmode ^{\prime \prime}\else $^{\prime \prime}$\fi}

\def\inch{\ifmmode ^{\prime \prime}\else $^{\prime \prime}$\fi}
\def\arcmin{\ifmmode ^{\prime}\else $^{\prime}$\fi}

\def\mjup{\ifmmode { M_J}\else $ M_J$\fi}
\def\rjup{\ifmmode { R_J}\else $ R_J$\fi}
\def\mearth{\ifmmode { M_{\oplus}}\else $ M_{\oplus}$\fi}
\def\rearth{\ifmmode { R_{\oplus}}\else $ R_{\oplus}$\fi}
\def\ldust{\ifmmode { L_d}\else $ L_d$\fi}
\def\ldstar{\ifmmode { L_d / L_{\star}}\else $ L_d / L_{\star}$\fi}
\def\lstar{\ifmmode { L_{\star}}\else $ L_{\star}$\fi}
\def\lsun{\ifmmode { L_{\odot}}\else $ L_{\odot}$\fi}
\def\mstar{\ifmmode M_{\star}\else $ M_{\star}$\fi}
\def\msun{\ifmmode M_{\odot}\else $ M_{\odot}$\fi}
\def\tstar{\ifmmode T_{\star}\else $ T_{\star}$\fi}
\def\rstar{\ifmmode R_{\star}\else $ R_{\star}$\fi}
\def\rsun{\ifmmode R_{\odot}\else $ R_{\odot}$\fi}
\def\mjup{\ifmmode M_{J}\else $ M_{J}$\fi}
\def\rjup{\ifmmode R_{J}\else $ R_{J}$\fi}
\def\mjupyr{\ifmmode { M_J~yr^{-1}}\else $ M_J~yr^{-1}$\fi}
\def\msunyr{\ifmmode { M_{\odot}~yr^{-1}}\else $ M_{\odot}~yr^{-1}$\fi}
\def\gyr{\ifmmode {\rm g~yr^{-1}}\else $\rm g~yr^{-1}$\fi}
\def\ergg{\ifmmode {\rm erg~g^{-1}}\else $\rm erg~g^{-1}$\fi}
\def\kms{\ifmmode {\rm km~s^{-1}}\else $\rm km~s^{-1}$\fi}
\def\ms{\ifmmode {\rm m~s^{-1}}\else $\rm m~s^{-1}$\fi}
\def\rhill{\ifmmode R_H\else $R_H$\fi}
\def\rfast{\ifmmode R_{fast}\else $R_{fast}$\fi}
\def\rgap{\ifmmode R_{gap}\else $R_{gap}$\fi}
\def\vhill{\ifmmode v_H\else $v_H$\fi}
\def\qdstar{\ifmmode Q_D^\star\else $Q_D^\star$\fi}
\def\mesc{\ifmmode m_{esc}\else $m_{esc}$\fi}
\def\rmin{\ifmmode r_{min}\else $r_{min}$\fi}
\def\mmin{\ifmmode m_{min}\else $m_{min}$\fi}
\def\rmax{\ifmmode r_{max}\else $r_{max}$\fi}
\def\mmax{\ifmmode m_{max}\else $m_{max}$\fi}
\def\rmind{\ifmmode r_{min,d}\else $r_{min,d}$\fi}
\def\rmaxd{\ifmmode r_{max,d}\else $r_{max,d}$\fi}
\def\mmaxd{\ifmmode m_{max,d}\else $m_{max,d}$\fi}
\def\qz{\ifmmode q_{0}\else $q_{0}$\fi}
\def\qi{\ifmmode q_{i}\else $q_{i}$\fi}
\def\ql{\ifmmode q_{l}\else $q_{l}$\fi}
\def\qs{\ifmmode q_{s}\else $q_{s}$\fi}
\def\rbrk{\ifmmode r_{brk}\else $r_{brk}$\fi}
\def\rdamp{\ifmmode r_{damp}\else $r_{damp}$\fi}
\def\r0{\ifmmode r_{0}\else $r_{0}$\fi}
\def\m0{\ifmmode m_{0}\else $m_{0}$\fi}
\def\M0{\ifmmode M_{0}\else $M_{0}$\fi}
\def\xm{\ifmmode x_{m}\else $x_{m}$\fi}
\def\gyr{\ifmmode {\rm g~yr^{-1}}\else ${\rm g~yr^{-1}}$\fi}
\def\cms{\ifmmode {\rm cm~s^{-1}}\else ${\rm cm~s^{-1}}$\fi}
\def\gcms{\ifmmode {\rm g~cm^{-2}}\else $\rm g~cm^{-2}$\fi}
\def\gcmc{\ifmmode {\rm g~cm^{-3}}\else $\rm g~cm^{-3}$\fi}

\def\2470{[24]--[70]}

\newbox\grsign \setbox\grsign=\hbox{$>$} \newdimen\grdimen \grdimen=\ht\grsign
\newbox\simlessbox \newbox\simgreatbox
\setbox\simgreatbox=\hbox{\raise.5ex\hbox{$>$}\llap
     {\lower.5ex\hbox{$\sim$}}}\ht1=\grdimen\dp1=0pt
\setbox\simlessbox=\hbox{\raise.5ex\hbox{$<$}\llap
     {\lower.5ex\hbox{$\sim$}}}\ht2=\grdimen\dp2=0pt

\begin{document}

\title{Variations on Debris Disks III. Collisional Cascades and Giant Impacts
in the Terrestrial Zones of Solar-type Stars}
\vskip 7ex
\author{Scott J. Kenyon}
\affil{Smithsonian Astrophysical Observatory,
60 Garden Street, Cambridge, MA 02138} 
\email{e-mail: skenyon@cfa.harvard.edu}

\author{Benjamin C. Bromley}
\affil{Department of Physics, University of Utah, 
201 JFB, Salt Lake City, UT 84112} 
\email{e-mail: bromley@physics.utah.edu}
%
%

\begin{abstract}

We analyze two new sets of coagulation calculations for solid particles
orbiting within the terrestrial zone of a solar-type star. In models of
collisional cascades, numerical simulations demonstrate that the total mass, 
the mass in 1~mm and smaller particles, and the dust luminosity decline 
with time more rapidly than predicted by analytic models, $\propto t^{-n}$ 
with $n \approx$ 1.1--1.2 instead of 1. Size distributions derived from 
the numerical calculations follow analytic predictions at $r \lesssim$ 0.1~km 
but are shallower than predicted at larger sizes. 
In simulations of planet formation, the dust luminosity declines more 
slowly than in pure collisional cascades, with $n \approx$ 0.5--0.8 
instead of 1.1--1.2.  Throughout this decline, giant impacts produce 
large, observable spikes in dust luminosity which last $\sim$ 
0.01--0.1~Myr and recur every 1--10~Myr.  If most solar-type stars 
have Earth mass planets with $a \lesssim$ 1--2~AU, observations of 
debris around 1--100~Myr stars allow interesting tests of theory. 
Current data preclude theories where terrestrial planets form out 
of 1000~km or larger planetesimals. Although the observed frequency 
of debris disks among $\gtrsim$ 30~Myr old stars agrees with our 
calculations, the observed frequency of warm debris among 5--20~Myr
old stars is smaller than predicted.

\end{abstract}


\section{INTRODUCTION}
\label{sec: intro}

Dense disks of debris surround many main sequence stars \citep[e.g.,][]
{bac1993,wyatt2008,matthews2014}. In these systems, particles with radii
$r \approx$ 1~\mum\ to 1~cm and temperatures $T \approx$ 50--500~K 
reradiate $\sim$ 0.001\% to 1\% of the light from the central star.  
Binary systems are almost as likely to harbor debris disks as apparently 
single stars \citep{trill2007,stauffer2010,kennedy2012,rodr2012,rodr2015}.
Among 
A-type stars, the frequency of debris disks declines from $\sim$ 50\% for 
ages of $\sim$ 10~Myr to $\lesssim$ 5\% at 0.5--1~Gyr \citep[e.g.,][]
{rieke2005,currie2008,carp2009a}. FGK stars have somewhat lower debris 
disk frequencies of 10\% to 20\%, with excesses that decline more slowly 
with stellar age \citep{carp2009a,carp2009b,kenn2013a,matthews2014}. Several 
older stars also appear to have substantial and very luminous disks or rings 
of debris which may decline rapidly with time \citep[e.g.,][]{song2005,
rhee2008,melis2010,melis2012,meng2015}. 

Debris disks are signposts of planet formation \citep[e.g.,][]
{kb2002b,zuck2004,kb2008,raymond2011,raymond2012}. In current theory, 
protoplanets grow through mergers of smaller objects. Once they reach 
sizes of 500--1000~km, protoplanets stir nearby small objects to large 
velocities. Continued growth enables stirring over larger and larger 
volumes, generating a cascade of destructive collisions among small 
leftovers. This collisional cascade grinds the leftovers into a ring 
or disk of very small particles with luminosity comparable to the dust
luminosities of many debris disks.  As the cascade evolves, occasional 
`giant impacts' between pairs of protoplanets produce clouds of small 
particles which interact with the rest of the disk \citep[e.g.,][]
{agnor1999,asphaug2006,grigor2007,genda2012}. In some giant impact 
models, the dust luminosity roughly matches the very luminous disks
associated with old G-type stars.

Interpreting observations of debris disks requires a robust theory of
planet formation which predicts the time evolution of luminosity and
the structure of solid material from $\sim$ 0.1~AU to beyond 100~AU. 
Analytic approaches specify an initial ensemble of solids with maximum 
size \rmax\ orbiting the central star at high velocity with eccentricity 
$e$, inclination $i$ and a range of semimajor axes $a$ \citep[e.g.,][]
{wyatt2002,dom2003,krivov2006,wyatt2007a,wyatt2007b,lohne2008,koba2010a,
kw2010,wyatt2011}. For an initial surface density $\Sigma_0$, several 
simplifying assumptions allow these models to predict the time evolution 
of the size distribution and total mass of the debris. Successful 
applications include matches to the time evolution of debris disks 
around A-type and G-type stars \citep{matthews2014}.  However, the 
approach fails to account for the frequency of giant impacts at late 
times.

Several numerical approaches place debris disks in the context of the
long-term evolution of a primordial disk of gas and dust \citep{kb2005,
kb2008,kb2010,weiden2010b,raymond2011,raymond2012,koba2014}. 
These studies adopt an initial mass and radial structure for the disk, 
establish prescriptions for the evolution and interactions of the gas 
and the solids, and derive $a$, $e$, $i$, $\Sigma$, and other properties 
of the debris as a function of the initial conditions and the final 
architecture of the planetary system.  With an accurate 
treatment for the long-term evolution of the cascade and the ability 
to follow dust production from giant impacts at late times, these
calculations can match observations of the time evolution of the 
brightness and frequency of debris disks. However, the cpu intensive 
nature of this approach often limits their ability to identify how the
evolution depends on uncertain input parameters.

Other numerical approaches extract plausible snapshots from detailed
planet formation simulations and model a well-defined subset of the
system \citep{grigor2007,booth2009,mustill2009,stark2009,jackson2012,
nesvold2013,kral2013}. Using \nbody\ dynamical calculations combined 
with semi-analytic results, these investigations derive the geometry 
and luminosity of the debris as a function of time. Aside from learning 
how these features depend on initial conditions, these investigations
excel at matching observations of particular systems. These algorithms
are also very cpu intensive and hinge on specifying a fairly small set 
of initial conditions from a very broad range of possibilities.

Weaving these investigations into a more robust theory for planet formation
requires calculations which test and unify the different approaches. Here
we consider results from two sets of coagulation calculations for swarms of
particles orbiting at 1~AU from a solar-type star. The first set of
simulations enables robust tests of analytic models for collisional cascades. 
The second set allows us to place collisional cascades in context with more
detailed planet formation simulations. Comparisons between the two sets yield
common features and differences which are useful for interpreting existing
observations.

Our study begins in \S\ref{sec: an-mod} with a brief discussion of the 
basic predictions of the analytic model. After describing the algorithms 
used in our numerical simulations (\S\ref{sec: model}), we report outcomes
for collisional cascades (\S\ref{sec: calcs-casc}) and the formation of
planets (\S\ref{sec: calcs-pf}) at 1~AU. After a brief discussion
(\S\ref{sec: disc}), we conclude with a short summary (\S\ref{sec: summ}).
Readers uninterested in the details can skim the background in
\S\ref{sec: an-mod}, glance through the figures in \S\ref{sec: calcs},
read the discussion in \S5, and finish with the summaries in 
\S\ref{sec: calcs-summ} and \S\ref{sec: summ}.

\section{ANALYTIC MODEL}
\label{sec: an-mod}

Building on the collisional cascade models of \citet{dohn1969} and \citet{hellyer1970}, 
\citet{wyatt2002} and \citet{dom2003} developed analytic models for the long-term 
collisional evolution of particles in a debris disk.  Objects with radius $r$, 
mass $m$, and mass density $\rho$ orbit within a cylindrical annulus with width $\delta a$ 
centered at a distance $a$ from a central star with mass \mstar\ and luminosity \lstar.
All particles have the same orbital eccentricity $e$ and inclination $i$.  Destructive 
collisions generate a collisional cascade where the mass in solids gradually diminishes 
with time.  For an upper mass limit $m_{max}$ of the largest solid objects participating 
in the cascade, the analytic model yields a simple formula for $N_{max}(t)$, the number 
of these large objects as a function of time.  Radiation pressure establishes $m_{min}$, 
a lower mass limit for solids with stable orbits around the central star.  Between $m_{min}$ 
and $m_{max}$, the cascade produces a power-law size distribution, $N(r) \propto r^{-q}$ 
\citep[e.g.,][]{dohn1969,hellyer1970, will1994,obrien2003,koba2010a,wyatt2011}.  Setting 
the slope $q$ of this size distribution yields another simple formula for the time evolution 
of the dust luminosity $L_d(t)$.

\subsection{Time Evolution of the Number Density}
\label{sec: an-mod-num}

To derive expressions for $N_{max}(t)$ and $L_d(t)$, we adopt a particle-in-a-box 
model, where kinetic theory sets the collision rate \citep{wyatt2002,dom2003}. For an 
annulus with volume $V$, we consider an ensemble of mono-disperse particles with number 
density $n_{max} = N_{max} / V$, collision cross-section $\sigma = 4 \pi \rmax^2$, and 
relative velocity $v$.  Every collision destroys two particles; the loss rate is
$\dot{N}_{max} = -2 N_{max} n_{max} \sigma v$. 

To express this rate in terms of basic disk parameters, we define the initial number of 
particles $N_{0,max}$, initial surface density $\Sigma_0 = N_{0,max} m_{max} / 2 \pi a \delta a$, 
the vertical height of the annulus $H$, and the volume $V = 4 \pi a \delta a H$.  For 
solids orbiting with angular frequency $\Omega$, orbital period $P = 2 \pi / \Omega$, 
vertical velocity $v_z$, and Keplerian velocity $v_K$ within the annulus, 
$H \approx v_z / \Omega$, $i \approx e/2$, $v_z \approx e v_K / 2$, and 
$v \approx e v_K$ \citep[e.g.,][]{weth1993,kl1998}.  

We set the collision time as
\begin{equation}
t_c = {\rmax \rho P \over 12 \pi \Sigma_0} ~ .
\label{eq: tc1}
\end{equation}
In this approach, $t_c$ is the time to destroy one of the largest particles in
the ensemble. Formally, collisions destroy two particles on the time scale $2 t_c$.
The collision rate then becomes
\begin{equation}
\dot{N}_{max} = -N_{max}^2 / N_{0,max} t_c ~ ,
\end{equation}
which has a simple solution
\begin{equation}
N_{max}(t) = { N_{0,max} \over {1 + t / t_c} } ~ .
\label{eq: nmax} 
\end{equation}

In this derivation, the collision time depends only on properties of the 
largest objects in the swarm \citep[see also][]{dom2003}. In a real cascade,
collisions with smaller particles in the swarm also remove mass from the largest 
particles \citep[e.g.,][]{wyatt2002,dom2003,wyatt2007a,wyatt2007b,koba2010a,
kw2011a,kw2011b,wyatt2011}. 
In addition to completely destructive collisions with objects having $m \gtrsim$ 
0.25 \mmax, cratering collisions with much smaller particles gradually chip away 
material from the largest objects.  These `extra' collisional processes modify 
the collision time by a factor $\alpha \lesssim$ 3--4.

To estimate $\alpha$ for an equilibrium cascade, we adopt results from 
\citet[][and references therein]{koba2010a}. If \qdstar\ is the collision 
energy required to eject half the mass from a pair of colliding objects
and if $Q_c$ is the center-of-mass collision energy, the largest object in 
the debris has mass $m_l \approx 0.2 ~ (Q_c / \qdstar)^{-1}$.  For simplicity, 
we assume $q \approx$ 3.5 for the entire swarm throughout the evolution and 
for the debris from a single collision.  Different choices have little impact 
on the results \citep{koba2010a}.  Using the $s_1 = 6.3$, $s_2 = 38.1$, and 
$s_3 = 5.6$ parameters defined following eq. 25 of \citet{koba2010a}, 
\begin{equation}
\alpha \approx 3.6 \left ( v^2 \over 2 \qdstar \right )^{-5/6} ~ .
\label{eq: alpha}
\end{equation}
For $(v^2 / 2 \qdstar) \gg$ 1, eqs. 9--13 of \citet{wyatt2007a} yield similar
results.  In most collisional cascades, $(v^2 / 2 \qdstar) \gtrsim$ 4; thus,
$\alpha \lesssim$ 1.8 \citep[see also][]{wyatt2011}.

To complete this discussion, it is useful to estimate the collision time in a
swarm of material at 1 AU.  For solids with $\rho$ = 3~\gcmc\ in a disk with 
$\Sigma_0$ = 10~\gcms\ \citep[comparable to the minimum mass solar nebula;
e.g.,][]{weiden1977b,hayashi1981,chiang2010}, 
\begin{equation}
\label{eq: tc2}
t_c \approx 8 \times 10^4 ~ \alpha ~
\left ( {r_{max} \over {\rm 100~km} } \right )
\left ( {\Sigma_0 \over 10~\gcms\ } \right )^{-1}
\left ( {P \over {\rm 1~yr}} \right ) ~ {\rm yr} ~ .
\end{equation}
Collision times are short, $\sim$ 0.1--1.0~Myr for swarms consisting of
particles with \rmax\ $\lesssim$ 100--1000~km at 1 AU. 

\subsection{Luminosity Evolution}
\label{sec: an-mod-lum}

With $N(t)$ known, the dust luminosity follows.  Destructive collisions of the
largest objects produce a power-law size distribution extending from the largest
objects with mass \mmax\ to the smallest objects with mass \mmin.  The total mass 
$M_d$ and cross-sectional area $A_d$ of particles within the annulus are simple 
functions of the minimum size, the maximum size, and the slope 
$q$ \citep[e.g.,][]{wyatt2002,dom2003,kcb2014}.  The stellar energy intercepted 
by the solids is $L_d = A_d / 4 \pi a^2$. If \mmin, \mmax, and $q$ never change, 
the initial dust luminosity $L_{d,0}$ is a simple function of $N_{max,0}$ 
and these parameters.  
The time evolution of the dust luminosity is then:
\begin{equation}
L_d(t) = { L_{d,0} \over {1 + t / t_c} } ~ .
\label{eq: ld} 
\end{equation}
Setting \rmin\ = 1~\mum\ and $q$ = 3.5 for the size distribution 
for an annulus with $\delta a = 0.2 a$ at 1~AU:
\begin{equation}
L_{d,0} = 7.9 \times 10^{-3}
\left ( {\rho \over {3~\gcmc} } \right )^{-1}
\left ( {r_{max} \over {\rm 100~km} } \right )^{-1/2}
\left ( {r_{min} \over 1~\mum } \right )^{-1/2}
\left ( {\Sigma_0 \over 10~\gcms\ } \right )
\lstar ~ .
\label{eq: ld0}
\end{equation}
The initial dust luminosity is independent of the collision time. 

At late times ($t \gg t_c$), $L_d(t) \approx L_{d,0} t_c / t$. 
Combining $L_{d,0}$ and $t_c$ into a new normalization factor 
$L^{\prime}_{d,0}$,
\begin{equation}
L_d(t) = L^{\prime}_{d,0} 
\left ( { t \over {\rm 1~Myr} } \right )^{-1} ~ ,
\label{eq: ld2}
\end{equation}
where 
\begin{equation}
\label{eq: ld0prime}
L^{\prime}_{d,0} = 5.6 \times 10^{-4} ~ \alpha 
\left ( {r_{max} \over {\rm 100~km} } \right )^{1/2}
\left ( {r_{min} \over 1~\mum } \right )^{-1/2}
\left ( {P \over {\rm 1~yr}} \right ) ~ \lstar ~ .
\end{equation}
Despite starting out with much smaller initial dust luminosity, old 
cascades with large \rmax\ are easier to detect than those with small 
\rmax\ \citep{wyatt2008,koba2010a}. At late times, the ability to detect 
a cascade is independent of the initial mass involved in the cascade.
Through the parameter $\alpha$, the dynamics of the swarm also set the
detectability.

\subsection{Deriving \rmin\ and \rmax}
\label{sec: an-mod-rmax}

Maintaining a power-law size distribution for particles with 
$m_{min} \lesssim m \lesssim m_{max}$ requires destructive collisions 
among roughly equal mass objects\footnote{Although cratering can also 
drive a cascade, the largest objects then probably grow with time. For 
simplicity, we focus here on systems where collisions reduce the masses
of all objects. We return to this issue in \S\ref{sec: calcs-casc-std}.}.
For impact velocity $v$, $Q_c = \mu v^2 / 2 (m_1 + m_2)$, where 
$\mu = m_1 m_2 / (m_1 + m_2)$ is the reduced mass for a pair of 
colliding planetesimals with masses 
$m_1$ and $m_2$. For equal mass objects, $Q_c = v^2 / 8$
\citep[see also][]{weth1993,will1994,tanaka1996b,stcol1997a,kl1999a,obrien2003,koba2010a}. 

Following standard practice,
\begin{equation}
\qdstar = Q_b r^{\beta_b} + Q_g \rho_p r^{\beta_g}
\label{eq: qd}
\end{equation}
where $Q_b r^{\beta_b}$ is the bulk component of the binding energy and
$Q_g \rho_g r^{\beta_g}$ is the gravity component of the binding energy
\citep[e.g.,][]{benz1999,lein2008,lein2009}.  For rocky objects, 
$Q_b \approx 3 \times 10^7$~erg~g$^{-1}$~cm$^{-\beta_b}$, 
$\beta_b \approx -0.40$, $Q_g \approx$ 0.3~erg~g$^{-2}$~cm$^{3-\beta_g}$, 
and $\beta_g \approx$ 1.35 \citep[see also][]{davis1985,hols1994,love1996,housen1999,
ryan1999,arakawa2002,giblin2004,burchell2005}.

Setting $Q_c \approx \qdstar$ establishes constraints on the particles destroyed by an
adopted collision velocity.  For large objects with $v \approx e v_K$ and $\beta_g$ = 1.35, 
collisions destroy equal-mass particles with $r \lesssim r_{c,max}$, where
\begin{equation}
r_{c,max} \approx 300 
\left ( {e \over 0.1} \right )^{1.48}
\left ( {v_K \over {\rm 30~\kms} } \right )^{1.48}
\left ( {\rho \over {\rm 3~\gcmc} } \right )^{-0.74}
\left ( {Q_g \over {\rm 0.3~erg~g^{-1}} } \right )^{-0.74} ~ {\rm km} ~ .
\label{eq: rc-max}
\end{equation}
For small objects, $\beta_b = -0.4$; collisions destroy equal mass objects with
$r \gtrsim r_{c,min}$, where
\begin{equation}
r_{c,min} \approx 0.02 
\left ( {e \over 0.1} \right )^{-2.5}
\left ( {v_K \over {\rm 30~\kms} } \right )^{-2.5}
\left ( {Q_b \over {\rm 10^7~erg~g^{-1}} } \right )^{2.5} ~ \mum ~ .
\label{eq: rc-min}
\end{equation}
The first relation establishes that objects with $r \lesssim$ 300~km participate in the 
cascade. Because radiation pressure typically ejects dust grains with $r \lesssim$ 1~\mum,
the second relation confirms that all objects smaller than 300~km participate in the 
cascade. Once the cascade begins, collisions maintain it forever.

Although collisions between equal-mass objects with $r > r_{c,max}$ result in some
net growth of one particle, cratering collisions with the rest of the swarm can
gradually reduce the mass of these objects \citep[e.g.,][]{koba2010a}. As with
estimates of the $\alpha$ factor for the collision time (eq.~\ref{eq: alpha}), 
deriving the maximum size of particles which slowly lose mass requires integrating
the collision rate over the mass distribution.  When we examine results of our
numerical simulations, we will illustrate how this maximum size depends on the
properties of the swarm.

Once the cascade begins, the slope $q$ of the power-law size distribution depends
on the details of the \qdstar\ relation \citep{obrien2003,koba2010a,wyatt2011}.
When \qdstar\ is independent of $r$, $q$ = 3.5. For the standard $\qdstar(r)$ in
eq.~\ref{eq: qd}, the size distribution consists of two power laws, with
$q_s = (7 - \beta_b/3) / (2 - \beta_b/3) \approx $ 3.68 for small particles and 
$q_l = (7 + \beta_g/3) / (2 + \beta_g/3) \approx $ 3.04 for large particles. The
slope of the double power law changes where $\qdstar(r)$ reaches a minimum, 
$r_Q \approx$ 0.1~km for rocky objects with the parameters quoted after 
eq.~\ref{eq: qd}.

\subsection{Collisional Cascade at 1 AU}
\label{sec: an-mod-1au}

Solving eqs.~\ref{eq: nmax}--\ref{eq: ld} for the long-term evolution of
the mass in solids and the dust luminosity requires initial conditions 
for the central star (\mstar, \lstar), the geometry of the annulus ($a$, 
$\delta a$, $e$, and $i$), and the properties of the solid particles
(\rmin, \rmax, $N_{0,max}$, $q$, and \qdstar).  To illustrate the long-term 
evolution, we consider a swarm of solid material orbiting a 1~\msun\ star 
at 1~AU. 

\begin{figure} 
\includegraphics[width=6.5in]{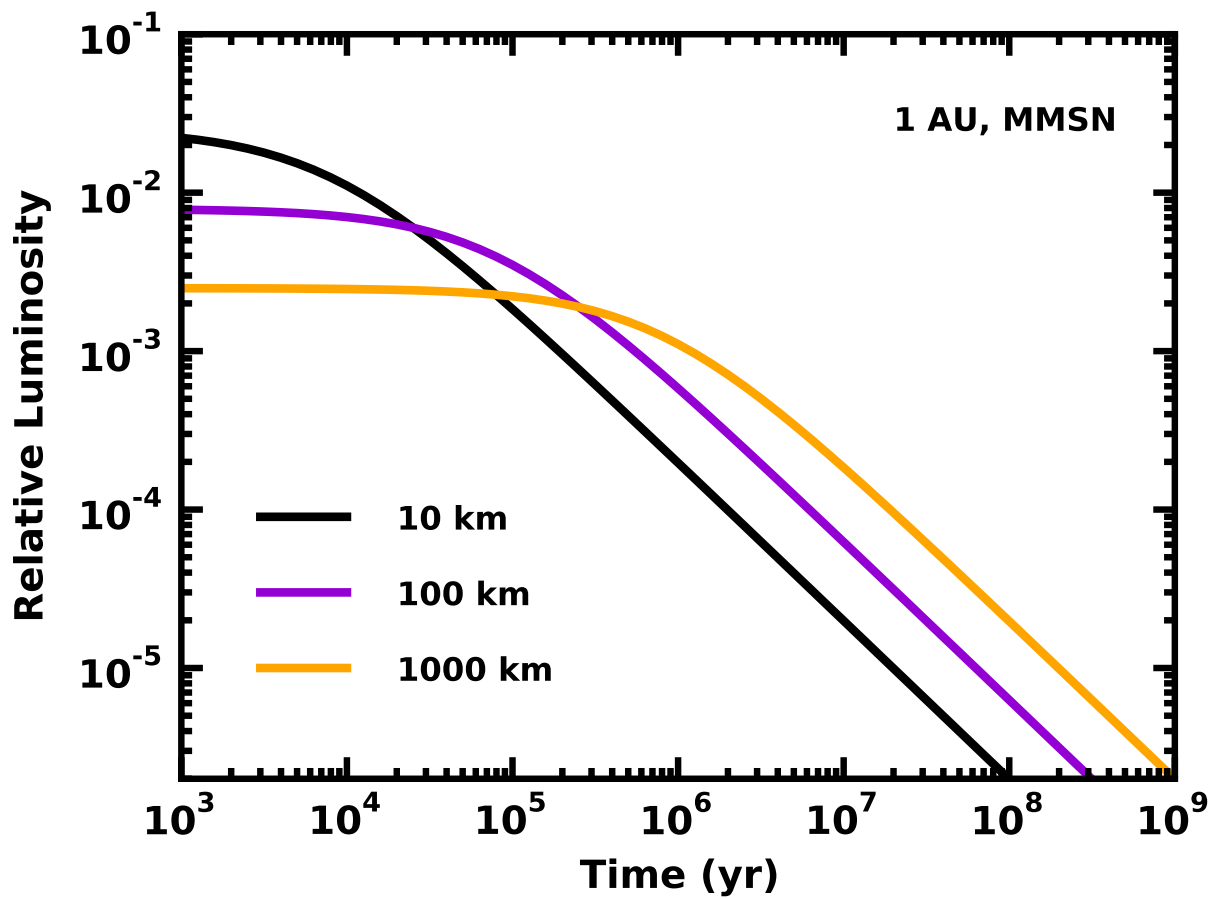}
\vskip 3ex
\caption{
Time evolution of the relative luminosity $L_d / \lstar$ derived 
from the analytic model for a collisional cascade. Material
orbits a 1~\msun\ star inside a cylindrical annulus with
$a$ = 1~AU, $\delta a$ = 0.2~AU, and $\Sigma$ = 10~\gcms.
Solid curves plot the evolution of $L_d/L_\star$ for
cascades with \rmax\ = 10~km (black curve), 100~km 
(violet curve), and 1000~km (orange curve). Cascades with
smaller \rmax\ have larger initial luminosity which declines
on much shorter time scales.
\label{fig: an1}
}
\end{figure}

Fig.~\ref{fig: an1} shows the time evolution of the dust luminosity for
particles with \rmin\ = 1~\mum, various \rmax\, $\Sigma_0$ = 10~\gcms, 
$e$ = 0.1, $i$ = $e$/2, and $\alpha$ = 1 orbiting in an annulus with 
$\delta a$ = 0.2 at $a$ = 1~AU around a central star with \mstar\ = 
1~\msun, \lstar\ = 1~\lsun.  When \rmax\ is 10~km, the initial luminosity 
is large, but the collision time is short. Within 1~Myr, the luminosity 
falls by a factor of $\sim$ 50. After $\sim$ 30~Myr, the relative luminosity 
falls below $10^{-5}$. When \rmax\ is much larger, longer collision times 
allow the luminosity to remain large at later times.  For \rmax\ = 1000~km, 
it takes 30--40~Myr (300--400~Myr) for the relative luminosity to fall 
below $10^{-4}$ ($10^{-5}$).

\section{NUMERICAL MODEL}
\label{sec: model}

To perform numerical calculations of collisional cascades, we use \orch, an 
ensemble of computer codes for the formation and evolution of planetary systems.  
In addition to other algorithms, \orch\ includes a multiannulus coagulation 
code which derives the time evolution of a swarm of solid particles orbiting 
a central object \citep{kb2004a,kb2008,kb2012}. Here, we use the coagulation 
code to follow the evolution of solids at 1~AU around a solar mass star.

\subsection{Numerical Grid}
\label{sec: model-grid}

To provide the closest representation of the analytic model, we conduct 
coagulation calculations within a single annulus with semimajor axis $a$
and width $\delta a$ around a star with mass \mstar\ = 1~\msun. Within
this annulus, there are $M$ mass batches with characteristic mass $m_k$ 
and radius $r_k$ \citep{spaute1991,weth1993,weiden1997b,kl1998}. Batches 
are logarithmically spaced in mass, with mass ratio 
$\delta \equiv m_{k+1} / m_{k}$.  Each mass batch contains $N_k$ particles 
with total mass $M_k$ and average mass $\bar{m}_k = M_k / N_k$. Particle 
numbers $N_k < 10^{15}$ are always integers.  Throughout the calculation, 
the average mass is used to calculate the average physical radius $\bar{r}_k$, 
collision cross-section, collision energy, and other necessary physical 
variables.  As mass is added and removed from each batch, 
the average mass changes \citep{weth1993}.

For any $\delta$, numerical calculations lag the result of an ideal 
calculation with infinite mass resolution (see \S\ref{sec: app-delta}). 
In most cases, 
simulations with $\delta$ = 1.05--1.19 yield better solutions to 
the evolution of the largest objects ($r \gtrsim$ 1~km) than calculations 
with $\delta$ = 1.41--2.00.  Although simulations with $\delta$ = 
1.05--1.10 allow better 
tracking of the gradual reduction in mass of the largest objects during the 
cascade, the evolution of the size distribution and the dust luminosity is
fairly independent of $\delta$.  Thus, we consider a suite of calculations 
with $\delta$ = 1.19 ($ = 2^{1/4}$).

In this suite of calculations, we follow particles with sizes ranging 
from a minimum size \rmin\ = 1~\mum\ to the maximum size \rmax. The 
algorithm for assigning material to the mass bins extends the maximum
size as needed to accommodate the largest particles.  When collisions 
produce objects with radii $r < \rmin$, this material is lost to the grid.

\subsection{Initial Conditions}
\label{sec: model-init}

All calculations begin with a swarm of particles with initial maximum size 
\r0\ and mass density $\rho_p$ = 3~\gcmc.  These particles have initial 
surface density $\Sigma_0$, total mass $M_0$, and horizontal and vertical velocities 
$v_{h,0}$ and $v_{z,0}$ relative to a circular orbit.  The horizontal velocity is related 
to the orbital eccentricity, $e^2$ = 1.6 $(v_h/v_K)^2$, where $v_K$ is the circular 
orbital velocity.  The orbital inclination is ${\rm sin}~i$ = $\sqrt{2} v_z/v_K$.

For the simulations in this paper, we consider two different initial size 
distributions for solid objects. To follow the analytic model as closely as 
possible, one set of calculations begins with a power law size distribution,
$N(r) \propto r^{-q}$ and $q$ = 3.5. To study whether our calculations produce
this equilibrium size distribution, we begin a second set of calculations with
a mono-disperse set of particles.

\subsection{Evolution}
\label{sec: model-evol}

The mass and velocity distributions of the solids evolve in time due to
inelastic collisions, drag forces, and gravitational encounters.  As summarized 
in \citet{kb2004a,kb2008}, we solve a coupled set of coagulation equations which
treats the outcomes of mutual collisions between all particles in all mass bins.
We adopt the particle-in-a-box algorithm, where the physical collision rate is 
$n \sigma v f_g$, $n$ is the number density of objects, $\sigma$ is the geometric 
cross-section, $v$ is the relative velocity, and $f_g$ is the gravitational focusing 
factor \citep{weth1993,kl1998}. The collision algorithm treats collisions in the
dispersion regime -- where relative velocities are large -- and in the shear regime --
where relative velocities are small \citep{kl1998,kb2014}. Within an annulus, 
massive protoplanets on nearly circular orbits are `isolated'; these objects can 
collide with smaller objects but cannot collide with other isolated objects 
\citep{weth1993,kl1998,kb2015a}.

In every time step, our algorithm requires an integral number of collisions 
$N_{kl,int}$ between mass bins $k$ and $l$. For each derived $N_{kl}$, the 
algorithm compares the fractional part to a random number drawn for each
$(k,l)$ pair. When the random number exceeds the fractional part, $N_{kl}$ 
is rounded down; otherwise $N_{kl}$ is rounded up. Comparisons with a
Runge-Kutta code (which allows fractional particles) and tests on systems
with analytic solutions verify this approach \citep[e.g.,][]{kl1998,kb2015a}.

For these simulations, we consider two approaches to collision outcomes. Adopting 
\qdstar\ = constant allows us to make the most direct comparison with analytic
models. Calculations with the more standard $\qdstar(r)$ measure the sensitivity 
of the evolution to changes in \qdstar\ and enable links to other published 
numerical simulations of collisional cascades. For the rocky solids in this study, 
we adopt $Q_b$ = $3 \times 10^7$~erg~g$^{-1}$~cm$^{-\beta_b}$, 
$\beta_b = -0.40$, $Q_g$ = 0.3~erg~g$^{-2}$~cm$^{3-\beta_g}$, and $\beta_g$ = 1.35.
These parameters are broadly consistent with published analytic calculations and 
numerical simulations \citep[e.g.,][]{davis1985,hols1994,love1996,housen1999}.  
At small sizes, they agree with results from laboratory experiments
\citep[e.g.,][]{ryan1999,arakawa2002,giblin2004,burchell2005}.

For each pair of colliding planetesimals, the mass of the merged planetesimal is
\begin{equation}
m = m_1 + m_2 - \mesc ~ ,
\label{eq: msum}
\end{equation}
where the mass of debris ejected in a collision is
\begin{equation}
\mesc =0.5 ~ (m_1 + m_2) \left ( \frac{Q_c}{Q_D^*} \right)^{b_d} ~ .
\label{eq: mej}
\end{equation}
The exponent $b_d$ is a constant of order unity 
\citep[e.g.,][]{davis1985,weth1993,kl1999a,benz1999,obrien2003,koba2010a,lein2012}. 
Here, we consider $b_d$ = 1 and 9/8.

To place the debris in the grid of mass bins, we set the mass of the 
largest collision fragment as 
\begin{equation}
\mmaxd = m_{l,0} ~ \left ( \frac{Q_c}{Q_D^*} \right)^{-b_l} ~ \mesc ~ ,
\label{eq: mlarge}
\end{equation}
where $m_{l,0} \approx$ 0.01--0.5 and $b_l \approx$ 0--1.25
\citep{weth1993,kb2008,koba2010a,weid2010}. When $b_l$ is large, 
catastrophic collisions with $Q_c \gtrsim \qdstar$ crush solids 
into smaller fragments.  Lower mass objects have a 
differential size distribution $N(r) \propto r^{-q_d}$. After placing a 
single object with mass \mmaxd\ in an appropriate bin, we place material 
in successively smaller mass bins until (i) the mass is exhausted or 
(ii) mass is placed in the smallest mass bin. Any material left over 
is removed from the grid.  

For calculations of collisional cascades, we assume that the orbital $e$ and $i$
are constant with time. When we consider how the growth of solids leads to a
collisional cascade, we derive orbital evolution due to collisional damping from 
inelastic collisions and gravitational interactions.  For inelastic and elastic 
collisions, we follow the statistical, Fokker-Planck approaches of \citet{oht1992} 
and \citet{oht2002}, which treat pairwise interactions (e.g., dynamical friction 
and viscous stirring) between all objects.  We also compute long-range stirring 
from distant oligarchs \citep{weiden1989}. 

Our solutions to the evolution equations conserve mass and energy to machine 
accuracy. Typical collisional cascade calculations require a single run on a
system with 16--32 cpus; over the $10^6$ timesteps in a typical 2~Gyr run, 
calculations conserve mass and energy to better than a part in $10^{10}$. 
Planet formation calculations usually require 3--4 runs on 64 cpus. 
Over $10^7 - 10^8$ timesteps, our calculations conserve mass and energy to 
better than a part in $10^7$.

\subsection{Other Approaches}
\label{sec: model-app}

Before describing our calculations, it is useful to place our numerical 
approach in context with other simulations. Starting with \citet{saf1969}, 
kinetic approximations to the particle collision rate and Fokker-Planck 
estimates of velocity evolution have a long history 
\citep[e.g.,][]{banderman1972,green1978,weth1980,weiden1980,green1984,
oht1988,weth1989,spaute1991,barge1991,weth1993,kok1996,weiden1997b,grogan2001,
oht2002,kokubo2006,chambers2008,morby2009b,koba2013,glaschke2014,shannon2015}. 
In addition to our studies \citep[e.g.,][]{kb2002b,kb2004a,kb2008,kb2010}, 
\citet{koba2010a}, \citet{weid2010}, and \citet{koba2014} follow the 
evolution of debris disks using a similar set of algorithms. Aside from
various details of the implementation -- e.g., single annulus 
\citep{koba2010a} or multi-annulus \citep{weid2010} -- these coagulation 
codes are similar to our multi-annulus coagulation code.

Several other techniques adopt the kinetic approximation for collisions but do 
not allow velocity evolution from gravitational interactions between mass bins. 
In addition to following the evolution of particle number in specific mass bins, 
\citet{krivov2006} track bins in semimajor axis and orbital eccentricity
\citep[see also][]{lohne2008,lohne2012,krivov2013}.  In these models, 
radiation pressure modifies the orbital parameters of grains ejected from 
collisions of larger particles \citep[e.g.,][]{burns1979}. As an interesting
variant of the particle-in-a-box model with discrete mass bins and integral
numbers of particles, \citet{gaspar2012a} derive the time evolution of the 
differential number density in a single annulus with collision rates inferred 
from the kinetic approximation. Although this technique does not track 
variations in $a$ and $e$, \citet{gaspar2012a} derive the impact of radiation
pressure as in \citet{krivov2006}.

A third group of codes employs \nbody\ techniques to evolve a collection of 
massless tracer particles \citep[e.g.,][]{grigor2007,booth2009,stark2009,
jackson2012,nesvold2013,kral2013}. 
These codes solve the equations of motion for a swarm of massless tracers 
responding to the gravity and radiation pressure from a central star and the 
gravity of a few other massive planets orbiting the central star. These tracers 
serve as proxies for a mass distribution of particles; the surface density of 
tracers is input for estimating the collision rates between particles of different 
masses. Although the algorithms in each implementation are different, aspects 
of these codes are similar to \nbody-based 
\citep[e.g.,][]{raymond2011,raymond2012,lev2012} 
and hybrid \citep{weiden1997b,bk2006,bk2011a,bk2011b,bk2013,glaschke2014} 
codes for calculating the formation and evolution of planetary systems.

In this paper, we compare results from single annulus coagulation calculations 
to predictions of the analytic model \citep[e.g.,][]{wyatt2011} and (where 
possible) other numerical calculations using the kinetic approximation 
\citep[e.g.,][]{krivov2006,koba2010a,gaspar2012a}.  Exploring outcomes as 
a function of \qdstar, $b_d$, $m_{l,0}$, $b_l$ and $q_d$ allow us to establish 
the sensitivity of observable quantities on unknown parameters.  These results 
serve as a foundation for future multi-annulus calculations with tracer 
particles, which will yield comparisons with \nbody\ approaches using tracer 
super-particles \citep[e.g.,][]{grigor2007,stark2009,nesvold2013,kral2013}.

\section{NUMERICAL CALCULATIONS}
\label{sec: calcs}

\subsection{Collisional Cascades}
\label{sec: calcs-casc}

\subsubsection{A Standard Model with Constant \qdstar}
\label{sec: calcs-casc-std}

To explore the long-term evolution of collisional cascades, we consider
a `standard' calculation of solids orbiting a solar-type star. Material
with maximum initial radius \r0\ = 100~km, \rmin\ = 1~\mum, and $\rho$ = 
3~\gcmc\ orbits with $e$ = 0.1 and $i$ = 0.05 ($\approx$ 3\deg) inside an 
annulus with $a$ = 1~AU and $\delta a$ = 0.2~AU. The initial size distribution 
is either a power law with $q$ = 3.5 or a mono-disperse ensemble with $r = \r0$. 
All objects have a constant \qdstar\ = $6 \times 10^7$~\ergg. To assign
collisional debris to mass bins, we set $q_d$ = 3.5 and examine results for
various combinations of $b_d$, $m_{l,0}$, and $b_l$. All calculations 
ignore Poynting-Robertson drag.

\begin{figure} 
\includegraphics[width=6.5in]{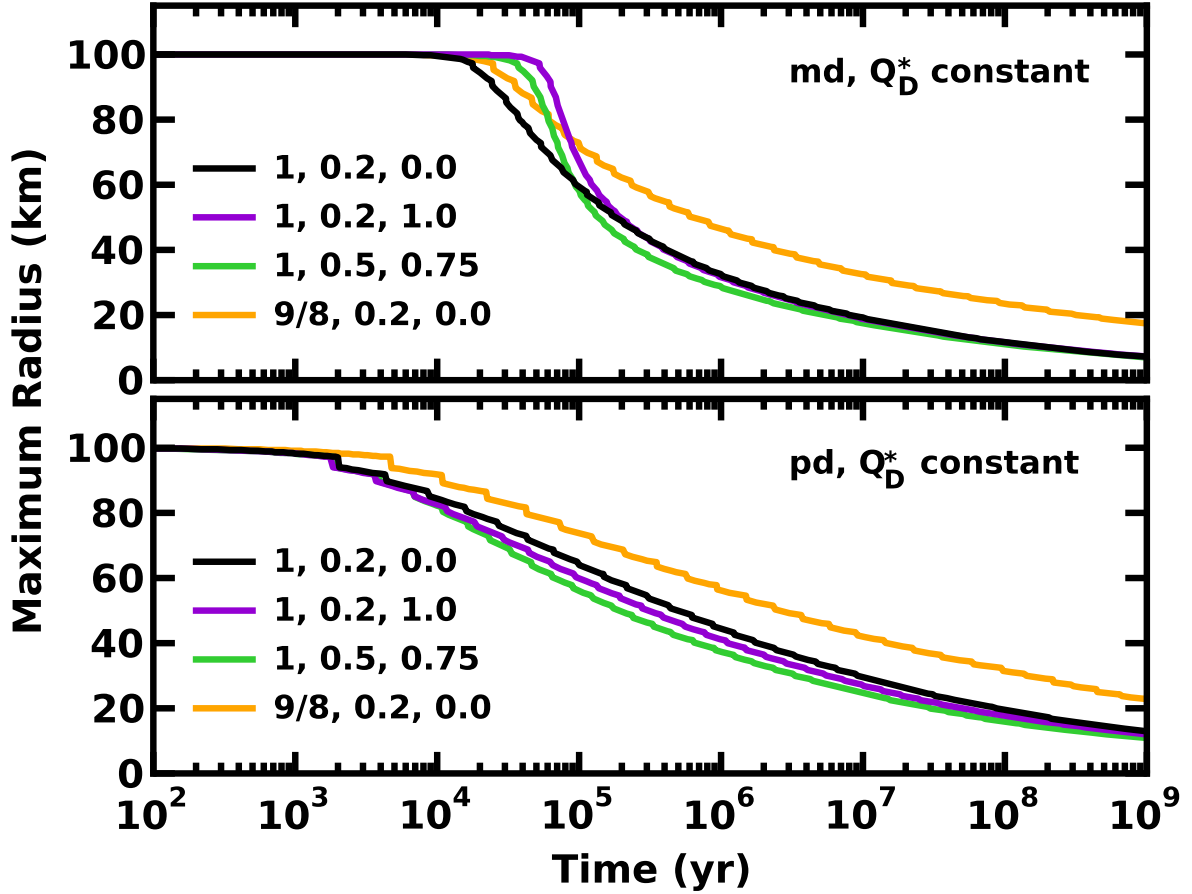}
\vskip 3ex
\caption{
Time evolution of \rmax\ in collisional cascades with 
$e$ = 0.1, \qdstar\ = $6 \times 10^7$~\ergg\ and various 
initial conditions and input parameters.  The legend 
indicates $b_d$, $m_{l,0}$, and $b_l$ for each curve.
{\it Upper panel:} results for calculations with a single initial
particle size (`md'; \r0\ = 100~km); after $10^4 - 10^5$~yr, 
the maximum
particle size rapidly evolves from 100~km to 10--20~km.
{\it Lower panel:} results for calculations with a power law 
initial size distribution having $q_0$ = 3.5 from 1~\mum\ to 
100~km (`pd'); the largest particles gradually diminish in 
size from 100~km to 20--30~km.
\label{fig: rmax1}
}
\end{figure}

For all calculations with constant $e$ and $i$, the size of the largest 
object gradually declines
with time (Fig.~\ref{fig: rmax1}).  The timing of this decline depends on 
the initial size distribution.  When the swarm is initially mono-disperse
(upper panel),
{\it all} collisions are catastrophic. As these fairly infrequent collisions 
produce smaller objects, cratering begins to chip away at the mass in all 
of the large objects.  When the initial size distribution of the swarm is 
a power-law (lower panel), cratering collisions begin immediately. Thus, 
\rmax\ initially declines more rapidly for calculations with an initial 
power law size distribution.  

Once \rmax\ begins to decline, the evolution is more dramatic in originally 
mono-disperse swarms. During the first 0.01--0.1~Myr of evolution, these 
swarms produce fewer particles with sizes smaller than 1~\mum\ which are 
lost to the grid. The swarms then have relatively more mass than swarms 
with an initial power law size distribution. More massive swarms have 
higher collision rates, resulting in a faster decline of \rmax\ with time.

The time evolution of \rmax\ also depends on the way we distribute debris 
into the mass bins. When $b_d$ = 1 (9/8), cratering (catastrophic) 
collisions with $Q_c \lesssim$ 1 ($Q_c \gtrsim$ 1) produce relatively
more debris. For swarms with a power law size distribution from 1~\mum\ to
100~km, cratering collisions produce more debris than catastrophic 
collisions. Thus, calculations with $b_d$ = 1 reduce \rmax\ more rapidly
than those with $b_d$ = 9/8. For any $b_d$, the mass of the largest object 
in the debris is smaller in calculations with $b_l > 0$ than in those with
$b_l$ = 0. Thus, \rmax\ declines more rapidly when $b_l$ is larger than
zero.

For any combination of $b_d$, $m_{l,0}$, and $b_l$, the dramatic decline 
in \rmax\ roughly follows a power-law. From $\sim$ 1~Myr to 1~Gyr, 
$\rmax \propto t^{-n}$ with $n$ = 0.1--0.2.  After 100~Myr, particles with 
\rmax\ = 100~km have lost from 97\% (power law size distribution, $b_l$ = 0) 
to 99.9\% (mono-disperse, $b_l$ = 1) of their initial mass. 
 
\begin{figure} 
\includegraphics[width=6.5in]{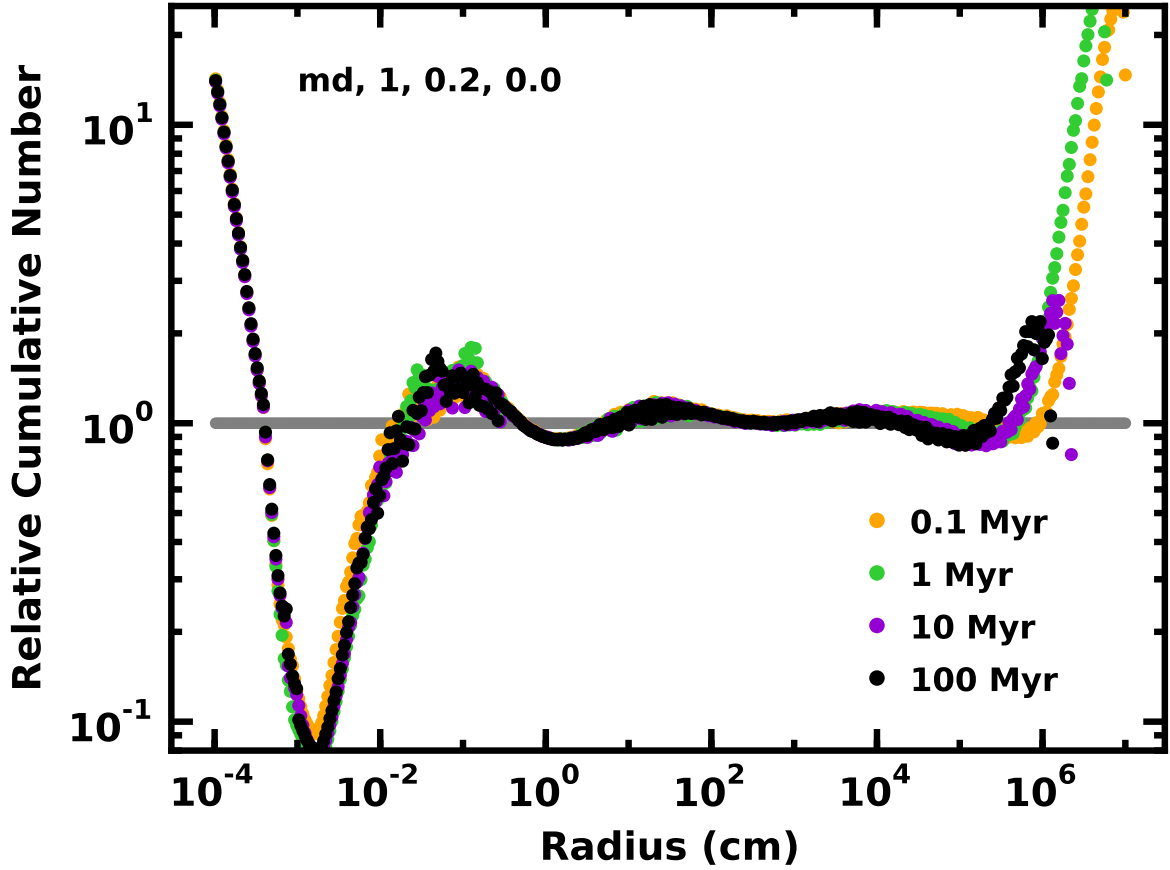}
\vskip 3ex
\caption{
Time evolution of the relative cumulative size distribution 
$N(>r) / r^{-2.5}$ for a collisional cascade starting with a 
mono-disperse (`md') set of planetesimals with \r0\ = 100~km, \qdstar\ = 
$6 \times 10^7$~\ergg, $b_d$ = 1, $m_{l,0}$ = 0.2, and $b_l$ = 0
(as indicated in the upper left corner for this and subsequent plots).
The legend in the lower right corner indicates the evolution time 
in Myr. The grey line shows the predicted relative size distribution 
when $N(>r) \propto r^{-2.5}$. 
Aside from the gradual loss of large objects ($ r \gtrsim$ 1~km) due 
to collisional erosion, the cumulative relative size distribution 
is nearly independent of time.
\label{fig: sd1}
}
\end{figure}

\begin{figure} 
\includegraphics[width=6.5in]{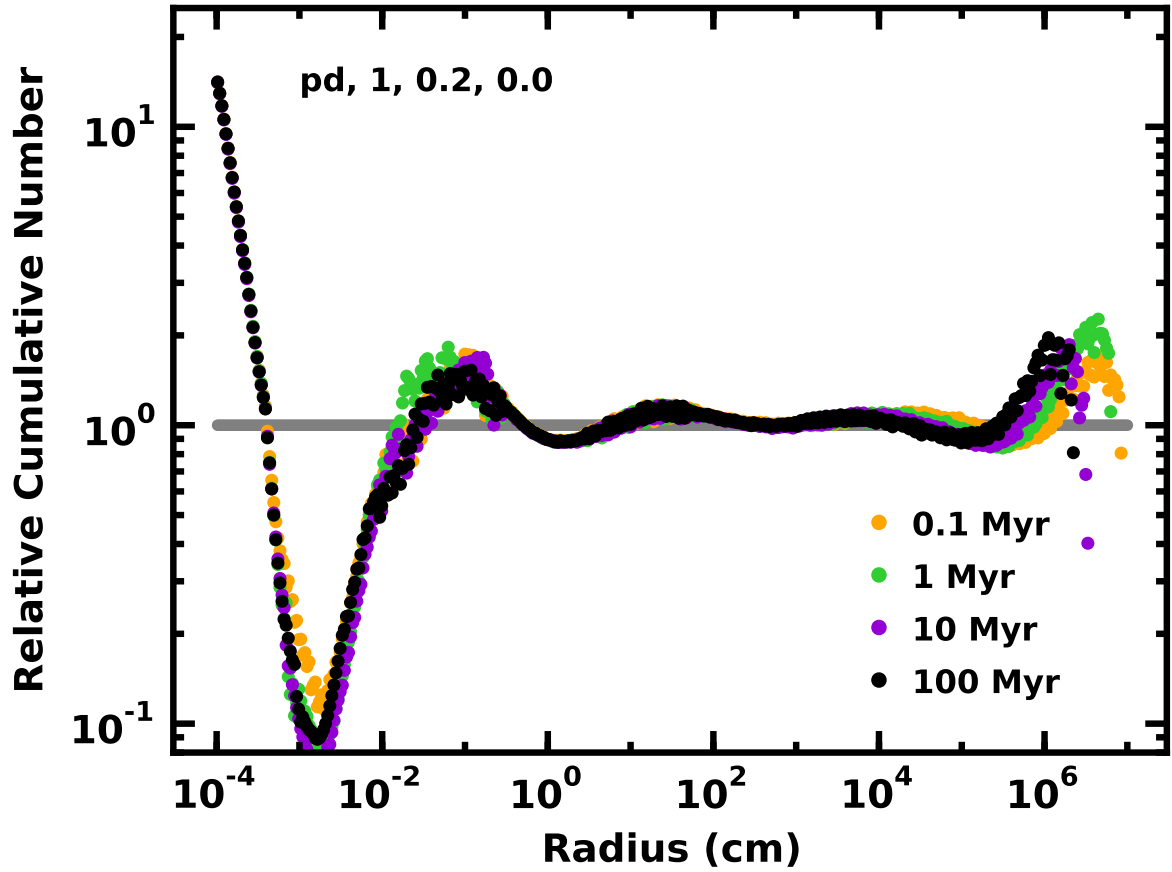}
\vskip 3ex
\caption{
As in Fig.~\ref{fig: sd1} for a cascade with an initial power-law
size distribution (`pd', $q_0$ = 3.5) from 1~\mum\ to 100~km. 
Compared to calculations starting with a mono-disperse set of 
large objects, the amplitude of the wave at $r \approx$ 10~km 
is smaller.
\label{fig: sd2}
}
\end{figure}

To describe the evolution of the size distribution, we focus first on 
results for $b_d$ = 1 and $b_l$ = 0 (Fig.~\ref{fig: sd1}--\ref{fig: sd2}). 
Based on previous analyses, we expect a roughly power law size distribution 
with $q$ = 3.5 from 1~\mum\ to \rmax. To minimize shot noise, we derive 
the cumulative size distribution, $N(>r)$, where the predicted slope is 
$q_c$ = 2.5.  To illustrate deviations from this prediction, we examine 
the relative cumulative size distribution, $N(>r) / r^{-2.5}$. The expected
power law slope of this relative size distribution is then $q_r = q_c - 2.5$
= 0.

When swarms begin with a power law size distribution, collisional evolution
rapidly transforms the system at all sizes 
\citep[Fig.~\ref{fig: sd1}, see also][]{campo1994}.
At the smallest sizes, we assume that
radiation pressure ejects very small grains with $r <$ 1~\mum\ from the grid. 
Compared to a model which includes these very small particles, grains with 
$r \approx$ 1--10~\mum\ experience {\it fewer} destructive collisions, 
leading to a clear excess of these particles. With an excess of 
1--10~\mum\ grains, particles with $r \approx$ 10--100~\mum\ experience 
{\it more} destructive collisions, producing a deficit of these particles
\citep[e.g.,][]{obrien2003,kral2013}. 

Collisional evolution also leads to excesses and deficits at $r \gtrsim$ 1~km.
For $r \approx$ 1~km to \rmax, destructive collisions gradually shift mass into
lower mass bins. Without a corresponding addition in mass from bins with 
$r \gtrsim$ \rmax, there is a deficit in the mass distribution at $r \approx$ 
\rmax.  This deficit enables an excess at smaller sizes, $r \approx$ 10~km.

Among intermediate sized particles (100~\mum\ to 1--3~km), there is a small
wave about the expected power law. This wave is a function of the deficits 
and excesses of particles at smaller and larger sizes 
\citep[see also \S\ref{sec: app-waves};][]{campo1994,durda1998,obrien2003}.  
As cratering reduces the sizes of the largest particles, the peak of the 
wave at 5--20~km gradually shifts to smaller sizes.  Over 100~Myr, the shape 
of the wavy power law remains roughly constant at smaller sizes.

For swarms starting with a mono-disperse size distribution, the system 
rapidly evolves to the `standard' size distribution (Fig.~\ref{fig: sd2}). 
After $\sim$ 0.1~Myr, the relative size distributions of swarms with
initially mono-disperse and power law size distributions are nearly 
identical from 1~\mum\ to 10~km.  At 0.1--100~Myr, there is a clear 
excess of 10--100~km objects which is not visible in calculations starting
with a power law size distribution of small objects. In mono-disperse
calculations, infrequent collisions among the largest objects prevents
reaching the equilibrium size distribution.

Overall, all approaches to fragmentation produce wavy structures in 
the relative cumulative size distribution (Fig.~\ref{fig: sd3}). 
However, the amplitudes and positions of the waves depend on $b_d$ and
$b_l$.  When $b_d$ = 9/8 (instead of 1), the deficits at 10~\mum\ and 
10--30~km are pronounced but somewhat smaller. Adopting $b_l$ = 
0.75--1.0 reduces the amplitude of the wave at small sizes but 
accentuates it at larger sizes. 

To conclude this discussion of the standard model, Fig.~\ref{fig: lum1}
shows the time evolution of the dust luminosity $L_d$. When collisional
cascades begin with a mono-disperse swarm of large objects, 
$L_d \approx$ 0. As collisions produce small particles, $L_d$ rises.
In models with $b_l > 0$, collisions produce and eject small particles 
more rapidly, leading to a more rapid rise in $L_d$ but a smaller peak
$L_d$. Although the dust luminosity at late times depends on $b_d$, 
$m_{l,0}$, and $b_l$, all models follow $L_d \propto t^{-n}$ with 
$n \approx$ 1.1--1.2. Thus, the decline is slightly steeper than 
predicted from the analytic model.

When the cascade starts with a power-law size distribution, the initial
$L_d$ is much larger. As the size distribution establishes the standard
wavy power law, $L_d$ varies slowly before settling in on some constant
value which depends on $b_d$, $m_{l,0}$, and $b_l$. At late times, 
$L \propto t^{-n}$ with $n \approx$ 1.1--1.2. Although the magnitude of 
$L_d$ at $t \gtrsim$ 1~Myr is sensitive to the fragmentation parameters, 
$L_d$ is independent of the initial size distribution.

\subsubsection{Evolution with Different \qdstar}
\label{sec: calcs-casc-qstar}

All analytic models for collisional cascades predict that the lifetime
depends on the ratio $v^2 / \qdstar$ \citep[e.g., eq.~\ref{eq: alpha}; see 
also][] {wyatt2007a,wyatt2007b,koba2010a,wyatt2011}. In our numerical approach, 
collisions between pairs of equal mass particles with $v^2 /8 \qdstar \gtrsim$ 1 
convert more than half of the combined mass into debris. To test the 
analytic model, we consider a suite of calculations with constant $v$,
\qdstar\ = $6 \times 10^7 - 6 \times 10^{10}$ \ergg, and $v^2 /8 \qdstar$ 
= 0.2--200. For comparison, eq.~\ref{eq: qd} with standard parameters for
rocky objects predicts $\qdstar\ \approx 10^6$~\ergg\ for 0.1~km particles. 

Fig.~\ref{fig: rmax2} illustrates the evolution of \rmax\ for calculations
with $b_d$ = 1, $m_{l,0}$ = 0.2, $b_l$ = 1, and either a mono-disperse 
(upper panel) or a power-law (lower panel) initial size distribution. 
When $v^2 /8 \qdstar \lesssim$ 1, the largest objects grow to sizes of 
500--1000~km in 10--100~Myr. For larger ratios, $v^2 /8 \qdstar \approx$ 
1--200, catastrophic and cratering collisions gradually reduce the size
of the largest objects. 

The time scale for the largest objects to lose mass depends on \qdstar\ and
the initial size distribution. In this set of calculations, the loss
time scales inversely with \qdstar\ for $v^2 /8 \qdstar \approx$ 5--200.
For smaller $v^2 /8 \qdstar$, the evolution time is fairly independent
of \qdstar. For any \qdstar, swarms with an initial power-law size
distribution begin to lose mass more rapidly than those with an initial 
mono-disperse set of large objects. At late times, however, originally
mono-disperse swarms lose more mass (see \S\ref{sec: calcs-casc-std}).

Despite the different loss rates, all of these calculations yield nearly
identical relative cumulative size distributions after 10--100 collision 
times. Systems with large \qdstar\ tend to have somewhat smaller waves than 
those with smaller \qdstar. Otherwise, the level and shape of the relative 
cumulative size distributions simply scale with the collision time.

Fig.~\ref{fig: lum2} demonstrates that the evolution of the dust luminosity 
also scales with \qdstar. When swarms begin with a power law size distribution,
they have substantial $L_d$. Destructive collisions among mono-disperse swarms 
gradually build a comparable $L_d$ over a few collision times. For this suite
of calculations, the peak $L_d$ is fairly independent of the ratio 
$v^2 /8 \qdstar$. 

At late times, all systems follow power law declines in $L_d$ with time.
Setting $L_d \propto t^{-n}$, this suite of calculations has $n$ = 1.1--1.2.
There is some tendency for larger $n$ at early times and smaller $n$ at late
times, but our sample size is too small to verify this behavior.

Throughout this evolution, systems with longer collision times have larger $L_d$. 
For $v^2 /8 \qdstar \gtrsim$ 10--20, $L_d \propto (v^2 /8 \qdstar)^{-1/2}$.
When $v^2 /8 \qdstar \approx$ 1--5, $L_d \propto (v^2 /8 \qdstar)^{-1}$.
Once $v^2 /8 \qdstar \lesssim$ 1, most of the solid mass ends up in large 
objects with negligible surface area. The dust luminosity is then close to
zero.

When \qdstar\ is independent of radius, all collisions either allow objects
to grow (large \qdstar) or produce debris (small \qdstar). In real systems,
\qdstar\ is a function of particle size with a distinct minimum at 
intermediate sizes $r \approx$ 10--100~m. To illustrate the evolution of 
\rmax\ in a real system, we consider standard calculations using the 
\qdstar\ relation in eq.~\ref{eq: qd}.

Fig.~\ref{fig: rmax3} shows results of calculations for initially 
mono-disperse and power-law
size distributions with $b_d$ = 1, $m_{l,0}$ = 0.2, $b_l$ = 1, and 
\r0\ = 100, 300, 500, and 1000~km. In swarms with a starting \r0\ = 
100~km and 300~km, the largest objects slowly lose mass with time. 
Systems with smaller \rmax\ evolve more rapidly and lose $\sim$
97\% (\r0\ = 100~km) to 70\% (\r0\ = 300~km) of their original
mass over 1~Gyr. As in \S\ref{sec: calcs-casc-std}, swarms with
power-law size distributions evolve more rapidly than those 
with mono-disperse size distributions at early times. At
later times, the mono-disperse swarms evolve more rapidly.

When $r$ is larger than 300~km, collisions among equal mass objects 
produce larger merged objects.  The evolution is then very sensitive 
to the starting size distribution of the swarm. Among originally 
mono-disperse swarms, the largest objects reach sizes of 4500--6000~km
fairly independently of the initial \rmax. In swarms with \r0\ = 500~km 
and a power law size distribution, cratering collisions are almost 
frequent enough to prevent growth.  After $\sim$ 10~Myr, growth overcomes 
cratering, but the largest objects only reach sizes of 1000--1500~km 
instead of 4000--5000~km. For swarms with a larger \r0, cratering is 
much less important (Fig.~\ref{fig: rmax3}, purple curves).

These results follow from eq.~\ref{eq: rc-max}. Among mono-disperse 
swarms, we expect growth when \r0\ exceeds 300~km; our numerical
results confirm this estimate. When swarms start with a power-law
size distribution, cratering collisions should prevent growth at some
larger \r0. The numerical calculations suggest this limit is roughly
400~km.

Compared to `normal' collisional cascades, systems where particles 
grow have much smaller $L_d$ (Fig.~\ref{fig: lum3}). In modo-disperse
swarms, the time scale for $L_d$ to rise from zero scales with \r0. 
When the largest objects cannot grow, collisions distribute debris 
among smaller objects.  Peak $L_d$ is then fairly independent of 
\r0\ (e.g., Fig.~\ref{fig: lum3}, thick orange and magenta lines).
In ensembles of growing large objects, collisions gradually concentrate 
more and more mass into larger and larger objects. Growth limits debris
production. Peak $L_d$ is then much smaller (e.g., Fig.~\ref{fig: lum3}, 
thick green and violet lines). At late times, the largest objects sweep
up {\it all} of the remaining small particles; $L_d$ then drops 
dramatically.

Systems with initial power-law size distributions exhibit similar behavior
(Fig.~\ref{fig: lum3}, thin lines). All swarms begin with large $L_d$.
As collisions destroy the largest objects (\r0\ = 100~km and 300~km), 
$L_d$ slowly declines with time. At late times ($t \gtrsim$ 10~Myr), the 
evolution of $L_d$ is independent of the initial size distribution. When
the largest particles grow with time (\r0\ = 500~km and 1000~km), all 
collisions produce some debris; continuous debris production maintains 
a slow decline in $L_d$. Eventually growth dominates debris production;
$L_d$ then plummets. For \r0\ = 1000~km, the substantial fall in $L_d$ 
begins at $\sim$ 300~Myr (somewhat later than the drop at $\sim$ 100~Myr 
for a mono-disperse system). When \r0\ = 500~km, a close balance between
growth and debris production maintains large $L_d$ past 1--2~Gyr. 
After a few more Gyr, the largest objects sweep up debris and $L_d$
rapidly falls to zero.

Despite the different evolution in \rmax\ and $L_d$, calculations with 
$\qdstar(r)$ still produce wavy size distributions about a standard
power-law (Fig.~\ref{fig: sd4}). For the fragmentation parameters
adopted here, the expected power law has slope $q_c \approx$ 2.68 
(\S\ref{sec: an-mod-rmax}). The relative cumulative size distribution 
is then
$n (>r) / r^{-2.68}$. Independent of the initial size distribution, 
$b_d$, $m_{l,0}$, and $b_l$, model results for $r \lesssim$ 0.1~km 
follow this power law fairly closely, with the usual large excess of 
small particles and small amplitude waves at 0.01~cm to 0.1~km.

At large sizes ($r \gtrsim$ 0.1~km), model size distributions have a
much shallower slope \citep[see also \S\ref{sec: app-waves};][]{wyatt2011}. 
For ensembles with \r0\ = 100~km, cumulative size distributions at 
10--100~Myr roughly follow a power-law from 0.1~km to 30--50~km with 
$n(> r) \propto r^{-q_c}$ and $q_c \approx$ 1.7--1.9 ($q_r \approx$ 
$-$1.0 to $-$0.8).  Oscillations about this power law have maxima at 
1--3~km and minima at 10--40~km.
 
\subsubsection{Evolution with Different $q_d$}
\label{sec: calcs-casc-qd}

In Figs.~\ref{fig: rmax1}--\ref{fig: lum1}, the evolution of the 
cascade clearly depends on the algorithms for deriving the amount
of debris and for placing this debris in mass bins. To examine 
how the size distribution of the debris impacts the evolution, we 
consider calculations with a mono-disperse initial size distribution,
$b_d$ = 1, $m_{l,0}$ = 0.2, $b_l$ = 1, and $q_d$ = 3.0, 3.5, 4.0, 
4.5, and 5.0.  When $q_d$ is small (large), most of the debris 
lands in bins with smaller (larger) average mass.  Based on results 
from the standard model, we expect the pace of the cascade to depend 
on $q_d$.

Fig.~\ref{fig: rmax4} shows the time evolution of \rmax\ as a function
of $q_d$. For an initially mono-disperse size distribution, catastrophic 
collisions between the largest objects place debris throughout the mass 
grid. When $q_d \approx$ 3, debris is concentrated among the smallest 
particles; collisions remove more mass from the grid. When $q_d \approx$
5, debris is concentrated among the largest particles with little mass 
lost from the grid.  Concentrating more debris among large particles 
enables a larger rate of mass loss from the largest objects. Thus, 
\rmax\ declines more rapidly when $q_d$ = 5 than when $q_d$ = 3. 

By the end of the calculation at 1~Gyr, catastrophic and cratering collisions
substantially reduce the mass in the largest objects. When $q_d$ = 3 (5), 
\rmax\ = 8~km (6~km). Most of this reduction occurs during the first 1~Myr 
of the evolution. Once most of the mass in the grid has been converted into 
1~\mum\ or smaller particles which are lost to the grid, the time variation 
of \rmax\ is slow.

For each of these calculations, it takes roughly 0.1~Myr to reach the `standard'
wavy power-law size distribution. By 1 Myr, calculations with different $q_d$
establish various wavy patterns at small particle sizes (Fig.~\ref{fig: sd5}). 
These patterns remain fixed for the rest of the evolution.  When $q_d$ = 3.5, 
there is a characteristic pattern of a large-amplitude wave at 1--100~\mum, 
several waves with diminishing amplitude at larger sizes, and a final wave of
modest amplitude at $r \lesssim$ \rmax\ \citep[see also][]{campo1994,
durda1998,obrien2003,kral2013}. 
For other values of $q_d$, the pattern for $r \gtrsim$ 0.1--1~cm is similar. 
At smaller sizes, however, there are a set of waves with smaller amplitude and
shorter wavelength. Although the amplitudes of these waves appear to depend on 
$q_d$, the wavelengths seem independent of $q_d$. 

Despite these curious differences in the size distribution at small sizes, the
mass in the grid is a simple function of $q_d$. At a fixed time, cascades with 
larger $q_d$ have less mass throughout the grid than cascades with smaller $q_d$.
The origin of this result is clear. When $q_d$ is small (large), collisions place 
more mass in smaller (larger) mass bins. Smaller particles remove less mass from
the largest particles, which contain most of the mass in the grid. By the end of
the calculations at 1~Gyr, cascades with $q_d$ = 3 have 35\% to 40\% more mass
than cascades with $q_d$ = 5.

These calculations also reveal interesting differences in the evolution of the
relative dust luminosity $L_d$ as a function of $q_d$ (Fig.~\ref{fig: lum4}). At 
the onset of the cascade, infrequent and random collisions (e.g., shot noise) set
the population of the small particles which contain most of the surface area. Thus, 
the timing of the abrupt rises in $L_d$ is random and fairly independent of $q$. 
During this phase of evolution, however, cascades with small $q_d$ lose more mass 
from the grid than cascades with large $q_d$. Once cascades establish the `standard'
size distribution, those with smaller $q_d$ have less mass than those with larger 
$q_d$. Thus, the peak $L_d$ is smaller in systems with smaller $q_d$.  This 
difference is substantial: cascades with $q_d$ = 5 are 2.5 times brighter at peak 
$L_d$ than cascades with $q_d$ = 3.

The timing of maximum $L_d$ also depends on $q_d$. When $q_d$ is large, erosion
of the largest particles is more rapid (Fig.~\ref{fig: rmax2}). More rapid 
erosion tends to produce a larger ensemble of small particles. Thus, cascades 
with large $q_d$ reach maximum $L_d$ earlier than cascades with small $q_d$. 
This conclusion is independent of the timing of the first rise in $L_d$. In
Fig.~\ref{fig: lum2}, the calculation with $q_d$ = 4 rises before those with
$q_d$ = 4.5 or 5, but still reaches a smaller maximum $L_d$. 

At late times, $L_d$ also correlates with $q_d$. Systems with larger $q_d$ have
larger $L_d$ at late times. Although somewhat counterintuitive, this outcome
has a simple physical explanation. When $q_d$ is large, more erosion of the 
largest particles yields a larger maximum $L_d$ with a somewhat smaller total
mass. Cascades with smaller masses and larger $q_d$ evolve more slowly (due to
fewer collisions) and retain debris for somewhat longer (due to larger $q_d$).
Thus, these systems tend to maintain their larger $L_d$ at later times.

In all calculations, $L_d$ declines somewhat more rapidly than predicted by
the analytic model. Our results follow $L_d \propto t^{-n}$ with $n$ = 
1.1--1.2 instead of the analytic $n$ = 1. The slow decline of \rmax\ with
time is responsible for this difference. When \rmax\ is fixed as in the
analytic model, $n$ must be unity as outlined in \S2. When \rmax\ declines
with time, the collision time becomes progressively shorter with time 
(eq.~\ref{eq: tc1}). As the collision time becomes shorter, the luminosity
declines more rapidly. 

\subsubsection{Model Comparisons}
\label{sec: calcs-casc-comps}

\begin{figure} 
\includegraphics[width=6.5in]{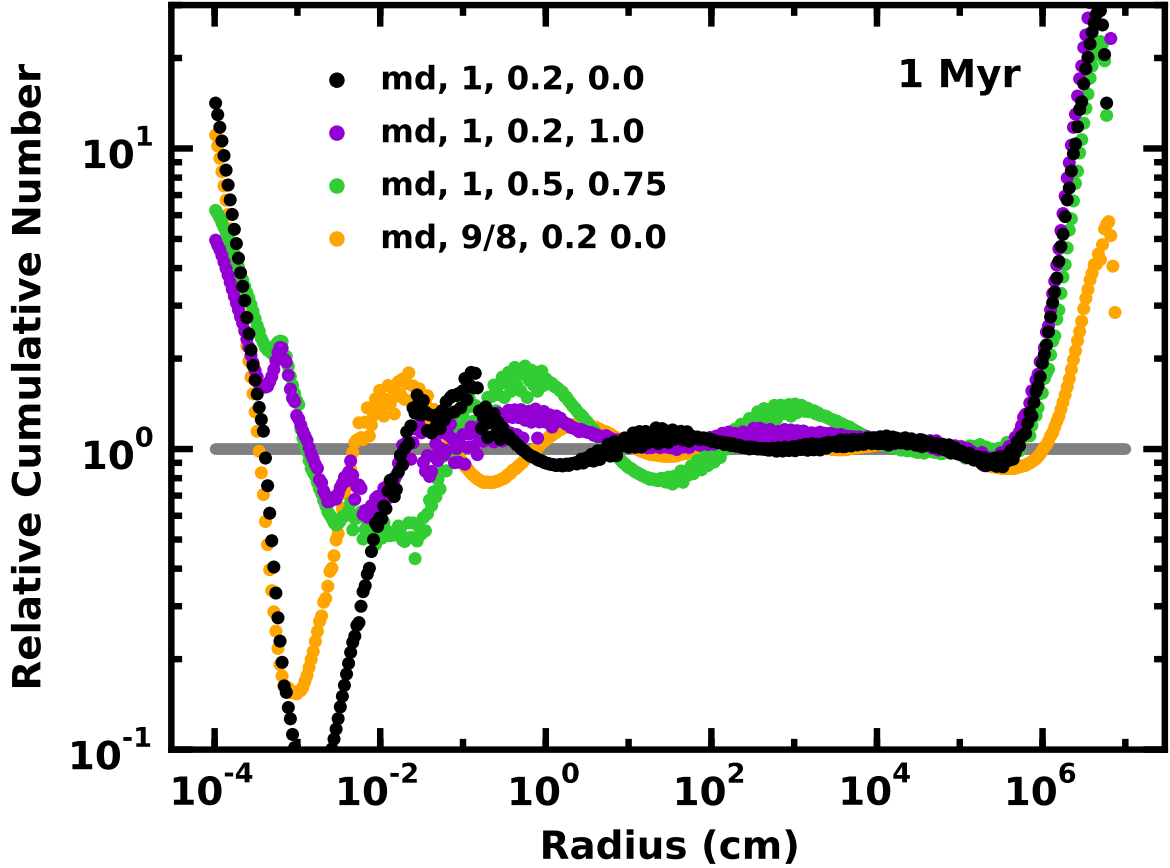}
\vskip 3ex
\caption{
Comparison of $n_{rel} (>r)$ at 1~Myr for calculations with a mono-disperse set 
of large planetesimals (\r0\ = 100~km) and \qdstar = $6 \times 10^7$~\ergg.
The legend indicates the type of initial size distribution and values for the 
input parameters $b_d$, $m_{l,0}$, and $b_l$. 
Relative to a power-law size distribution with $N(>r) \propto r^{-2.5}$,
all curves have a similar morphology at 1 Myr:
an excess of 1--10~\mum\ particles, 
a deficit of 10--100~\mum\ particles, 
a wavy behavior from 100~\mum\ to 10~km, 
an excess of 10--30~km particles, and
a deficit of 30--100~km particles.
Although the fine details of the shape depend on the input parameters,
the shape for each set of input parameters is independent of time
(see also Figs.~\ref{fig: sd1}--\ref{fig: sd2}).
\label{fig: sd3}
}
\end{figure}

\begin{figure} 
\includegraphics[width=6.5in]{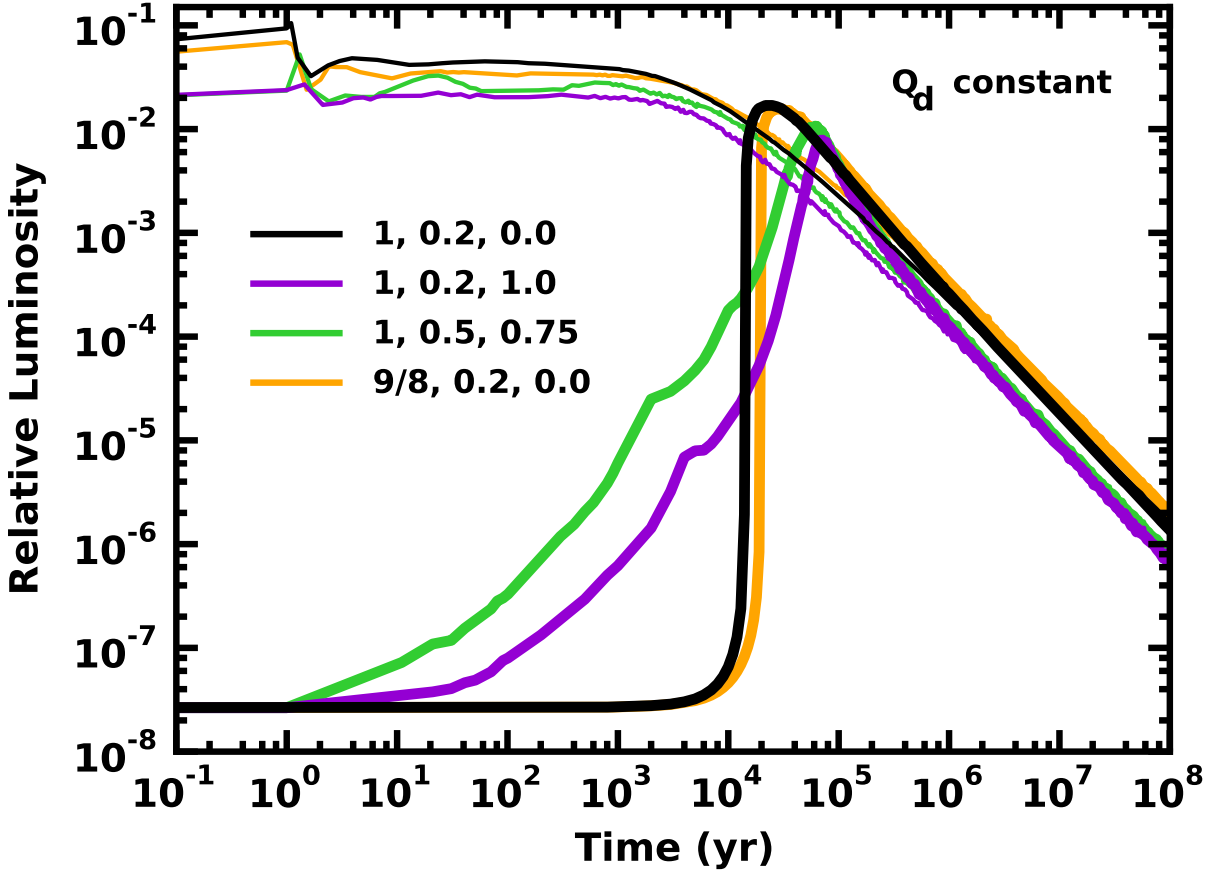}
\vskip 3ex
\caption{
Evolution of the relative dust luminosity \ldstar\ for the 
calculations in Fig.~\ref{fig: rmax1}.
Starting from a mono-disperse ensemble of planetesimals (thick 
lines), \ldstar\ rises rapidly from roughly zero to $\sim 10^{-2}$ 
and then declines roughly linearly with time.
When calculations begin with a power-law size distribution of 
planetesimals (thin lines), \ldstar\ fluctuates about a constant 
as the size distribution reaches an equilibrium and then declines 
roughly linearly with time. Although the evolution of \ldstar\ at 
late times depends on $b_d$, $m_{l,0}$, and $b_l$ (as indicated in
the legend), it is insensitive to the shape of the initial size 
distribution.
\label{fig: lum1}
}
\end{figure}

\begin{figure} 
\includegraphics[width=6.5in]{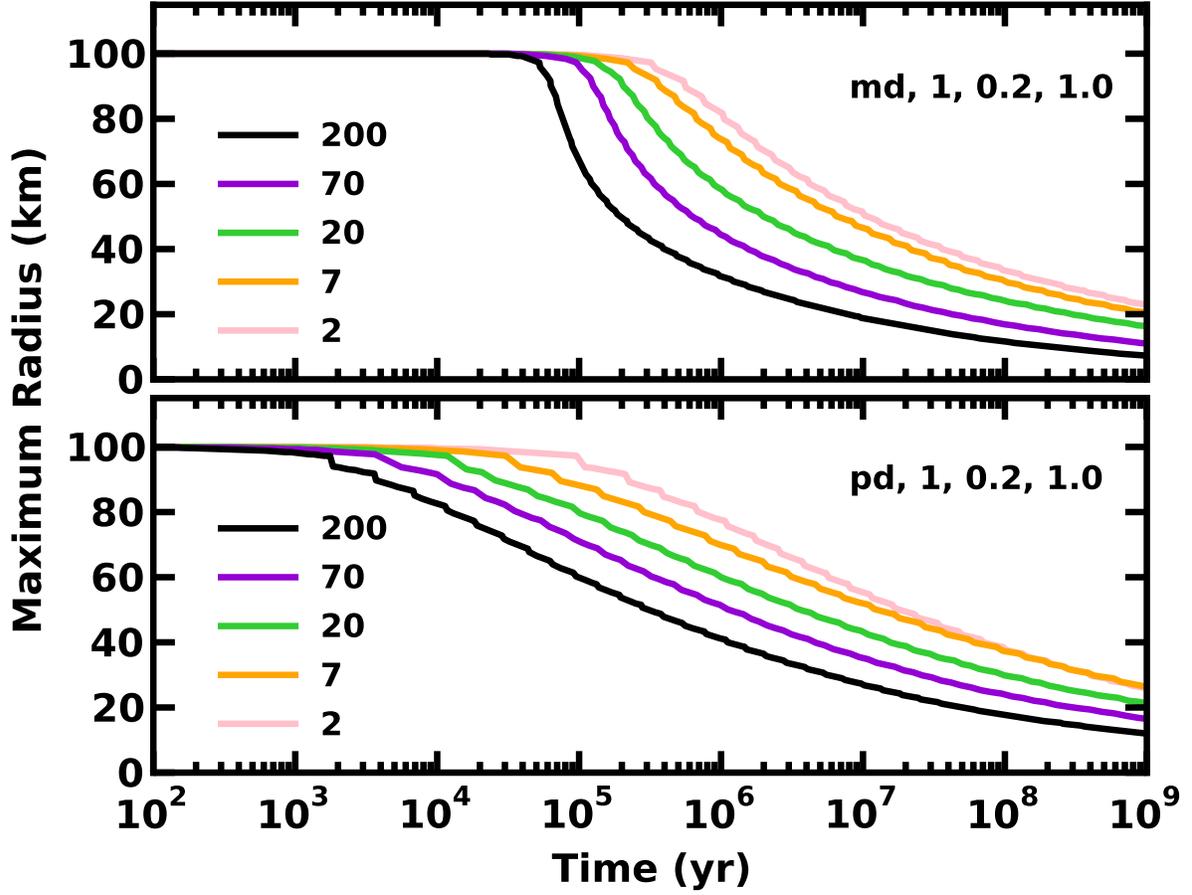}
\vskip 3ex
\caption{
Evolution of \rmax\ for cascade models with constant \qdstar, $b_d$ = 
1, $m_{l,0}$ = 0.2, $b_l$ = 1.0, initially mono-disperse (`md', upper 
panel) or power-law (`pd', lower panel) size distributions, and various 
ratios $v^2 / 8 \qdstar$ as indicated in the legend on the left side of
each panel.  Systems with 
larger $v^2 / 8 \qdstar$ have shorter collision time scales. 
\label{fig: rmax2}
}
\end{figure}

\begin{figure} 
\includegraphics[width=6.5in]{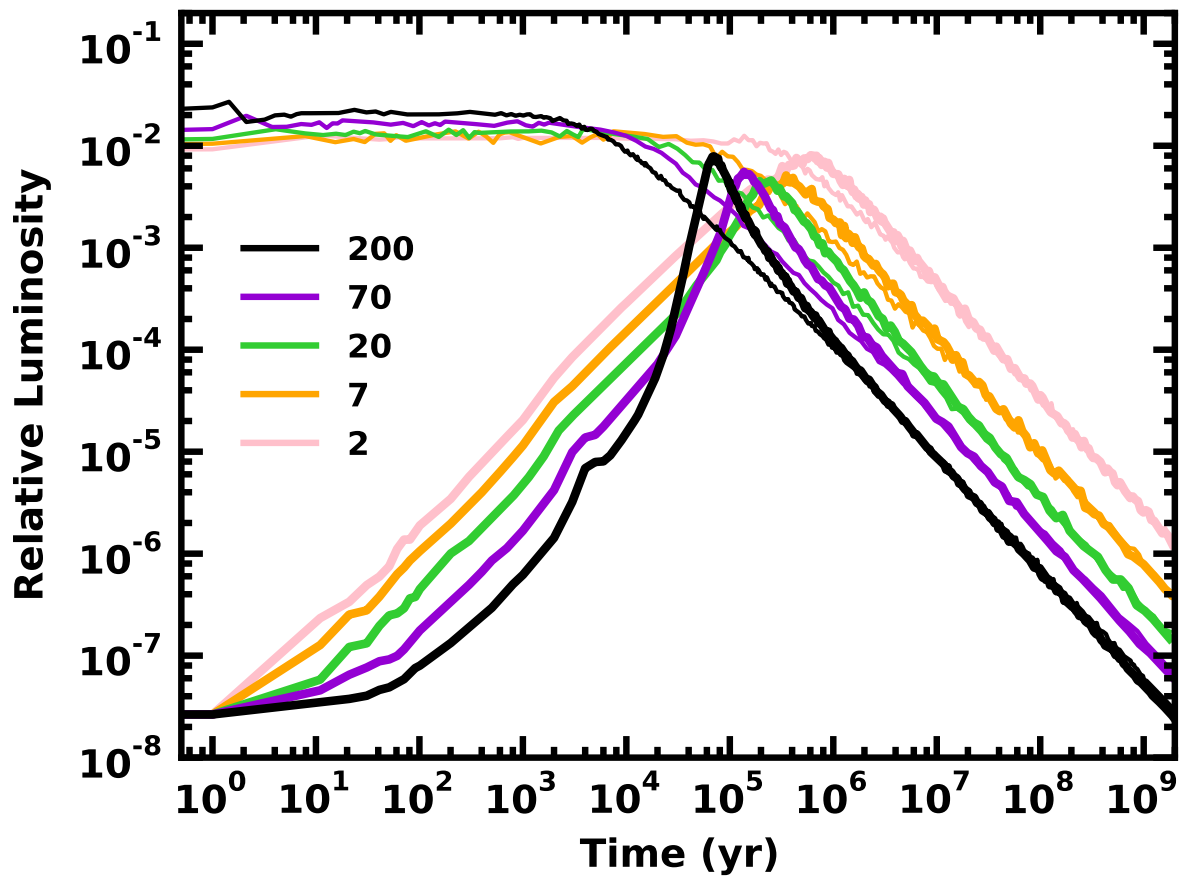}
\vskip 3ex
\caption{
Evolution of $L_d / \lstar$ for cascade models with various ratios 
$v^2 / 8 \qdstar$ as in Fig.~\ref{fig: rmax2}.  Although the peak 
luminosity is fairly independent of $v^2 / 8 \qdstar$, systems 
with larger ratios reach peak luminosity earlier in time. Independent 
of the initial size distribution, systems with smaller $v^2 / 8 \qdstar$
ratios have larger relative luminosity at late times.
\label{fig: lum2}
}
\end{figure}

\begin{figure} 
\includegraphics[width=6.5in]{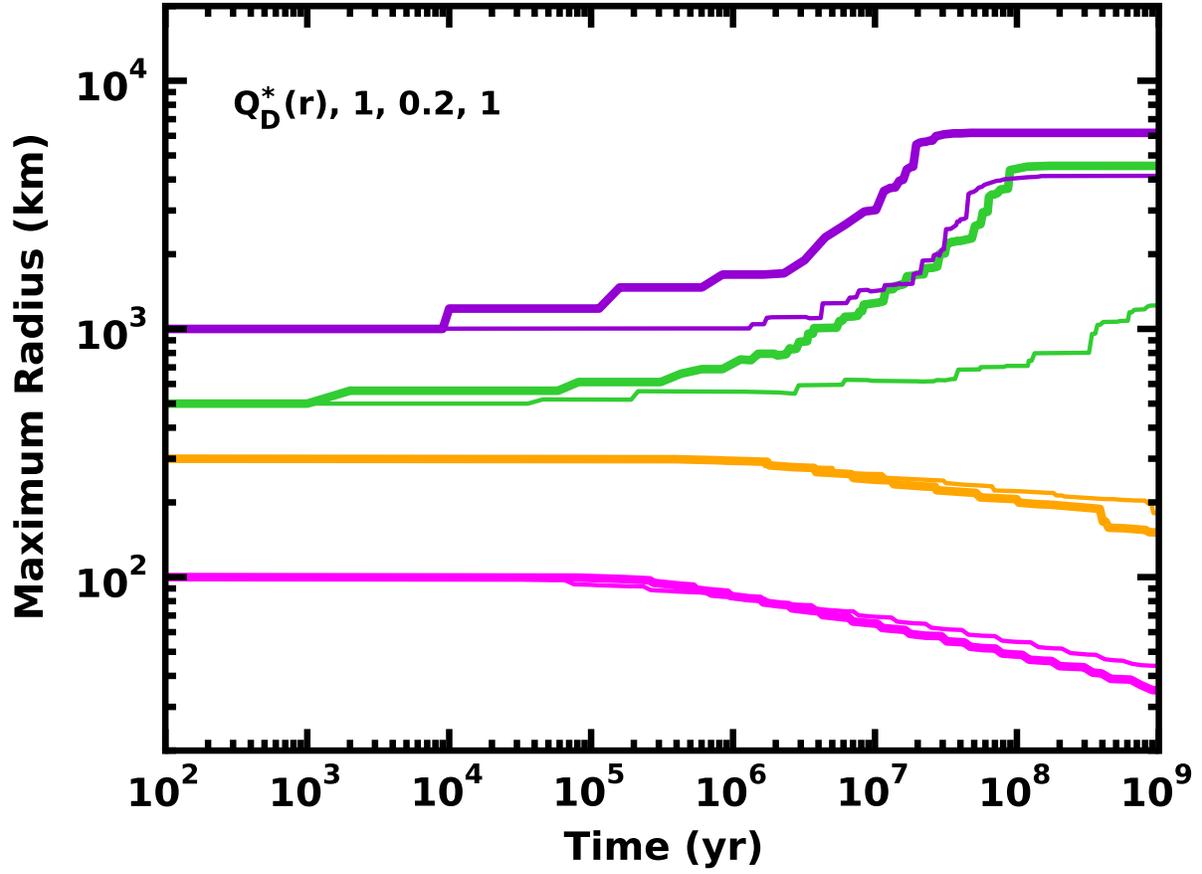}
\vskip 3ex
\caption{
Evolution of \rmax\ for cascade models with $\qdstar(r)$, $b_d$ = 1, 
$m_{l,0}$ = 0.2, $b_l$ = 1, initially mono-disperse (thick lines) or 
power-law (thin lines) size distributions, and various \r0. The boundary 
between growth and destructive evolution is at \r0\ $\approx$ 400~km.
\label{fig: rmax3}
}
\end{figure}

\begin{figure} 
\includegraphics[width=6.5in]{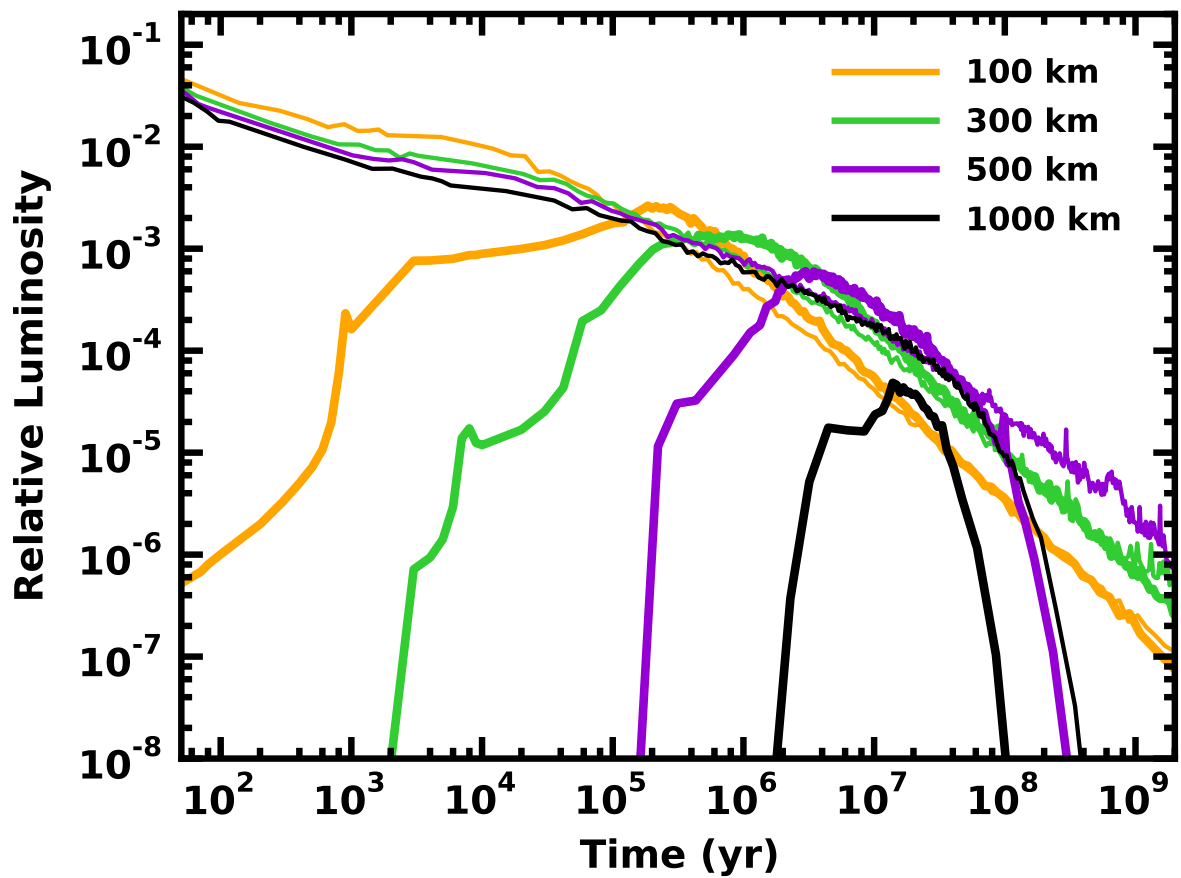}
\vskip 3ex
\caption{
Evolution of $L_d/\lstar$ for cascade models with various \r0\ (in km) as 
indicated in the legend and mono-disperse (thick lines) or power-law
(thin lines) initial size distributions.  When the largest objects 
grow through mergers, originally mono-disperse swarms have minimal 
$L_d$ for short periods of time. In swarms with an power law initial
size distribution, $L_d$ is fairly well-correlated with \r0.
\label{fig: lum3}
}
\end{figure}

\begin{figure} 
\includegraphics[width=6.5in]{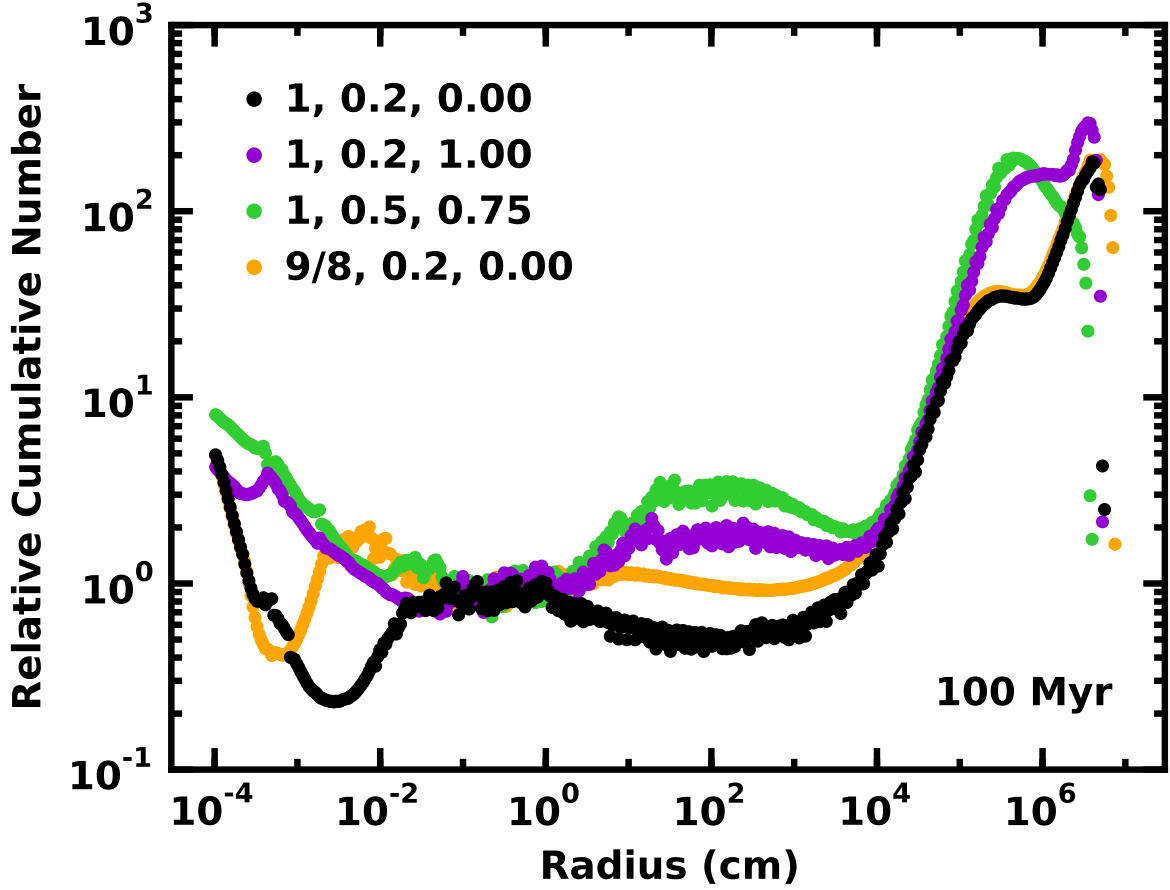}
\vskip 3ex
\caption{
Relative cumulative size distributions at 100~Myr for cascade models 
with a mono-disperse initial size distribution, $\qdstar(r)$, 
\r0\ = 100~km, and various $b_d$, $m_{l,0}$, and $b_l$ as indicated 
in the legend.  Relative to a 
power-law size distribution with $N(>r) \propto r^{-2.68}$, all curves 
have a wavy morphology with distinct excesses of particles at 
$r \approx$ 1--10~\mum\ and $r \gtrsim$ 0.1~km. The rise in the relative
cumulative size distribution occurs at the minimum in \qdstar\ at 0.1~km.
Although the fine details of the shape depend on the input parameters,
the shape for each set of input parameters is independent of time for
$t \gtrsim$ 1~Myr (see also Figs.~\ref{fig: sd1}--\ref{fig: sd3}).
\label{fig: sd4}
}
\end{figure}

\begin{figure} 
\includegraphics[width=6.5in]{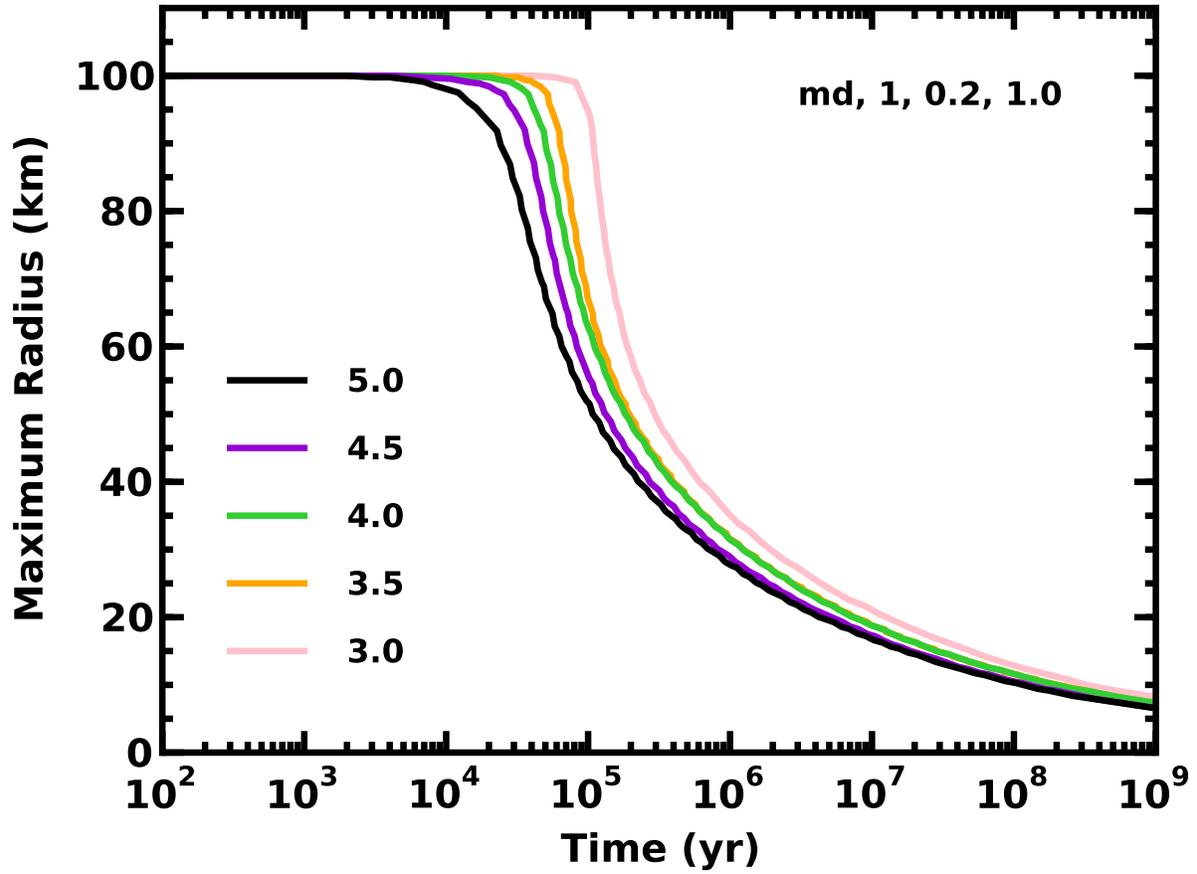}
\vskip 3ex
\caption{
Evolution of \rmax\ for cascade models with a mono-disperse initial size
distribution, $b_d$ = 1, $m_{l,0}$ = 0.2, and $b_l$ = 1.0 (as indicated
in the legend in the upper right corner) and various $q_d$ as indicated 
in the legend in the lower left corner. In simulations with larger $q_d$, 
the largest objects reach smaller sizes at late times.
\label{fig: rmax4}
}
\end{figure}

\begin{figure} 
\includegraphics[width=6.5in]{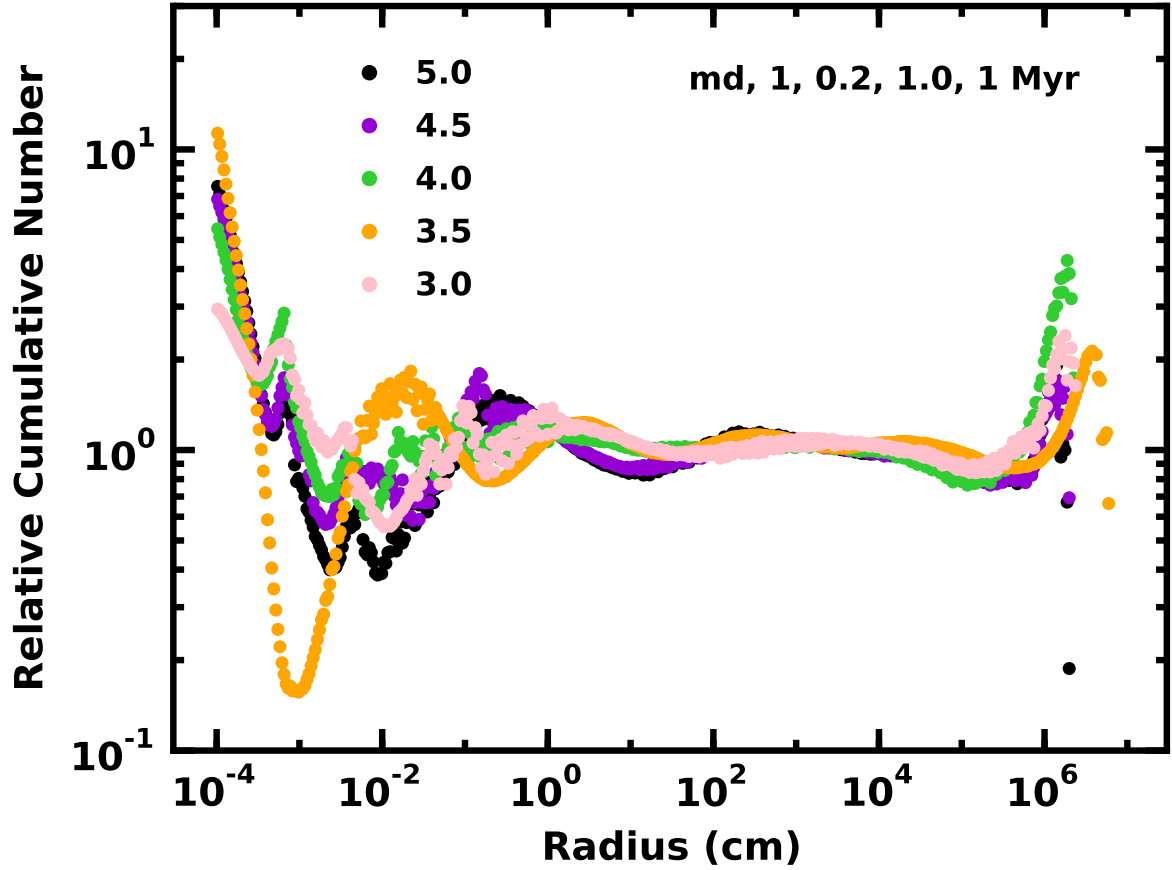}
\vskip 3ex
\caption{
Relative cumulative size distributions at 1~Myr for mono-disperse cascade models 
with $b_d$ = 1, $m_{l,0}$ = 0.2, $b_l$ = 0.1 and various $q$ as indicated 
in the legend. Aside from the morphology of the waviness for $r \approx$
1~\mum\ to 10--100~cm, the size distribution is independent of $q$.
\label{fig: sd5}
}
\end{figure}

\begin{figure} 
\includegraphics[width=6.5in]{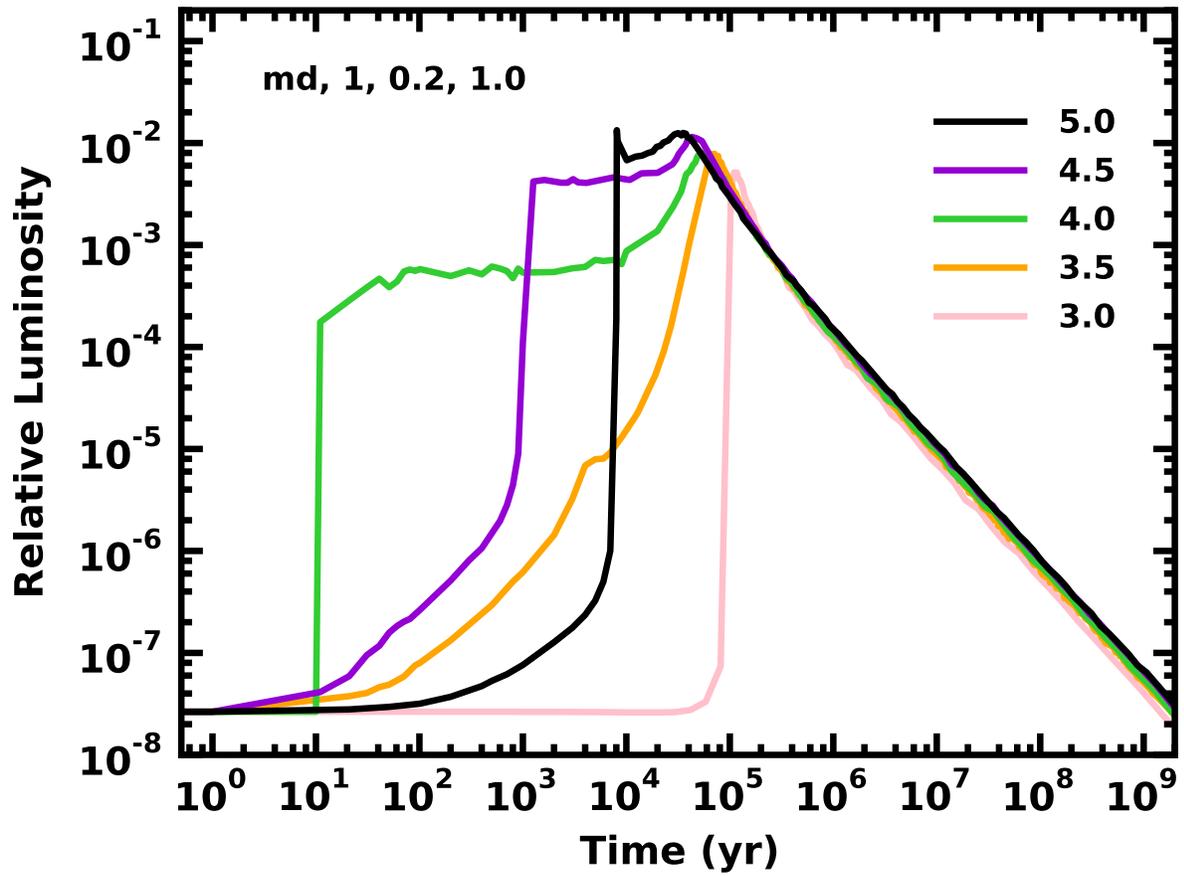}
\vskip 3ex
\caption{
Evolution of $L_d/\lstar$ for the mono-disperse cascade models in 
Figs.~\ref{fig: rmax4}--~\ref{fig: sd5}.  Systems with larger $q_d$ 
reach larger $L_d$ and decline more slowly than systems with smaller 
$q_d$.
\label{fig: lum4}
}
\end{figure}

Our analysis suggest several clear differences between predictions 
of the standard analytic model and the results of numerical simulations 
with \qdstar\ = constant (Figs.~\ref{fig: rmax1}--\ref{fig: lum2}, 
\ref{fig: rmax4}--\ref{fig: lum4}). For material orbiting with $e$ = 0.1 
inside an annulus with $\Sigma$ = 10~\gcms\ at 1~AU, 
(i) the radius (mass) of the largest object declines by a factor of 12--16
(2000--4000) in 1~Gyr,
(ii) the size distribution is a power law which generally follows the 
$N(r) \propto r^{-3.5}$ of the analytic model but has large amplitude 
waves at large and small sizes, and
(iii) the evolution of the luminosity at late times is somewhat steeper 
than the predicted $L_d \propto t^{-1}$. 

When \qdstar\ is a function of $r$, the numerical calculations provide other
tests of the analytic model. For $e$ = 0.1 and $i/e$ = 0.5 at $a$ = 1~AU,
swarms with \r0\ = 100--400~km follow the evolution of the constant
\qdstar\ models fairly closely. Over 2~Gyr, (i) the largest objects lose 
from 75\% to 99\% of their initial mass and (ii) the dust luminosity declines
as $L_d \propto t^{-n}$ with $n \approx$ 1.1--1.2. At late times, the 
size distribution for $r \lesssim r_Q =$ 0.1~km roughly tracks a power law with 
the $q \approx$ 3.68 predicted by the analytic model. Dynamical ejection of
particles with $r \lesssim$ 1~\mum\ leads to an excess (deficit) of particles 
with $r \approx$ 1--10~\mum\ (10--100~\mum).  Among larger particles with 
$r \gtrsim r_Q$, the power law slope is smaller and has a broader range, 
$q \approx$ 2.7--2.9, relative to the predicted $q \approx$ 3.04.  With no 
supply of debris from much larger particles with $r \gtrsim \r0$, there are 
large-amplitude waves about the average power law at large sizes. 

Calculations with \r0\ = 500--1000~km evolve differently. When $e \approx$ 0.1,
the largest particles grow with nearly every collision. Although collisions
remove mass from smaller particles, growth concentrates the debris from these
collisions into the largest objects.  Thus, the mass (dust luminosity) of the 
swarm declines more slowly (rapidly) with time. For this set of calculations,
the mass (luminosity) varies with time as $M_d \propto t^{-n}$ with $n \lesssim$ 0.2 
($L_d \propto t^{-n}$ with $n \approx$ 1.5--2.5). Although these results allow
no test of the analytic model, they provide useful context for the planet formation
calculations in \S4.2.

To evaluate other results in the context of a specific analytic model, we 
consider the size distribution, the mass in small particles, the mass loss 
rate through the grid, and the dust luminosity in more detail. In analytic 
models for collisional cascades, each range of sizes loses mass at the same 
rate which declines as $\dot{M} \propto t^{-2}$ 
\citep[e.g.,][]{koba2010a,wyatt2011}.  This behavior allows the cascade to 
maintain an invariant size distribution. In our calculations, the shape of 
the size distribution is nearly fixed in time 
(Fig.~\ref{fig: sd1}--\ref{fig: sd2}), matching a basic prediction of the 
analytic model.

When \qdstar\ = constant over a finite range of particle sizes, analytic 
models predict a set of waves superposed on a power-law size distribution 
\citep{wyatt2011}. Our calculations match the predicted slope and the 
amplitude of the primary wave close to the small size cutoff at 
1~\mum\ (see \S\ref{sec: app-waves}). The numerical simulations produce
(i) waves with longer wavelengths and smaller amplitudes at 0.1~mm to 1~km
and (ii) a larger wave at 1--100~km which are not present in the analytic
model.

When \qdstar\ is a function of radius, the numerical calculations match 
(i) the predicted slope of the power law from 1~\mum\ to 0.1~km, 
(ii) the amplitude of the wave at the smallest sizes,
(iii) the distinctive rise in the relative cumulative size distribution 
at 0.1~km, and 
(iv) the general shape of the relative size distribution from 0.1~km to
10--30~km (see \S\ref{sec: app-waves}). However, the level of waviness 
at 0.1~mm to 0.1~km is larger than predicted; the slope of the size 
distribution for $r \gtrsim$ 0.1~km is shallower than predicted. 

Overall, this good level of agreement is encouraging. The gradual reduction
in the sizes of the largest objects probably produces differences between 
the analytic model and the simulations at the largest sizes.  We suspect 
that cratering collisions change the level of waviness at smaller sizes. 
We plan to investigate these possibilities in future studies.

The mass in 1~mm and smaller particles provides another test
\citep[e.g.,][and references therein]{wyatt2011}. We define the relative
dust mass $\xi$ as the ratio of the mass in small particles to the total mass.  
In models with fixed \qdstar, \r0\ = 100~km, \rmin\ = 1~\mum, and $q$ = 3.5, 
$\xi = 10^{-4}$. When \qdstar\ is a function of particle size, $q$ = 3.68 (3.04) 
for $r \lesssim r_Q$ ($r \gtrsim r_Q$); with $r_Q$ = 0.1~km, $\xi \approx 10^{-4}$. 

\begin{figure} 
\includegraphics[width=6.5in]{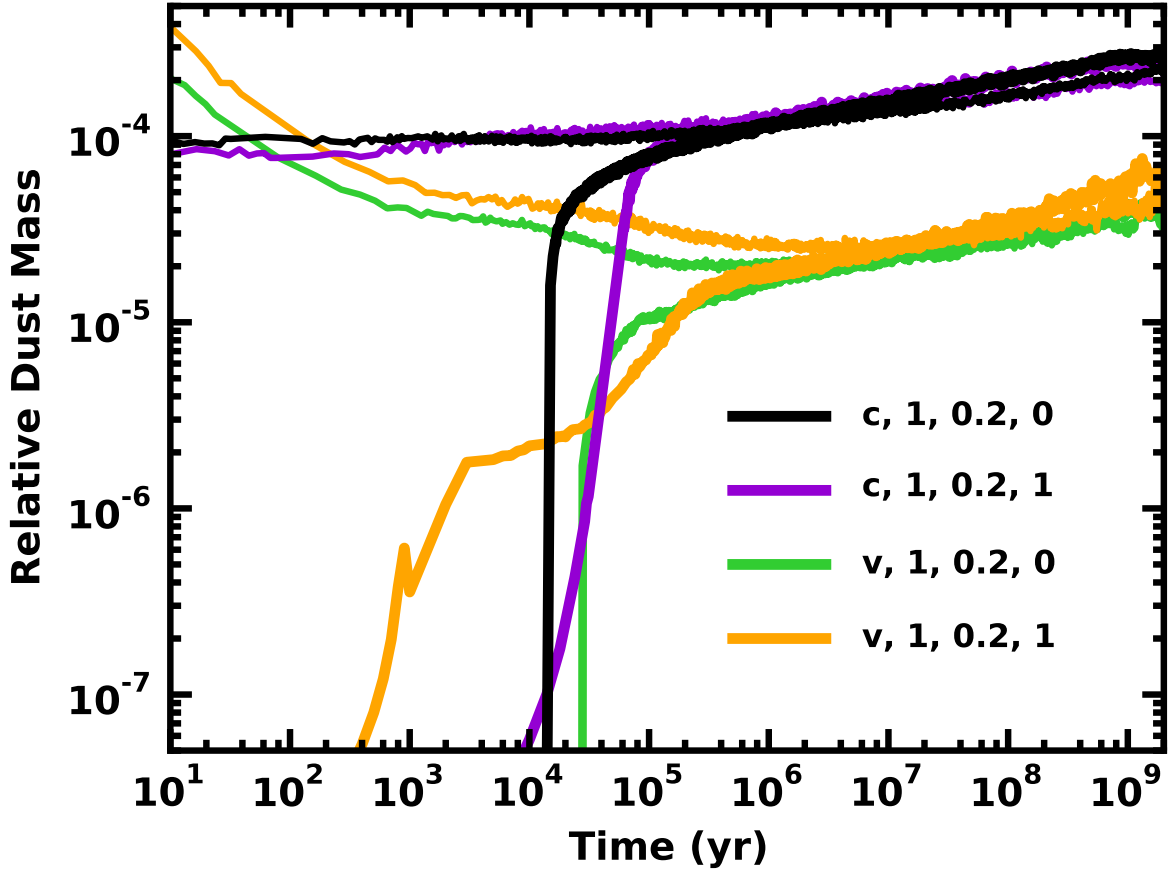}
\vskip 3ex
\caption{
Time evolution of the relative dust mass $\xi$ for collisional cascades
with mono-disperse (thick lines) and power law (thin lines) initial size 
distributions. The legend indicates the prescription for \qdstar\ (`c' for 
constant and `v' for $\qdstar(r)$) and values for $b_d$, $m_{l,0}$, and $b_l$.
On time scales longer than the collision time, the relative dust mass slowly 
grows with time.  Although $\xi$ depends on \qdstar, it is fairly independent 
of the initial size distribution and other fragmentation parameters.
\label{fig: mdust1}
}
\end{figure}

Fig.~\ref{fig: mdust1} illustrates the behavior of $\xi$ for a set of cascade 
models with \r0\ = 100~km. When the swarm has an initial power law size 
distribution, $\xi \approx 10^{-4}$. As systems with constant \qdstar\ evolve
(Fig.~\ref{fig: mdust1}, thin black and green curves), the relative dust mass
gradually grows with time and reaches $\xi \approx 2 - 3 \times 10^{-4}$ at
$t$ = 1--2~Gyr. When \qdstar\ is a function of particle size (thin orange and
magenta lines), $\xi$ declines to $2 - 3 \times 10^{-5}$ at 0.1--1~Myr and then 
slowly recovers to $4 - 8 \times 10^{-5}$ at $t$ = 1--2~Gyr. Swarms with 
mono-disperse initial size distributions follow similar paths. Once collisions
begin to produce dust, the relative dust mass grows to 
$\xi \approx 10^{-4}$ (for constant \qdstar\ swarms) or
$\xi \approx 2 - 3 \times 10^{-5}$ (for swarms with $\qdstar(r)$). The relative
dust mass then grows slowly with time.

For swarms with constant \qdstar, gradual reductions in \rmax\ explain the 
evolution of the relative dust mass. After 1--2 Gyr of evolution, \rmax\ $\approx$ 
6--10~km.  For a system with $q$ = 3.5, the expected relative dust mass is 
$\xi \approx 3 - 4 \times 10^{-4}$. Allowing for small differences in dust mass 
and total mass due to the wavy size distribution (Fig.~\ref{fig: sd3}), the
$\xi \approx 2 -3 \times 10^{-4}$ from our calculations is close to the expected
value.

When \qdstar\ is a function of $r$, the change in the size distribution at
large sizes is as important as the reduction of \rmax\ with time. For swarms 
with a double power-law size distribution and a break at $r_Q \approx$ 0.1~km, 
\begin{equation}
\xi \approx  \left ( { 4 - q_l \over 4 - q_s } \right )
\left ( { r_d^{4-q_s} \over \rmax^{4-q_l} } \right )
r_Q^{q_s - q_l} ~ ,
\label{eq: xi}
\end{equation}
where $r_d$ = 1~mm is the radius of the largest dust particle and
$q_s, q_l$ are the slopes of the size distribution for large
and small particles. In our calculations, the slope of the
size distribution for small particles -- $q_s \approx 3.68$ -- 
is close to expectation. At large sizes, however, the slope is
shallower than predicted, $q_l \approx$ 2.7--2.9 instead of 
$q_l \approx$ 3.04. The shallower slope lowers the predicted 
relative dust mass to 
$\xi \approx 1 - 4 \times 10^{-5}$ for \rmax = 100~km and 
$\xi \approx 3 - 9 \times 10^{-5}$ for \rmax = 50~km. Our
calculations match these expectations.

For the time evolution of $\dot{M}$ and $L_d$ in the analytic model, we 
derive an estimate for the collision time from eq.~\ref{eq: tc2} with
$\alpha$ = 1.  Adjusting $\alpha$ allows us to match the analytic 
$\dot{M}_0 = M_0 / t_c$ at $t = 0$ with the initial production rate for 
1~\mum\ and smaller objects from our numerical simulations. 
This mass loss rate then declines with time as
\begin{equation}
\label{eq: mdot}
\dot{M} = t_c^{-1} 
{M_0 \over ( 1 + t/t_c)^2 } ~ .
\end{equation}
Adopting an initial power law size distribution with $q$ = 3.5 for 
\qdstar\ = constant and $r \le$ \rmax\ yields the initial $L_d$ which 
then evolves as in eq.~\ref{eq: ld}. 

\begin{figure} 
\includegraphics[width=6.5in]{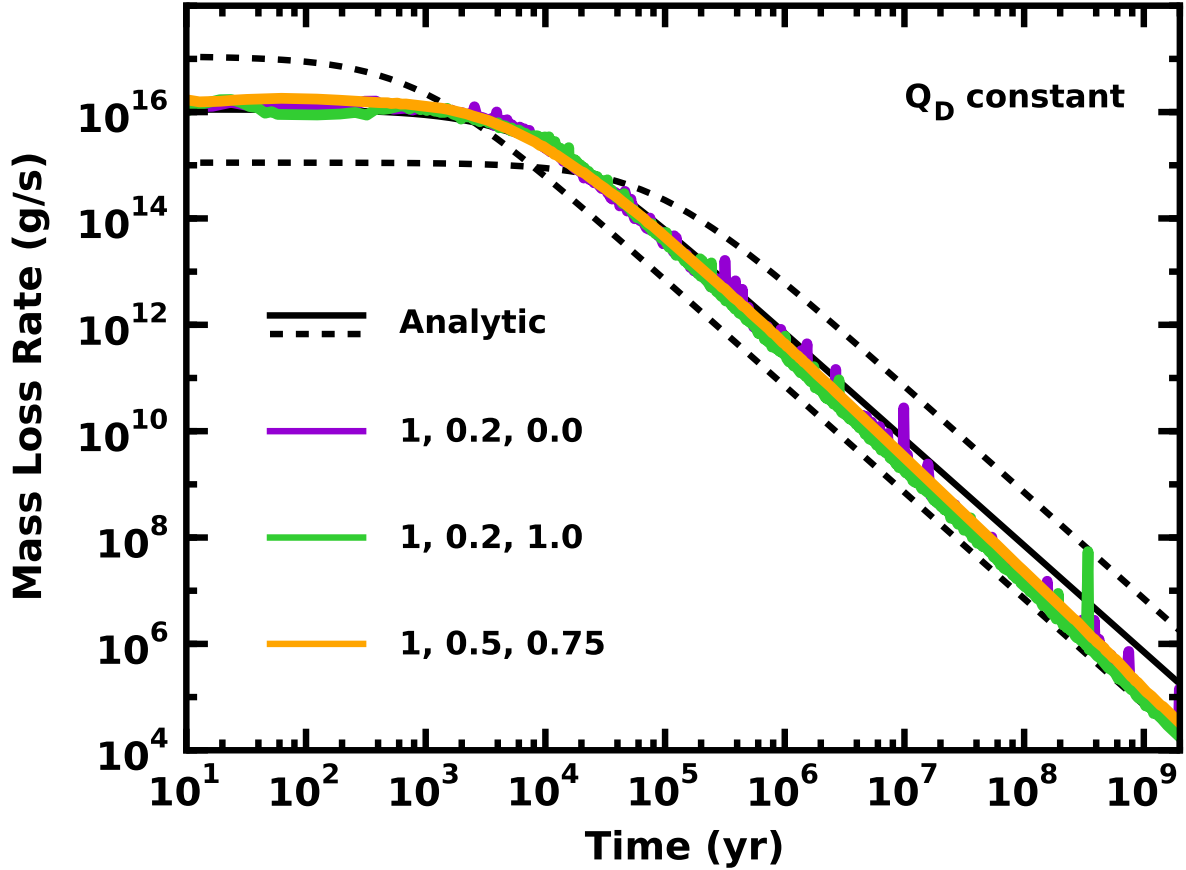}
\vskip 3ex
\caption{
\label{fig: mdot1}
Comparison of the mass loss rate from numerical simulations with the
analytic model. The legend indicates values of $b_d$, $m_{l,0}$, and 
$b_l$ for simulations with a constant \qdstar\ = $6 \times 10^7$ 
erg~g$^{-1}$. The black lines plot predictions from the analytic model
for $\alpha$ = 1 (upper dashed), 0.1 (solid), and 0.01 (lower dashed).  
Colored lines plot results for calculations with an initial power law
size distribution.
Although the numerical estimates match the analytic prediction for the 
mass loss rate with $\alpha$ = 0.1 at early times, they decline with 
time more rapidly than the analytic result.  Occasional spikes in the 
mass loss rate result from occasional collisions of large objects.
}
\end{figure}

Fig.~\ref{fig: mdot1} compares $\dot{M}$ derived from our baseline calculations
with predictions of the analytic model for $\alpha$ = 0.01, 0.1, and 1.0. At
$t$ = 0, the numerical calculations match the analytic model for $\alpha \approx$ 
0.1.  As the calculations proceed, the derived $\dot{M}$ declines slightly more
rapidly than the analytic prediction. By the end of the calculation, the mass
loss rate is similar to analytic model predictions for $\alpha \approx$ 0.01.

To test the analytic model in more detail, we consider a set of calculations 
with a range in \qdstar\ (e.g., Fig.~\ref{fig: rmax2}).
At late times, $\dot{M} \approx M_0 t_c t^{-2}$. From eq.~\ref{eq: tc2},
$t_c \approx t_0 \alpha$ with $t_0 \approx 8 \times 10^4$~yr. Substituting
this result into the expression for $\dot{M}$ yields
$\alpha \approx \dot{M} t^2 / M_0 t_0$. Numerical results for $\dot{M}(t)$
thus yield $\alpha (t)$.  For models with $\qdstar$ = $6 - 600 \times 10^7$~\ergg, 
we infer:
\begin{equation}
\alpha \approx 12 
\left ( v^2 \over 2 \qdstar \right )^{-1} 
\left ( t \over {\rm 1~Gyr} \right )^{-0.13} ~ .
\label{eq: alphac}
\end{equation}
At 0.1--1~Gyr, the numerical calculations evolve roughly 3 times more slowly than 
the analytic model (e.g., eq.~\ref{eq: alpha}) when $v^2 / 2 \qdstar \approx$ 1 and 
approximately in step with the analytic model when $v^2 / 2 \qdstar \approx$ 200.

In addition to the steady decline, the calculations exhibit clear spikes in the
mass loss rate. Stochastic variations in the collision rate cause this behavior.
To try to mimic real collisional systems, we require an integral number of 
collisions every time step. For infrequent collisions between the largest objects, 
our algorithm rounds a fractional number of collisions down or up to the nearest 
integer. When the algorithm rounds up, it produces a clear spike in $\dot{M}$.

\begin{figure} 
\includegraphics[width=6.5in]{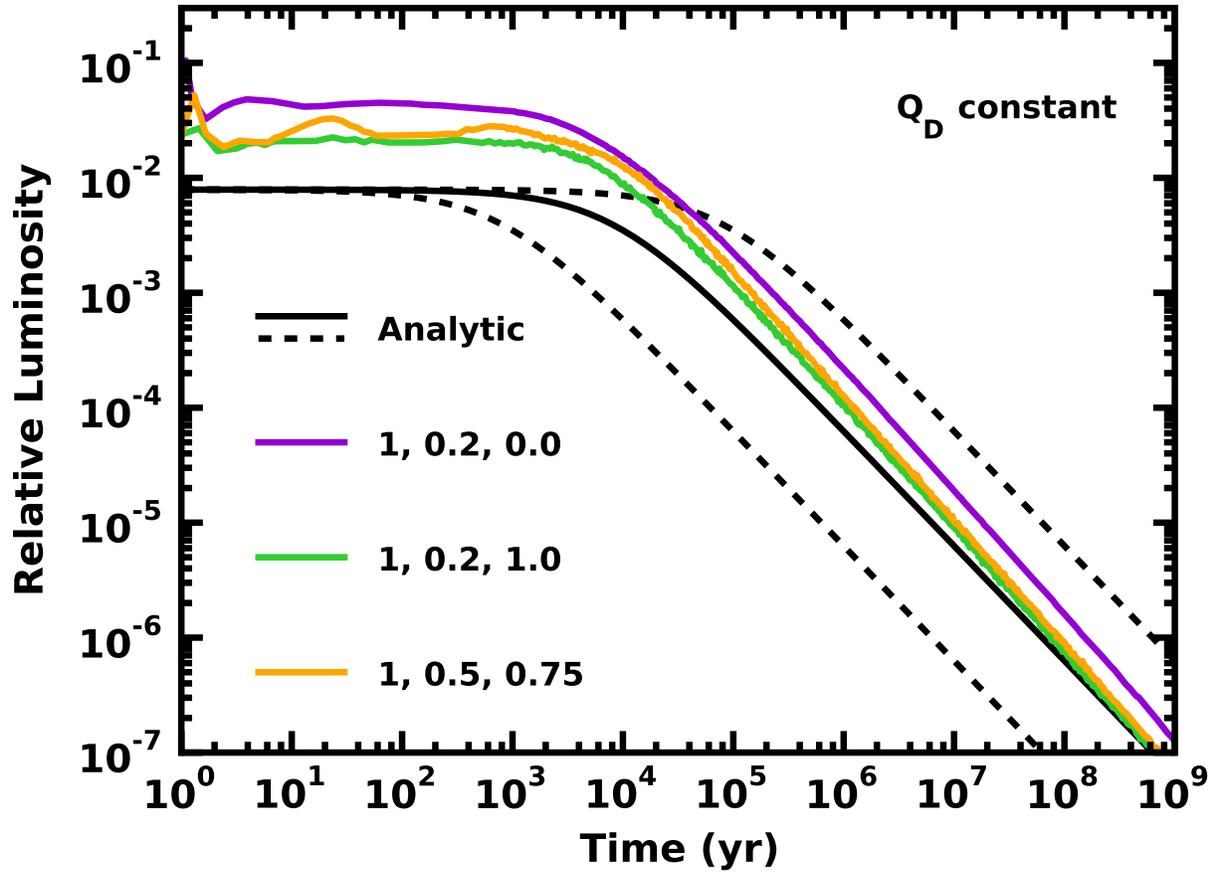}
\vskip 3ex
\caption{
\label{fig: lum5}
Comparison of the relative luminosity $L_d / \lstar$ from numerical 
simulations with the analytic model.  Despite the random spikes in 
the mass loss rate (see Fig.~\ref{fig: mdot1}), the luminosity
declines smoothly with time.
}
\end{figure}

Fig.~\ref{fig: lum5} compares the baseline $L_d$'s for the simulations with the
analytic model. At early times, the number of small particles in the calculations
oscillates about a constant value as the system establishes an equilibrium size
distribution. This constant is a factor of 4--5 larger than the prediction of the
analytic model. At $\sim$ 0.01~Myr, the luminosity in the analytic and numerical 
calculations starts to fall. Because $L_d$ in the numerical model declines faster
than predicted by the analytic model, $L_d$ from the numerical models eventually
falls below the analytic prediction.

In the numerical models, the excess of small particles at $r \approx \rmin$ produces
the larger $L_d$. Although these particles remove larger particles from the size
distribution, the excess surface area from the small particles more than compensates
for the deficit in surface area at somewhat larger sizes. Thus, the numerical models
have larger $L_d$ than the analytic model. 

Despite the spikes in the mass loss rate, the numerical models show no spikes in $L_d$.
When \rmax\ $\approx$ 100--300~km,
every spike in $\dot{M}$ sprinkles debris throughout the grid, with no impact on the 
shape of the size distribution and little impact on the total mass in the grid. Thus,
there is negligible change in surface area and $L_d$. In \S\ref{sec: calcs-summ}, we
describe how collisions in systems with larger \rmax\ produce clear spikes in $L_d$.

Results for models where \qdstar\ depends on particle size yield similar results. The
dust luminosity and the mass loss rate from the grid generally follow the trends 
established in calculation with constant \qdstar: (i) a plateau in $\dot{M}$ or
$L_d$ followed by a decline and (ii) occasional spikes in $\dot{M}$ (but not $L_d$)
as the system evolves.

\subsection{Planet Formation}
\label{sec: calcs-pf}

Our analysis pinpoints several clear differences between the analytic model and 
detailed coagulation calculations of collisional cascades. However, real cascades 
do not occur in isolation. Within planet formation theory, cascades begin as the 
gravity of growing or fully-formed planets stirs up surrounding smaller particles 
\citep[e.g.,][]{kb2002b,mustill2009}. As these systems evolve, the collisional 
cascade converts a fraction $f_c$ of these small particles into dust grains which 
are ejected by radiation pressure from the central star.  Massive planets accrete 
or dynamically eject the rest of the small particles.  Deriving $f_c$ and making 
a more robust link between the theories of collisional cascades and planet formation 
requires more detailed sets of calculations.

To begin to make this link, we consider a suite of planet formation calculations. 
Within a single annulus, the initial swarm consists of a set of mono-disperse 
particles with total mass $M_d$ = 0.5~\mearth, surface density $\Sigma_0$ = 
10~\gcms, and initial radius \r0\ = $10^n$ ($n$ = 0, 1, 2, $\ldots$, 8). 
Particles have $e_0$ = $10^{-5}$ (\r0\ = 1--10~cm), $10^{-4}$ (\r0\ = 1--100~m), 
$10^{-3}$ (\r0\ = 1--10~km), $10^{-2}$ (\r0\ = 1--100~km), and $i_0$ = $e_0$/2. 
As particles grow, collisional damping, dynamical friction, and viscous stirring 
modify $e$ and $i$ for each mass bin. All calculations ignore gas drag and 
Poynting-Robertson drag. 

Although the initial conditions strongly favor growth through mergers, every 
collision produces a modest amount of debris. To specify the amount and size
distribution of the debris, we adopt the standard $\qdstar(r)$ relation with
parameters specified after eq.~\ref{eq: qd} and set $b_d$ = 1, $m_{l,0}$ = 0.2, 
and $b_l$ = 1 in eqs.~\ref{eq: mej}--\ref{eq: mlarge}. Modest changes in these 
assumptions have little impact on the results.

To quantify the diversity of outcomes, we perform $\sim$ 5 calculations for each
set of initial conditions. Within a set of calculations, each run begins with a
different random number seed. During each timestep, \orch\ uses a random number
generator to decide whether to round up (or down) a fractional number of collisions
to an integer.  Randomness in the collision rates generates a dispersion in outcomes. 

Despite some intrinsic diversity, all calculations follow a standard pattern. 
Initially, particles grow through mergers. As particles grow, collisional 
damping and gravitational interactions modify $e$ and $i$ for each mass bin. 
Damping reduces $e$ and $i$ for small particles, $r \lesssim$ 1--10~cm. Dynamical
friction transfers orbital kinetic energy from the largest particles to the
smallest particles.  Once particles have $r \gtrsim$ 1--10~km, viscous stirring
raises $e$ and $i$ for all particles.

As $e$ and $i$ evolve, the largest particles step through several stages of growth.
At the start of each calculation, gravitational focusing factors $f_g$ are small.
Growth is slow and steady. Damping and dynamical friction enhance $f_g$, enabling 
a phase of runaway growth where the largest particles gain mass much more rapidly
than smaller particles. Once viscous stirring dominates collisional damping and
dynamical friction, gravitational focusing factors for the largest particles drop. 
Oligarchic growth -- where large particles add mass more slowly -- begins. 

Rising viscous stirring rates also initiate the collisional cascade. Collisions 
among particles close to the minimum in \qdstar\ -- at roughly $r_Q \approx$ 
0.1~km -- eject copious amounts of debris into the swarm. As viscous stirring 
continues to raise $e$ and $i$ for small particles, collisions destroy particles 
which are smaller and larger than $r_Q$. With $\qdstar \propto r^{-0.4}$ 
for $r < r_Q$ and $\qdstar \propto r^{1.35}$ for $r > r_Q$, it is easier for 
collisions with fixed kinetic energy to destroy particles with $r \ll r_Q$ than 
those with $r \gg r_Q$. These collisions produce debris with sizes $r \lesssim$ 100~cm.

As the debris accumulates, there are two possible outcomes 
\citep[][and references therein]{kb2015a}. If collisional damping dominates the loss
of particles from destructive collisions, small particles have very small $e$ and $i$.
Damping then limits the collisional cascade.  Large gravitational focusing factors 
initiate a second phase of runaway growth, where the largest objects accrete most of 
the mass in small particles. If damping is ineffective, small particles remain at 
large $e$ and $i$; the collisional cascade proceeds unabated.

\begin{figure} 
\includegraphics[width=6.5in]{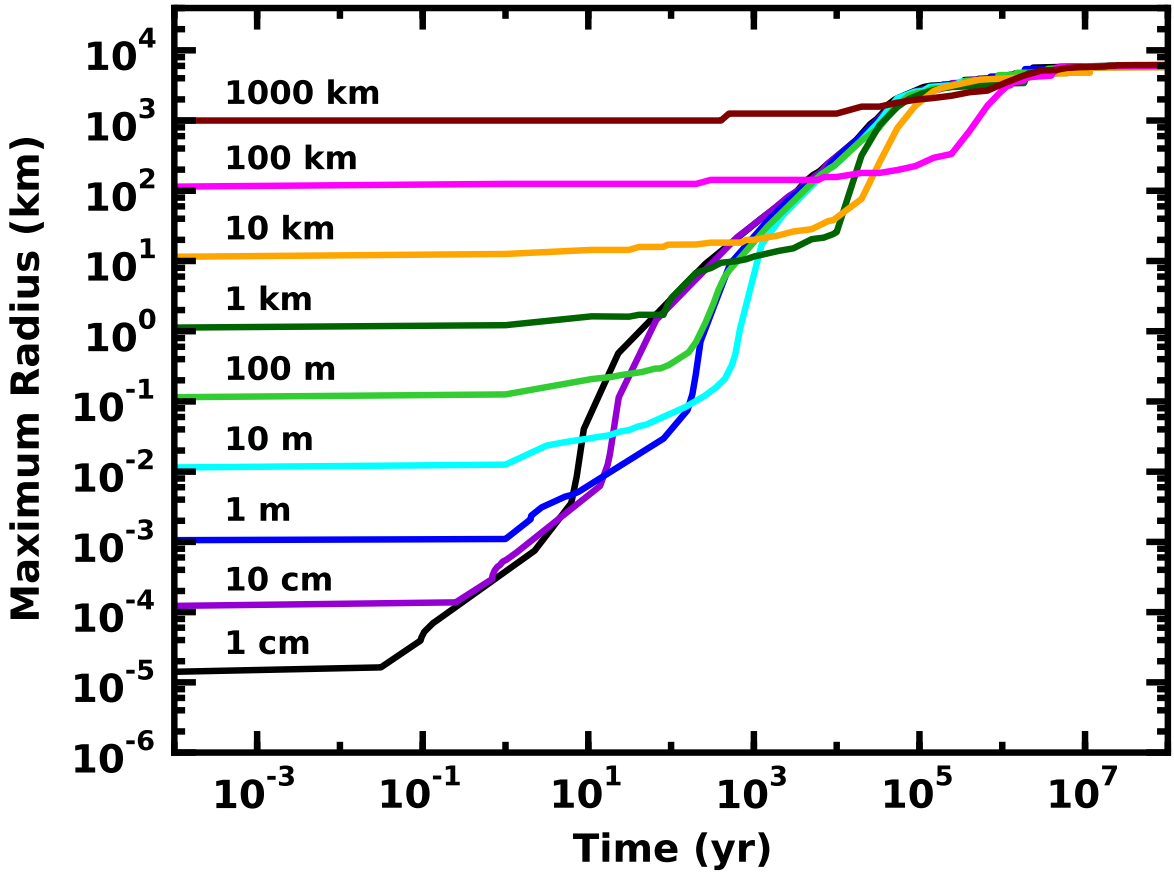}
\vskip 3ex
\caption{
\label{fig: rad1000}
Growth of the largest object in a suite of planet formation calculations with 
$\Sigma_0$ = 10~\gcms\ in an annulus with $\delta a$ = 0.2~AU at $a$ = 1~AU.
Labels indicate the initial radius \r0\ of a mono-disperse set of objects.
Swarms with small \r0\ evolve more rapidly than those with large \r0. After 
0.1--1~Myr, all swarms produce a few objects with radii exceeding 3000~km. 
This final radius is independent of \r0.
}
\end{figure}

Fig.~\ref{fig: rad1000} illustrates the growth of the largest particles as a function
of \r0. From eq.~\ref{eq: tc1}, small particles evolve more rapidly than large particles. 
The ensemble of 1~cm particles requires less than a year to reach sizes of 30~cm; after 
another 20~yr, the largest particles have radii of 1~km. By $\sim 10^4$~yr, the radii 
of the largest protoplanets surpass $10^3$~km. As the evolution slows, occasional 
collisions among these objects eventually yields a single planet with a radius of 
roughly 6000~km. In contrast, sets of 100~km or 1000~km particles take more than 
0.1~Myr for the largest objects to double in mass. Despite the delay, protoplanets 
still achieve radii of roughly 6000~km on a time scale of 1 Myr.

\begin{figure} 
\includegraphics[width=6.5in]{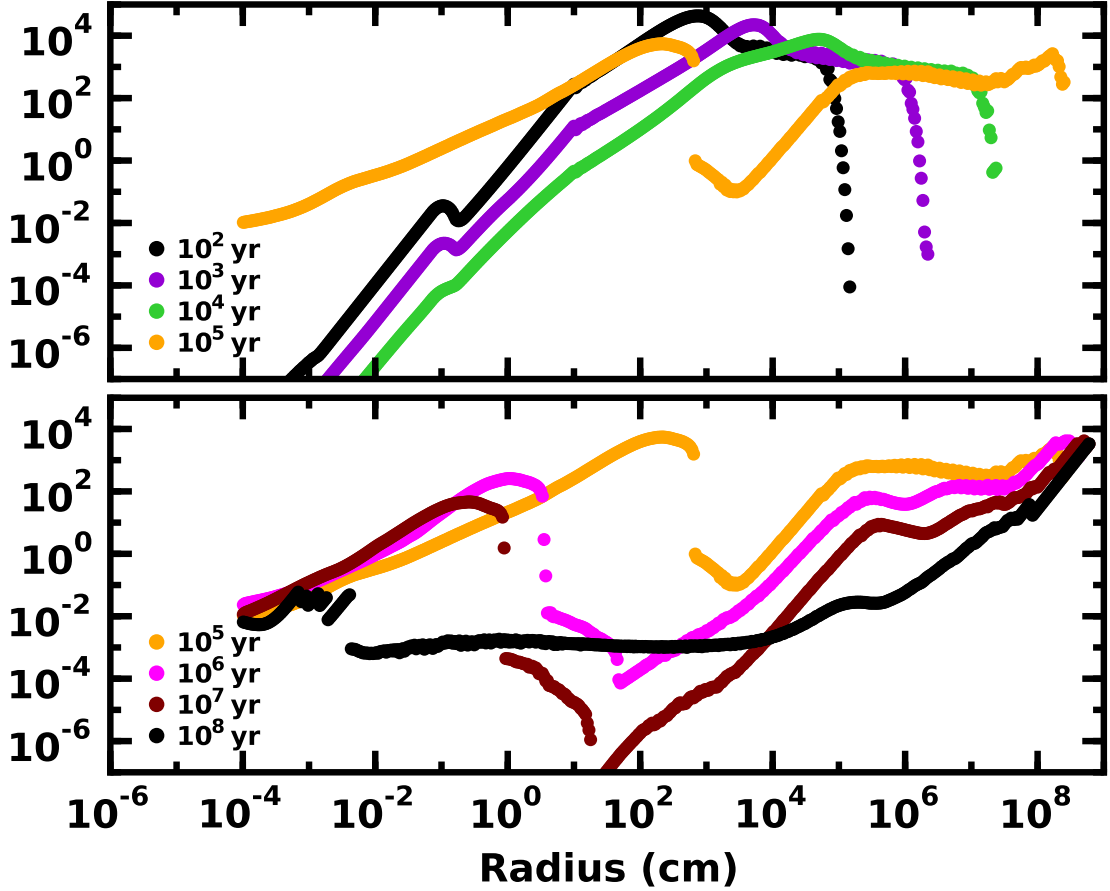}
\vskip 3ex
\caption{
\label{fig: sd6}
Evolution of the relative cumulative size distribution ($N(>r) / n_0 r^{-2.5}$, 
with $n_0$ = $10^{20}$) for a planet formation calculation with 
$\Sigma_0$ = 10~\gcms\ and \r0\ = 10~cm.
In each panel, the legend indicates the evolution time in~yr. To aid in the
comparison at late times, both panels include the size distribution at $10^5$~yr
(orange points).  From $10^2$~yr to $10^4$~yr (upper panel; black, violet, and
green points), the largest object grows from 1~km to 100~km and sweeps up smaller 
particles in the debris tail.  As objects grow from 100~km to 1000~km (0.01--1~Myr,
green, orange, and magenta points), collisions produce a prominent debris tail at 
sizes of 100~cm and smaller. Continued evolution produces a significant deficit 
of particles at sizes ranging from 1~cm to 1~km (1--10~Myr; magenta and maroon
points). Eventually collisions remove most particles smaller than $\sim$ 100~km 
(100~Myr; black points).
}
\end{figure}

Throughout this evolution, changes in the relative cumulative size distribution follow 
a fairly standard 
pattern (Fig.~\ref{fig: sd6}).  For a swarm of particles with \r0\ = 10~cm, mergers 
rapidly produce a swarm of particles with $r \approx$ 1--10~m. The largest particles 
then begin to grow faster than their smaller counterparts. After $\sim$ 100~yr, the 
size distribution has three main pieces: (i) an exponential at large sizes, 
(ii) a plateau at intermediate sizes, and (iii) a debris tail at small sizes. 
Over the next $10^4$~yr, the exponential and the small peak at the small size 
end of the plateau shift to larger and larger sizes. Large particles sweep up the 
smallest particles, gradually diminishing the population of the debris tail.

From $\sim 10^4$~yr to $\sim 10^5$~yr, the system makes a transition from runaway growth 
to oligarchic growth. During oligarchic growth, the size distribution develops a small
rise at $r \approx$ 1000~km.  Once viscous stirring raises the velocities of the smallest 
particles, collisions destroy particles with $r \approx r_Q \approx$ 0.1~km.  The 
relative cumulative size distribution develops a sharp dip at these sizes.  Debris 
from destructive collisions adds mass to the debris tail. 

As the evolution proceeds, the largest objects gradually accumulate most of the mass in
the annulus. With most mass already concentrated in 1-100~km objects, this behavior produces
(i) an abrupt rise in the size distribution from 1000~km to larger sizes and (ii) a smaller
plateau at 1--1000~km. For somewhat smaller particles ($r \gtrsim$ 1~cm), stirring 
continues to raise collision velocities, enabling the destruction of smaller and smaller
objects. The minimum in the relative cumulative size distribution expands from 0.01--1~km 
at 0.1~Myr to 5~cm to 2~km at 1~Myr to 1~cm to 3~km at 10~Myr. The minimum of this dip 
shifts to smaller sizes and deepens considerably. After $\sim$ 100~Myr, the size distribution 
consists of a steeply rising piece for $r \gtrsim$ 10~km and a fairly flat piece for 
$r \lesssim$ 10~km. By 1~Gyr, accretion by 1--2 large objects and destruction by 
the collisional cascade remove all small particles from the annulus.

The sharp minimum at $r \approx$ 1--100~m is a newly identified feature in the
size distributions of evolving swarms of particles \citep{kb2015a}. At the small size 
end of the minimum, there is a discontinuity where the relative number of particles 
increases by 2--4 orders of magnitude. This collisional damping front separates 
the swarm into two distinct velocity regimes at a radius $\rdamp$.  Defining 
$e_H = (\mmax / 3 \msun)^{1/3} $ as the Hill eccentricity of the largest object
with mass $\mmax$, we express particle eccentricities as
\begin{equation}
e_{rel} = e / e_H ~ .
\label{eq: erel}
\end{equation}
Particles with $r \gtrsim \rdamp$ have $e_{rel} \approx$ 5--30; small objects with
$r \lesssim \rdamp$ have $e_{rel} \approx$ 1.

This large difference in $e_{rel}$ between small and large particles tempers the
collisional cascade. Although collisions between particles with $r \gtrsim \rdamp$
produces substantial debris, collisions among smaller particles often yield larger 
merged objects. As material cycles between small and large objects, the largest
objects accrete solids from both reservoirs. Eventually, $\sim$ 90\% of the 
initial mass ends up in the largest objects. The collisional cascade converts 
only $\sim$ 10\% of the initial mass into 1~\mum\ and smaller particles.

\begin{figure} 
\includegraphics[width=6.5in]{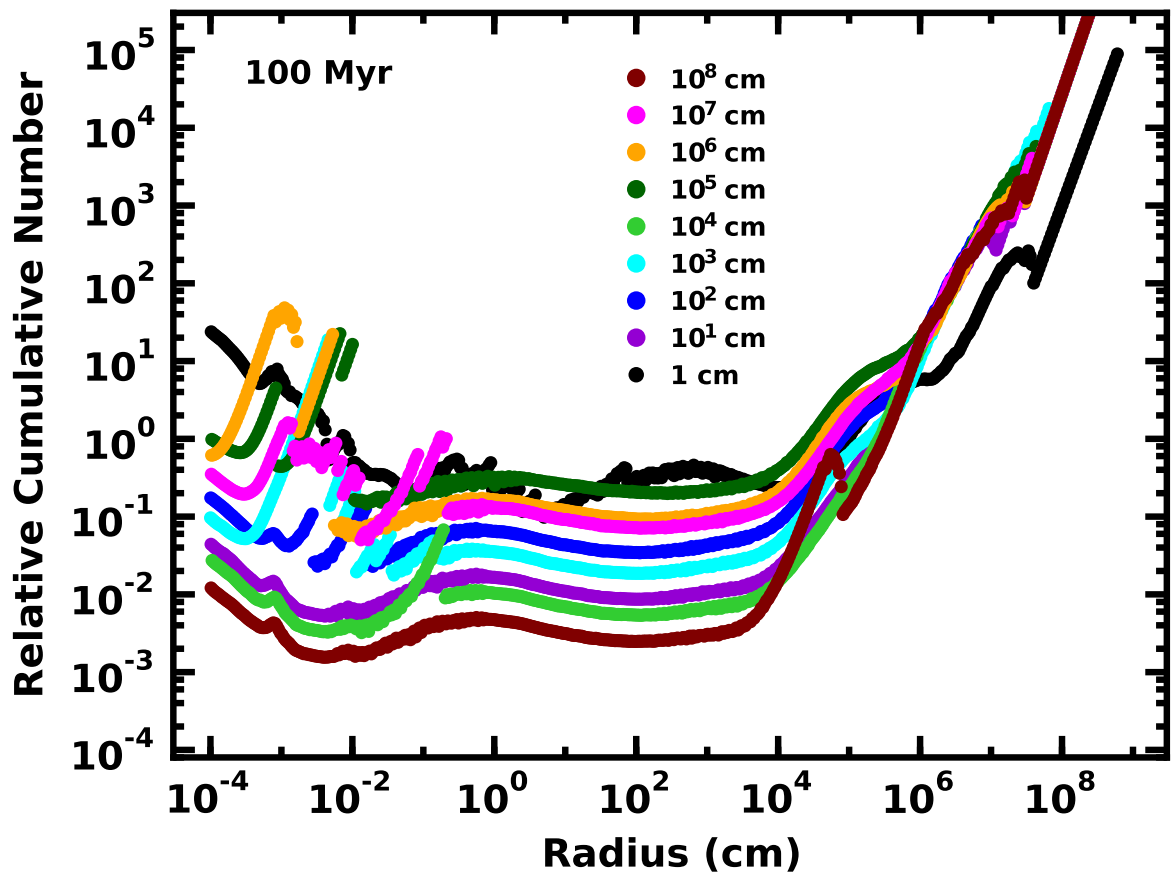}
\vskip 3ex
\caption{
\label{fig: sd7}
Comparison of relative cumulative size distributions at 100~Myr for calculations
with $\Sigma_0$ = 10~\gcms\ at 1~AU for various \r0\ (in cm) as indicated in the 
legend. At $r \gtrsim$ 10~km, all starting points lead to the same size distribution. 
For 0.1~cm to 10~km, the size distributions are nearly identical except for a modest 
offset which depends on the complete collision history but not on \r0. The relative 
abundance of particles with sizes smaller than 0.1~cm depends on the recent history 
of collisions between large objects with radii of 10--1000~km. 
}
\end{figure}

Despite significantly different evolution times for systems with different 
\r0\ (Fig.~\ref{fig: rad1000}), all systems have similar size distributions
at 0.1--1~Gyr (Fig.~\ref{fig: sd7}). Every swarm has one or two objects which 
contain nearly all of the mass. For $r \gtrsim$ 10~km, the cumulative size 
distribution approximately follows a power law with $N(>r) \propto r^{-q_c}$ 
and $q_c \approx$ 4.0--4.3. At intermediate sizes ($r \approx$ 0.1~cm to 10~km),
the cumulative size distribution has a much shallower slope with a waviness
that is similar to the size distributions of collisional cascades
(e.g., Fig.~\ref{fig: sd3}). Among smaller particles with $r \approx$ 1~\mum\ to
1~cm, the size distributions are steeper with more pronounced waviness. This
feature depends on the timing of the last large collision between objects with 
$r \gtrsim$ 100~km. Systems with more recent large collisions have larger rises
to small sizes and more waviness along this rise.

\begin{figure} 
\includegraphics[width=6.5in]{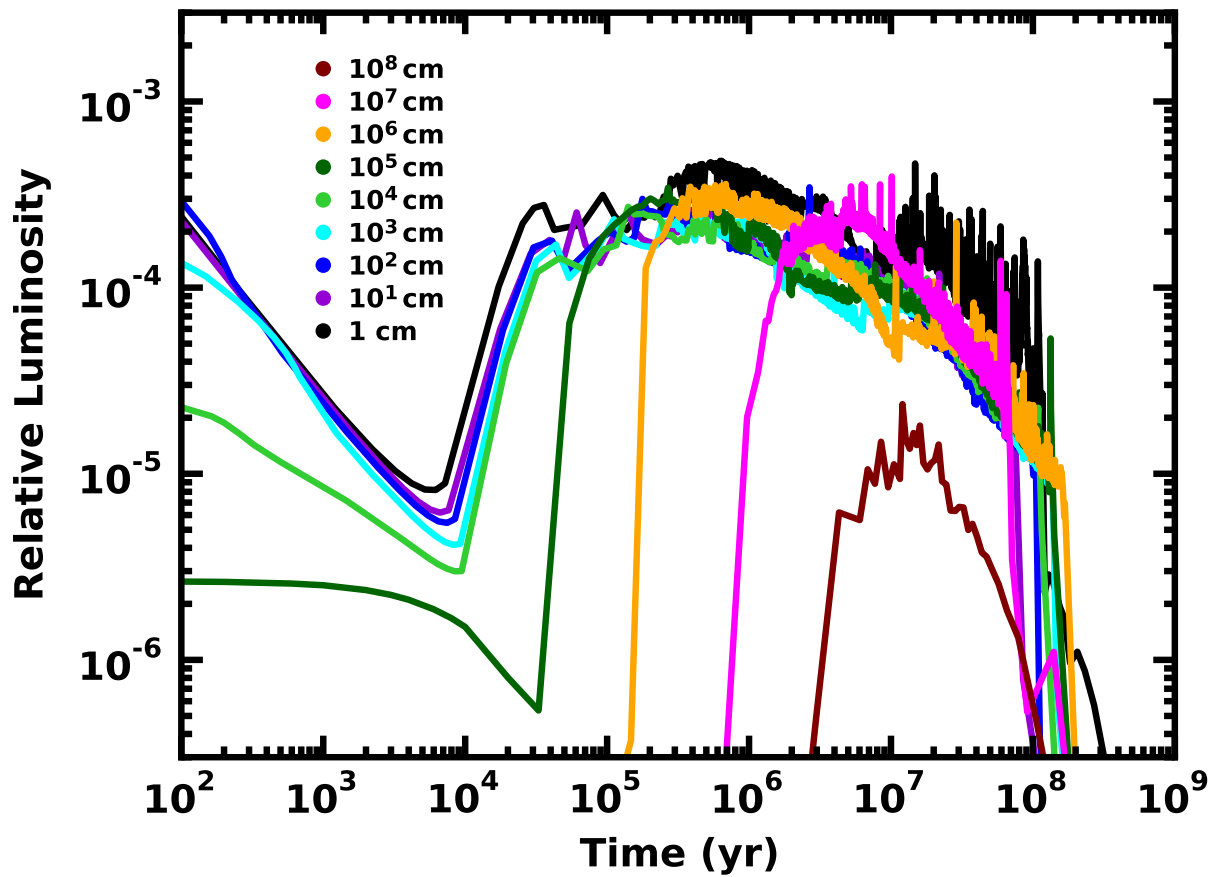}
\vskip 3ex
\caption{
\label{fig: lum6}
Evolution of the dust luminosity for the calculations in Fig.~\ref{fig: rad1000}.
The legend indicates \r0\ (in cm) for each calculation. At early times, particle 
growth dramatically reduces the ratio of the surface area to the mass for the 
entire swarm. The dust luminosity declines. Once the collisional cascade begins, 
the dust luminosity rises rapidly. Despite different rise times for calculations 
with different \r0, the maximum dust luminosity is fairly independent of \r0. 
At late times, the evolution consists of spikes from giant impacts superposed 
on a slow decline in $L_d$.
}
\end{figure}

To examine the observable consequences of this evolution, we consider the dust 
luminosity $L_d$ (Fig.~\ref{fig: lum6}). At early times, growth concentrates 
mass into objects with smaller surface area; $L_d$ declines with time. Once the 
collisional cascade produces sufficient debris, $L_d$ rises. For systems with 
\r0\ $\lesssim$ 0.1~km, dust production rates rise at $\sim 10^4$~yr. Among 
swarms with larger \r0, the rise in $L_d$ occurs at progressively later times, 
reaching $\sim$ 1~Myr when \r0\ = 100~km and 5--10~Myr when \r0\ = 1000~km.

For our calculations, the peak relative dust luminosity,
$L_d / \lstar \approx 2 - 3 \times 10^{-4}$ is relatively 
independent of \r0. Close to the peak, stochastic variations in 
the collision rate produce 50\% variations in $L_d$. The magnitude of these
variations is independent of \r0. Repeat calculations with identical initial
conditions (but different random number seeds) yield similar variations and 
similar peak $L_d$.

When \r0\ $\gtrsim$ 1000~km, peak dust luminosity is much smaller. In these 
systems, most of the initial mass ends up in protoplanets with $r \gtrsim$ 
2000~km which are hard to break. Binary collisions then produce little debris. 
Within this debris, collisional damping of small objects is negligible. Thus, 
the collisional cascade grinds all of the debris into small particles which 
are eventually ejected from the swarm.

After 1--10~Myr, the $L_d$ for all swarms gradually declines. As the systems
decline, fluctuations -- including large spikes in $L_d$ -- become more 
pronounced.  The magnitude and frequency of the spikes is independent of \r0; 
however, the spikes are clearly more obvious at later times.  Although the 
overall evolution of $L_d$ results from countless collisions of 1--100~km 
objects, binary collisions between 300--1000~km objects produce the large 
spikes. These giant impact events release dust masses comparable to (and 
sometimes exceeding) the total dust mass in the rest of the swarm. 

With few large objects in the swarm, impact events are rare and random
\citep[see also][]{genda2015}.  The typical spacing between large collisions 
is $\sim$ 1~Myr at 10~Myr and $\sim$ 1~Myr at 100~Myr. Collisions between the
largest objects produce more dust but are less common.  As summarized in KB05, 
it takes 25--50~yr for the debris from a giant impact to spread into a ring
which merges with the rest of the swarm.  As the debris spreads, we expect 
variations in the emission on time scales of months to years \citep[e.g.,][]
{melis2012,meng2012,meng2014,meng2015}.  Relative to the long-term evolution
of the collisional cascade, the elevated emission from the debris ejected 
during a single giant impact lasts 0.01--0.1~Myr.  At later times, when the 
mass in the rest of the swarm is relatively small, elevated levels of dust 
emission last much longer, $\gtrsim$ 1~Myr \citep[see also][]{genda2015}. 
Overall, the likelihood of observing a giant impact in the terrestrial zone
of any solar-type star is small, $\lesssim$ 0.1\% to 0.2\% for detecting any
dust from a giant impact and $\lesssim$ 0.001\% of detecting emission within
25--50~yr of the impact.  In a large ensemble of stars, however, detecting
debris from giant impacts is possible. For example, KB05 estimate that the
{\it Kepler} satellite might identify 1--2 transit events from impact 
debris during the lifetime of the satellite.

To develop a better understanding of the spikes in $L_d$, we define the break
radius \rbrk\ as the maximum size for a particle destroyed in a collision
between equal mass objects. In our approach, $Q_c = v^2/8$ is the center of 
mass collision energy for a pair of equal mass objects and \qdstar\ is the 
collision energy required to eject half the mass of the colliding pair to 
infinity. As the largest objects grow, their gravity stirs up smaller objects
to larger and larger $e$ and $i$. Particles with larger $e$ collide with 
larger velocities. For fixed $r$, the ratio $Q_c / \qdstar$ grows with time;
collisions among pairs of objects with identical $r$ eject more mass at later
times. Fixing $Q_c / \qdstar$ = 1, larger collision velocities result in 
collisions that destroy pairs of objects with larger and larger $\rbrk$ as the 
system evolves.

\begin{figure} 
\includegraphics[width=6.5in]{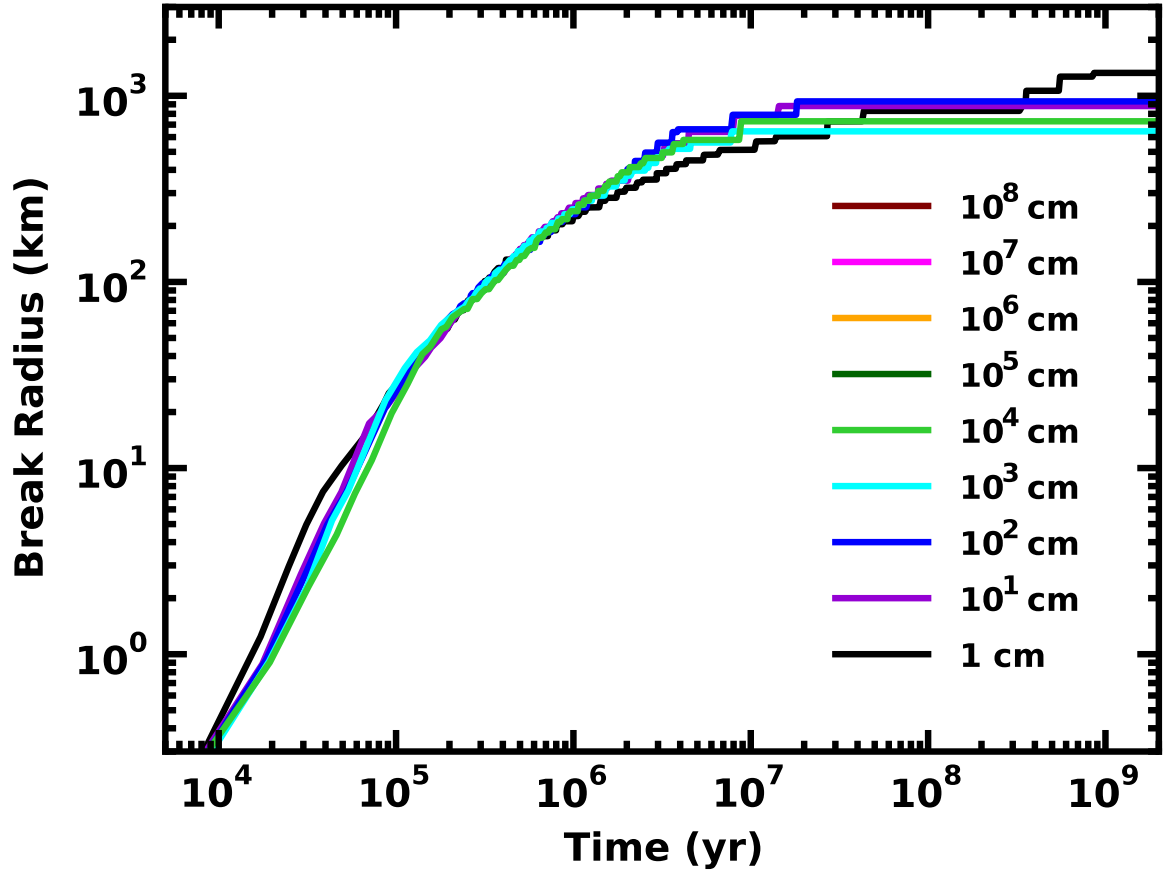}
\vskip 3ex
\caption{
\label{fig: rbreak}
Time evolution of \rbrk\ for planet formation simulations with various \r0\ as indicated
in the legend. Independent of \r0, \rbrk\ gradually increases with time and then reaches
a plateau at \rbrk\ $\approx$ 1000~km.
}
\end{figure}

Fig.~\ref{fig: rbreak} shows the time evolution of \rbrk\ for swarms
with \r0\ = 1~cm to 1000~km. After $10^4$~yr, all systems follow the same 
pattern: \rbrk\ gradually grows from roughly 1~km at 0.01~Myr to roughly 
1000~km at 10~Myr. After 10~Myr, one or two objects contain nearly all of 
the mass. Stirring of $e$ and $i$ effectively ceases; \rbrk\ remains 
roughly constant with time.

This behavior has a profound impact on $L_d$ (Fig.~\ref{fig: lum6}). 
When \rbrk\ is small, every collision among roughly equal mass objects 
yields a merger (for large objects) or a small amount of debris (for 
small objects). This debris has little impact on $L_d$. As \rbrk\ grows, 
binary collisions among objects with $r \approx \rbrk$ produce more and 
more debris. At peak $L_d$, these collisions cause modest fluctuations 
in $L_d$. As $L_d$ declines, collisions among larger and larger objects 
disperse more and more debris, resulting in more prominent spikes in 
$L_d$. By the end of the calculation, giant impacts among 1000~km 
objects lead to factor of 5--10 changes in $L_d$.

Repeat calculations confirm this behavior. Modest changes in $e_0$ and $i_0$ 
have little impact on the results. Over 2~Gyr of evolution, the collisional
cascade converts 10\% $\pm$ 3\% of the initial mass into 1~\mum\ and smaller 
particles. The rest of the initial mass lies in 1--2 large objects with radii
of 5000--6000~km. During the early stages of the cascade (0.1--10~Myr), the
peak dust luminosity is $L_d / \lstar \approx 2 - 3 \times 10^{-4}$.  Near 
peak brightness, factor of 1.5--2 fluctuations in $L_d$ are typical. For 
older systems (10--100~Myr), a slow decline in $L_d$ is occasionally
interrupted by factor of 2--10 spikes in the dust luminosity. As collisions
destroy larger and larger objects, the amplitudes of these spikes gradually 
grow with time. After $\sim$ 100~Myr, the dust luminosity drops by several 
orders of magnitude. The timing of this decline has a broad range,
100--300~Myr, and is fairly independent of \r0\ and $e_0$.

\subsection{Summary}
\label{sec: calcs-summ}

Our suite of numerical calculations yields several interesting insights into
the evolution of collisional cascades. 

For `pure' cascade models with constant $e$ and $i$, we confirm several 
aspects of the analytic model. Throughout the evolution of systems where
\qdstar\ in independent of $r$, the size distribution generally follows 
the expected $N(r) \propto r^{-3.5}$.  When \qdstar\ is a function of $r$, 
the size distribution follows the predicted $N(r) \propto r^{-3.68}$ for
$r \lesssim r_Q \approx$ 0.1~km. For $r \gtrsim r_Q$, the typical slope
of $q \approx$ 2.7--2.9 is smaller than the analytic result 
$q \approx$ 3.04.  As in previous studies, the numerical results show a 
waviness of modest amplitude about these power laws 
\citep[\S\ref{sec: app-waves}, see also][]{campo1994,obrien2003,wyatt2011,kral2013}. 

The long-term evolution of cascades is close to the predictions of the 
analytic model.  In our calculations, the decline is somewhat faster -- 
$L_d, M_d \propto t^{-n}$ and $n$ = 1.1--1.2 -- than the analytic solution, 
where $L_d, M_d \propto t^{-1}$. The ratio $\xi$ of the dust mass to the 
total mass slowly grows with time, $\xi \propto t^n$ and $n$ = 0.05--0.15.  
For fixed $e$ and $i$, the collision time scale for the numerical 
simulations -- $t_c \propto (v^2 / 2 \qdstar)^{-1}$ -- is reasonably close 
to the analytic result where $t_c \propto (v / 2 \qdstar)^{-5/6}$.

A major difference between analytic and numerical models for collisional 
cascades is the evolution of \rmax. Analytic models assume constant \rmax. 
In the numerical calculations, \rmax\ either grows 
($v^2 / 8 \qdstar\ \lesssim$ 1) or declines 
($v^2 / 8 \qdstar\ \gtrsim$ 1) with time. After 1~Gyr of evolution,
factor of 2--10 changes in \rmax\ are typical. In systems where 
\rmax\ declines with time, our results yield $\rmax \propto t^{-n}$ 
with $n$ = 0.1--0.2. With $t_c \propto \rmax$, we then expect 
$L_d \propto t^{-n}$ with $n$ = 1.1--1.2. Thus, the gradual decline
of \rmax\ with time accounts for the more rapid decline of $L_d$ with 
time.  A changing \rmax\ is likely responsible for other, more subtle 
differences between the analytic and numerical model.

Planet formation simulations yield important insights into the onset and
evolution of real collisional cascades. At 1 AU, systems of small particles
with $\Sigma_0 \approx$ 10~\gcms\ evolve into a few large oligarchs in 
0.1--1~Myr. The time scale to produce 2000--4000~km oligarchs is fairly
independent of \r0. For $t \gtrsim$ 10--100~Myr, the size distribution of
small particles and the dust luminosity also have little memory of \r0. 

For collisional cascades, the unique feature of planet formation simulations
is the steady evolution of \rbrk, the maximum particle size destroyed in 
collisions of equal mass objects. In analytic models, \rbrk\ = \rmax\ =
constant. In numerical simulations of planet-forming disks, growing oligarchs
gradually stir up the orbits of smaller solid particles; \rbrk\ grows
persistently with time. As \rbrk\ evolves, more and more material becomes
involved in the collisional cascade. Thus, $L_d$ and $M_d$ decline more
slowly with time.

The relative dust mass $\xi$ also evolves differently in planet formation 
calculations. In cascades with \r0\ $\approx$ 100--300~km, $\xi$ gradually
grows over time (Fig.~\ref{fig: mdust1}). In cascades with \r0\ = 500--1000~km
and in planet formation calculations, the relative dust mass is usually much
smaller ($\xi \approx 10^{-6} - 10^{-5}$) and fluctuates by a factor of 2--20
when giant impacts add debris to the swarm. As the largest objects sweep up
debris, planet formation calculations exhibit large drops in $\xi$ which do
not occur in cascade calculations.

\begin{figure} 
\includegraphics[width=6.5in]{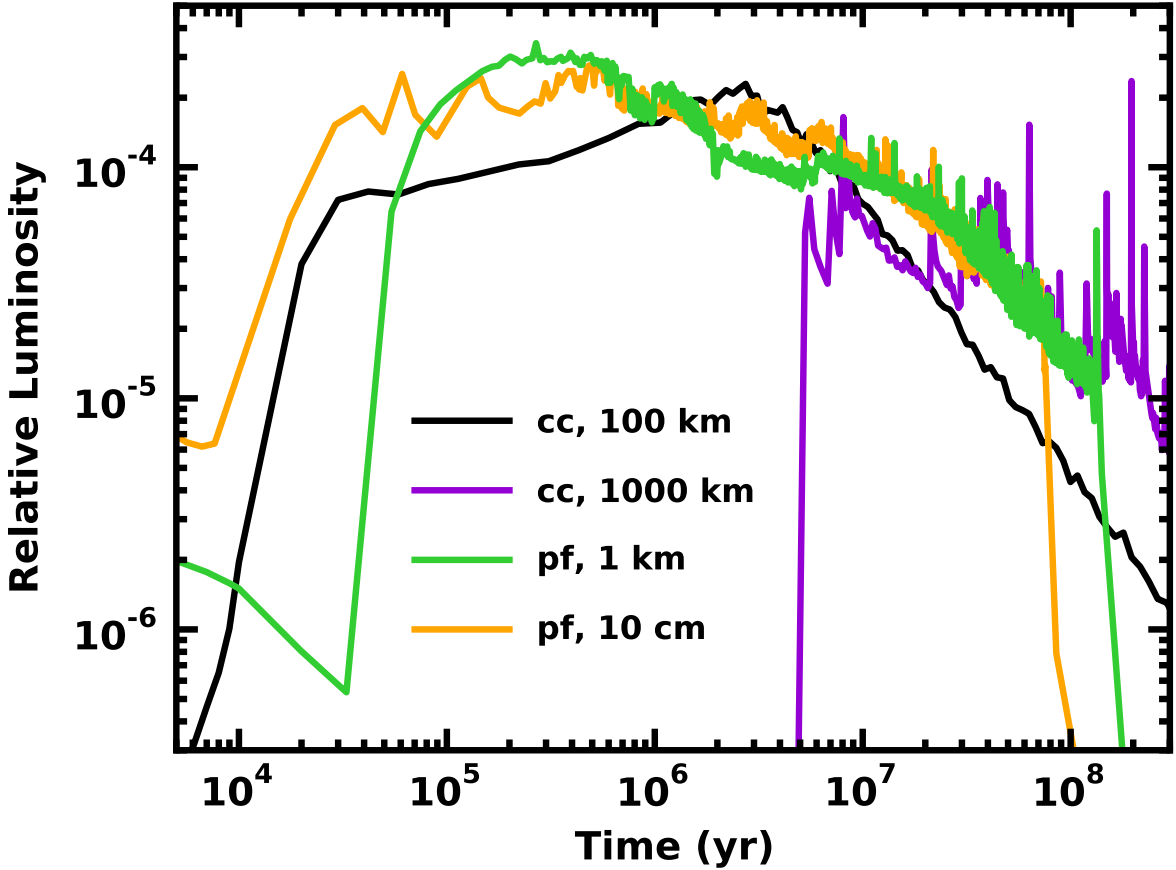}
\vskip 3ex
\caption{
\label{fig: lum7}
Comparison of $L_d/\lstar$ for two collisional cascade calculations (black and violet 
curves, with \r0\ listed in the legend) and two planet formation calculations from 
Fig.~\ref{fig: lum6} (green and orange curves, with \r0\ listed in the legend). 
At early times ($\lesssim$ 10~Myr), three calculations produce similar $L_d$. 
When $t$ = 10--100~Myr, $L_d$ from the cascade with \r0\ = 100~km is considerably 
smaller than from the two planet formation models. Despite negligible $L_d$ at 
early times, the cascade with \r0\ = 1000~km achieves a dust luminosity similar to
the planet formation calculations after 30--40~Myr.  Once $t \gtrsim$ 100~Myr, 
planets have removed all of the smaller particles from the annulus; $L_d$ for 
these simulations then lies well below the predictions of cascade models.
}
\end{figure}

To focus on observable differences between numerical calculations of planet
formation and pure cascades, Fig.~\ref{fig: lum7} compares the evolution of 
$L_d / \lstar$ for two collisional cascade calculations and two planet formation 
calculations. For the cascade calculations, we set \rmax\ = 100~km and 
$e_0$ = 0.1 (black curve) and \rmax\ = 1000~km and $e_0$ = 0.25 (violet
curve). For these cascades, we set the initial mass \M0\ = 0.05~\mearth, 
the mass typically lost from the planet formation calculations.  Tracks for 
the two planet formation calculations repeat those from Fig.~\ref{fig: lum6} 
for \r0\ = 10~cm (magenta curve) and \r0\ = 1~km (cyan curve).

The four sequences show several similarities in the time evolution of $L_d$.
All have roughly the same peak luminosity, 
$L_d / \lstar \approx 1 - 3 \times 10^{-4}$. The rise time for the 100~km
collisional cascade model is similar to rise times for the two planet 
formation calculations. The large spikes in $L_d$ for the 1000~km collisional
cascade model are similar to those in the two planet formation calculations.

Several differences in the evolution are also apparent. When \rbrk\ $\approx$
100~km in the two planet formation calculations (0.1--1~Myr), $L_d$ shows
significant fluctuations not visible in the collisional cascade calculation
with \r0\ = 100~km. When \rbrk\ evolves well past 100~km in the planet formation
calculations, the decline in $L_d$ is much slower than in the collisional
cascade calculation with \r0\ = 100~km. Finally, when the largest object in
the planet formation calculations has accreted most of the mass in the annulus, 
$L_d$ plummets. In contrast, both collisional cascade calculations maintain 
a $t^{-1.1}$ decline for much longer time scales.

\section{DISCUSSION}
\label{sec: disc}

\subsection{Theoretical Issues}
\label{sec: disc-theory}

To follow the evolution of a swarm of planetesimals into a planetary system, 
coagulation calculations include a broad set of physical processes 
\citep[e.g.,][and references therein]{green1978,weth1993,weiden1997b,kl1999a,
kb2008, koba2010b,ormel2010b,raymond2011,glaschke2014,shannon2015,johan2015}. 
Modeling each physical process involves several assumptions and choices for 
various input parameters.  Here, we consider how these decisions impact our 
results and conclusions.

For this study, we ignore interactions between solid particles and a gaseous
disk.  Although gaseous circumstellar disks are a major component in planet 
formation models \citep[e.g.,][]{youdin2013}, all analytic models of collisional 
cascades assume the gas has already dissipated.  Neglecting the gas allows us 
to pinpoint common features in numerical calculations of collisional cascades 
and planet-forming disks.

Although we consider radiation pressure from the central star to set \rmin, we 
do not include Poynting-Robertson drag. For 1~\mum\ particles at 1~AU, the 
collision time is shorter than the time scale for Poynting-Robertson drag when 
$L_d / \lstar \gtrsim 10^{-7}$ \citep[e.g.,][]{bac1993,wyatt2008}. In most of 
the calculations reported here, the dust luminosity meets this limit. 

Unlike several numerical algorithms \citep[e.g.,][]{krivov2006,gaspar2012a,kral2013}, 
our calculations also neglect the radiation force on fragments ejected from a 
collision.  Defining $\beta$ as the ratio of the radiation force to the 
gravitational force, debris particles have orbits with semimajor axis 
$a^{\prime} = ( { 1 - \beta \over 1 - 2 \beta } ) a $ and 
eccentricity $e^{\prime} = \beta / (1 - \beta)$ \citep[e.g.,][]{burns1979}. 
Unbound particles with $\beta \gtrsim$ 1/2 and $r \lesssim$ 1~\mum\ leave the 
swarm on a time scale comparable to the orbital period. For swarms with 
$\Sigma$ = 10~\gcms\ at 1~AU, collisions between bound and unbound particles 
are rare.  Particles with $\beta \approx$ 0.08--0.5 ($r \approx$ 1--6~\mum) 
remain bound to the central star but lie outside the single annulus in our
calculations. 

Despite its absence from current analytic models, it is worth estimating the
impact of high-$\beta$ particles on our planet formation simulations.  With 
larger $a$ and $e$, these particles have longer lifetimes than particles 
inside our single annulus \citep{wyatt2010}.  Fewer collisions tend to increase 
the population of particles with (i) $r \gtrsim$ 6~\mum\ inside the annulus 
and (ii) $r \lesssim$ 6~\mum\ outside the annulus. Despite producing a steeper
size distribution at small sizes, the impact on $L_d$ is probably small
\citep{wyatt2010}. With negligible total mass, the abundance and orbits of
this population have little impact on the growth of the largest objects. We
plan multiannulus calculations to investigate these possibilities in more detail.

To calculate collisional processes, we specify a variety of parameters for 
collisional damping, the cross-section, gravitational stirring, and fragmentation. 
Comprehensive analytic and \nbody\ calculations verify the algorithms for 
collisional damping, the collision cross-section, and stirring \citep[e.g.,][]
{green1990,spaute1991,oht1999,lee2000,oht2002,gold2004,lev2007,koba2010a,kb2015a}. 
Within the fragmentation algorithm, we derive material lost from `cratering' and 
`catastrophic disruption' using the expressions in eq.~\ref{eq: qd}, 
eqs.~\ref{eq: mej}--\ref{eq: mlarge}, and a power law size distribution for the 
debris. For cratering collisions where the mass of one component (the `projectile')
is much smaller than the other (the `target'), results for the ejected mass agree 
with detailed analyses of laboratory measurements \citep[see the discussion in
\S3.2.1 of][and references therein]{kcb2014}. Criteria for catastrophic disruption
are consistent with comprehensive \nbody\ simulations \citep[e.g.,][]{benz1999,
lein2009,lein2012}. At present, these algorithms are state-of-the-art.

For cascade models, Fig.~\ref{fig: lum2} illustrates the impact of changing 
\qdstar\ on the collision time and the dust luminosity. For fixed $e$ and $i$, 
systems where \qdstar\ is
small evolve more rapidly than systems where \qdstar\ is large. Within a suite
of planet formation calculations, the impact of \qdstar\ is less dramatic 
\citep[e.g.,][]{kb2008,kb2010,kb2012}. Although small particles with 
$r \lesssim$ 1~km can be strengthless rubble piles 
\citep[e.g.,][]{weiden1994,asp1996,hols2007}, large reductions in $Q_b$ and 
$\beta_b$ probably have little impact on the evolution \citep[e.g.,][]{kb2008}. 
Among larger objects with $r \gtrsim r_Q$, self-gravity limits the ability to 
change \qdstar\ \citep[e.g.,][]{benz1999,lein2012}. However, even large changes 
in $Q_g$ and $\beta_g$ may not change $L_d$ significantly \citep[e.g.,][]{kb2012}.

Once a few large objects contain most of the mass in the annulus, the approximations
used in our statistical estimates for collision rates and stirring begin to break 
down. As the largest protoplanets continue to accrete material from the rest of the
swarm, their dynamical interactions become stronger and more chaotic. Among these 
large objects, our calculations then tend to overestimate the growth rate and 
underestimate $e$ and $i$. 

To avoid these issues in several previous investigations, we promote the largest 
objects into the \nbody\ component of \orch, where we follow the orbits of each 
large protoplanet \citep[e.g.,][]{kb2006,bk2011a,kb2014}.  In this study, our 
main focus is on the evolution of the collisional cascade instead of the final 
orbits of the largest objects. To save cpu time for a broad suite of simulations, 
we therefore chose not to promote the largest protoplanets.  When these protoplanets 
contain nearly all of the mass, calculations for the rest of the swarm -- including 
the flow of mass from \rbrk\ to \rmin\ and the accretion of small objects by the 
largest objects -- remain accurate. The evolution of the largest objects then have 
little impact on the evolution of the mass and luminosity of the small particles. 

In multiannulus calculations of terrestrial planet formation, promotion into the
\nbody\ code is more important \citep[e.g.,][]{kb2006,raymond2011,raymond2012}.
When protoplanets contain more than half the mass, they undergo a period of 
chaotic growth where the largest protoplanets scatter smaller objects throughout
the terrestrial zone. Giant impacts are then more frequent and occur at higher
velocity than in pure coagulation calculations \citep[see also][]{asphaug2006,genda2015}.
Thus, our single annulus calculations probably somewhat underestimate the number 
and magnitude of large spikes in dust luminosity on time scales $\gtrsim$ 10~Myr. 
Our main conclusion -- that the dust luminosity in coagulation calculations 
declines more slowly than in collisional cascade calculations -- is unchanged.

\subsection{Comparison with Other Calculations}
\label{sec: disc-other}

In the past 10--15~yr, there have been many numerical studies of collisional cascades 
and planet-forming disks using coagulation and related techniques.  Most investigations 
concentrate on the evolution at large distances, 10--200~AU, from the host star
\citep[e.g.,][]{kb2002a,krivov2006,theb2007,lohne2008,gaspar2013,kral2013,koba2014}.
Where possible, we compare our results with these studies and the few analyses which 
focus on the long-term evolution of fairly massive disks in the terrestrial zone.  

In systems with large $v^2 / 2 \qdstar$ and short $t_c$, the linear decline of 
mass with time is a basic feature of the analytic model for collisional cascades 
\citep[e.g.,][]{wyatt2002,dom2003,wyatt2008,koba2010a}. Aside from 
\citet{koba2010a}, results from other studies do not test this prediction: 
the largest objects (i) are allowed to grow or (ii) have very long collision 
times \citep[e.g.,][]{krivov2006,lohne2008,gaspar2012b,gaspar2013,kral2013}. 

For all of our calculations, the disk mass declines with time somewhat more rapidly 
than the analytic prediction: $M_d \propto t^{-n}$ with $n$ = 1.1--1.2 instead of 1. 
In every calculation, catastrophic and cratering collisions slowly reduce the size 
of the largest objects.  As the radius of the largest objects grows smaller, the 
collision time also decreases (eq.~\ref{eq: tc1}). Shorter collision times speed up 
the evolution, resulting in a larger mass loss rate per unit time.  Thus, the slow
reduction in \rmax\ produces a faster evolution of $M_d$. Calculations with other
codes \citep[e.g.,][]{lohne2008,koba2010a,kral2013} could test this result.

Tests of the variation of $t_c$ with \qdstar\ are currently ambiguous. The numerical 
simulations of \citet{koba2010a} confirm the analytic result, 
$t_c \propto (v^2 / \qdstar)^{-5/6}$. However, \citet{lohne2008} quote 
$t_c \propto (v^2 / \qdstar)^{-9/8}$. Our result, $t_c \propto (v^2 / \qdstar)^{-1}$,
lies roughly midway between these two extremes.

All calculations produce wavy size distributions \citep[e.g.,][]{krivov2006,lohne2008,
gaspar2012b,kral2013}. For small particles and intermediate mass particles with 
$r \lesssim$ 1~km, various approaches yield fairly similar results \citep[see also][]
{campo1994,obrien2003}. Among larger particles, different treatments of the evolution
of the largest particles produce waves with somewhat different amplitudes and positions.
However, these differences seems small compared to changes in \rmax\ and the slope of
the size distribution as a function of time.

When numerical simulations consider the long-term evolution of objects with 
$r \gtrsim$ 300--500~km and $v^2 / 8 \qdstar \gtrsim$ 1, giant impacts which release 
copious amounts of dust are inevitable \citep[e.g.,][]{agnor1999,agnor2004,genda2015}.
The giant impact events in our calculations are similar to those in \citet{weiden2010b}
and \citet{raymond2011}. 

To conclude this section, our results for planet formation time scales in 
the inner solar system are similar to those reported previously 
\citep[e.g.,][and references therein]{kominami2002,kb2004b,naga2005,raymond2005,
kb2006,obrien2006b,kokubo2006,chambers2008,raymond2014a}. Typically, it takes
$10^4 - 10^5$~yr to produce lunar mass protoplanets.  Collisions combine these
objects into a few 3000--6000~km objects over 10--100~Myr. Because we neglect 
gas drag, our calculations take a factor of $\sim$ 2 longer to form lunar mass 
objects \citep[e.g.,][]{youdin2013}. Gas drag has little impact on the evolution
of 100--1000~km objects; the growth of these objects in our simulations agree with
rates established by other investigations.

\subsection{Observational Issues}

Based on a set of multiannulus coagulation calculations at 0.68--1.32~AU,
\citet[][hereafter KB04]{kb2004b} first predicted the detectability of 
warm dust (T $\approx$ 200--300~K) from terrestrial planet formation. 
\citet[][hereafter KB05]{kb2005} later described results for a larger 
disk covering 0.4--2~AU and suggested the possibility of detecting debris 
ejected from individual giant impacts of 1000~km and larger 
objects\footnote{KB05 also predicted levels for warm dust emission in the
terrestrial zones of A-type stars.}.  In 
these studies, the predicted mid-IR (10--25~\mum) emission from dust reaches 
a peak relative flux $F_d / F_\star \approx$ 10--20 at 0.1--1~Myr and gradually
falls to $F_d / F_\star \approx$ 2--3 at $\sim$ 10~Myr. Superimposed on 
this decline, spikes from giant impacts raise the relative flux by factors
of 3--10. Rapid variability of dust production on 1--100~yr time scales is 
another characteristic of these calculations.

Recent investigations confirm these results \citep[e.g.,][]{weiden2010b,raymond2011,
raymond2012,jackson2012,genda2015}. Once 500--1000~km objects form, destructive 
collisions produce copious amounts of debris throughout the terrestrial zone. Giant 
impacts, including those capable of producing the Earth-Moon system, generate additional
debris. As the overall level of debris declines, sporadic spikes in emission from 
the giant impacts become more prominent. At late times, these spikes dominate dust
production.

Although the new calculations in \S\ref{sec: calcs-pf} yield luminous debris disks 
at 0.1--100~Myr as in KB04 and KB05, the long-term evolution of systems with
\r0\ $\lesssim$ 100~km is somewhat 
different. By explicitly following the dynamical evolution of 1~\mum\ to 1~m particles,
we now identify an evolutionary phase at 1--10~Myr where collisional damping among
mm- and cm-sized particles overcomes viscous stirring by large oligarchs. During
this epoch, the lack of destructive collisions among small particles effectively
halts the cascade. With a slower cascade, these systems lose a smaller fraction of
their initial mass and are 2--3 times fainter at peak brightness than the debris 
disks in KB04 and KB05. However, the disks also remain brighter for a factor 
of 2--3 longer, allowing the debris from terrestrial planet formation to remain 
detectable among older stars. 

Scaling the results of the single 0.9--1.1~AU annulus in \S\ref{sec: calcs-pf} to 
the KB05 multiannulus calculations at 0.4--2 AU, the new calculations predict 
typical peak relative fluxes 
$F_d / F_\star \approx$ 2--3 at 10--12~\mum\ and 
$F_d / F_\star \approx$ 3--5 at 20--25~\mum.
Occasional bright spikes from giant impacts can raise the peak flux by a factor
of 3--10.
Although the bulk emission usually declines below detectable levels at 10--20~Myr, 
giant impacts occasionally raise the system above current detection limits until 
$\sim$ 100~Myr \citep[see also][]{raymond2011,raymond2012,jackson2012,genda2015}.

Calculations with $\r0 \gtrsim$ 1000~km yield very different outcomes. Viscous
stirring from the largest objects dominates the evolution; collisional damping of 
small particles is unimportant. Although the largest protoplanets still reach sizes
of 4000--6000~km, the dust luminosity always remains very low. The systems are not
detectable with \spitz\ or \wise.

Despite the identification of many cold debris disks around solar-type stars, 
initial attempts to detect warm emission with \iras\ and ground-based data
were discouraging \citep[e.g.,][]{wein2004,mama2004}.  \citet{song2005} 
then discovered a very luminous debris disk associated with the old field 
star BD+$20 \deg\ 307$. \iras\ also measured a significant 12~\mum\ excess
associated with HD~23514, a member of the Pleiades cluster; confirmation as 
a warm debris disk required data from \iso, \spitz, and ground-based 
telescopes \citep{spangler2001,rhee2008}.

Over the past decade, data from \akari, \spitz, and \wise\ have enabled many 
surveys for 12--25~\mum\ emission from debris orbiting solar-type main 
sequence stars \citep[e.g.,][]{stauffer2005,silver2006,beich2006,siegler2007,
gorlova2007,
currie2007b,meyer2008,trill2008,carp2009a,carp2009b,moor2009,stauffer2010,
beich2011,chen2011,smith2011,zuck2011,luhman2012b,ribas2012,urban2012,kw2012,
zuck2012,fujiwara2013,kenn2013a,clout2014}.  Current compilations include 
stars with ages ranging from $\sim$ 10~Myr to $\gtrsim$ 1~Gyr and relative 
fluxes of $F_d / F_\star \approx$ 0.5--100 at 8--24~\mum.  In some systems,
high quality spectra and imaging data confirm the presence of silicate dust 
orbiting within 1--3~AU of the central star \citep[e.g.,][]{smith2008,
currie2011,olofsson2012,smith2012,schneider2013,ballering2014}.  For several
others, short time scale variations in $F_d / F_\star$ suggest rapid changes 
in the typical sizes and total mass of small particles 
\citep[e.g.,][]{melis2012,meng2012,meng2014,meng2015}. Interferometric 
observations further reveal 300~K or hotter dust in some systems
\citep[e.g.,][]{absil2013,ertel2014,menne2014,defrere2015}.

Taken together, the complete set of data suggest a fairly low warm dust 
frequency for solar-type stars \citep[see also][]{stauffer2005,gorlova2007,
carp2009a,carp2009b,chen2011,luhman2012b,ribas2012,fujiwara2013,kenn2013a,
patel2014,clout2014}.  
Among older stars with ages $\gtrsim$ 100~Myr, only $\sim$ 0.01\% have 
debris disks with $F_d / F_\star \gtrsim$ 0.2--0.5 at 10--12~\mum. For
younger stars, the frequency is $\sim$ 1\% for 100 Myr-old stars and 
$\sim$ 2\% to 3\% for 10--20~Myr stars.  

These statistics are generally consistent with the standard theoretical 
picture where a collisional cascade and occasional giant impacts generate 
a large population of small particles in the terrestrial zone 
\citep[see also][]{raymond2011,fujiwara2013,kenn2013a,genda2015}.  
As in Fig.~\ref{fig: lum7}, emission from the cascade gradually declines 
with time from 1--10~Myr to $\sim$ 100--150~Myr. Throughout this period and 
beyond, giant impacts are responsible for large spikes in dust emission 
\citep[see also Fig. 6 of][]{genda2015}. Typical predictions for the 
level of dust emission are close to those observed.

To make a basic quantitative comparison of our calculations with observations, 
we scale the dust emission from the 0.9--1.1~AU annulus to an annulus at 
0.4--2~AU and derive the fraction of time spent above the \spitz\ and 
\wise\ detection limits ($F_d / F_\star \approx$ 0.2--0.5 at 8--12~\mum). 
Based on current planet detection statistics, we assume 50\% of solar-type 
stars have at least one Earth-mass planet at 0.4--2~AU 
\citep[e.g.,][]{youdin2011b,petig2013b,foreman2014,burke2015}.  For these 
systems, the initial surface density distribution adopted in our calculations  
is a reasonable choice \citep[e.g.,][]{chambers2001a,kominami2004,naga2005,
kb2006,kokubo2006,obrien2006b,chambers2008,fogg2009,lunine2011,
raymond2011,raymond2012,hansen2012,hansen2013}.  

Among this set of calculations, the expected detection frequency $f_d$ 
depends on stellar age and \r0, the maximum size of solid objects at 
the start of the calculation. For any \r0, $f_d \lesssim$ 1\%--10\% for 
stellar ages of 30--100~Myr; $f_d \lesssim$ 1\%--2\% for stars older than
100~Myr. These predictions are consistent with observations derived from
\spitz\ and \wise. 

Observations of dust emission for younger stars appear to rule out the 
possibility that a large fraction of planetary systems produce Earth-mass 
planets starting from 1000~km or larger objects. When $\r0 \gtrsim$ 1000~km, 
our calculations predict $f_d \lesssim$ 1\%--2\% for all stellar ages.  
Thus, these systems simply produce too little dust at ages when observations 
suggest at least some systems produce dust.  Data for the size distribution 
of asteroids in the solar system also favor swarms with smaller initial 
planetesimal sizes \citep[e.g.,][]{morby2009b,weiden2011,johan2015}. 

Compared to observations of stars with ages of 1--30~Myr, calculations with 
$\r0 \lesssim$ 500~km tend to produce too much dust. Among the youngest stars 
(1--5~Myr), the predicted $f_d$ is a strong function of \r0, ranging from 
$f_d \approx$ 15--50\% when $\r0 \lesssim$ 100~km to $f_d \lesssim$ 5\% for 
$\r0 \approx$ 500~km.  For older stars with ages of 5--10~Myr (10--30~Myr), 
$f_d \approx$ 30\%--40\% (5\%--10\%) independent of \r0. Given current 
constraints on warm dust for stars with ages of 1--30~Myr, our predictions 
agree with observations for 20--30 Myr old stars but are too large by factors 
of 2--5 for younger stars. The differences between theory and observations 
for these young stars are fairly independent of \r0.

Among younger stars, the difference between theory and observations 
is probably less extreme than indicated by these statistics.  In our 
simulations, the typical level of dust emission at 1--10~Myr is only 
a factor of $\sim$ 2--3 larger than current detection limits.  Modest 
reductions in the population of small particles are possible if 
(i) the smallest particles emit less efficiently than blackbodies
\citep[e.g.,][]{bac1993},
(ii) the total mass in solids is somewhat smaller at 0.4--2~AU,
(iii) the radiation field or the stellar wind of young solar-type stars 
removes particles with $r \gtrsim$ 1~\mum\ \citep{plavchan2005,wurz2012}, 
(iv) small particles are much weaker than assumed,
(v) collisional damping is more effective among small particles, or
(vi) protoplanets accrete small particles more efficiently. Although
investigating these possibilities is beyond the scope of this paper, 
we plan to analyze their likelihood in future studies.  Better limits 
on the frequency of debris disks with relative fluxes 3--5 times smaller 
than current detection limits would help to guide this analysis.

\section{SUMMARY}
\label{sec: summ}

We analyze a suite of coagulation calculations to examine the long-term 
evolution of collisional cascades in a ring of solid material with
surface density $\Sigma_0 \approx$ 10~\gcms\ at 1~AU from a solar-mass star. 
In pure cascade models where the orbital $e$ and $i$ of the solids remain 
constant with time, destructive collisions slowly reduce the size of the 
largest objects by factors of 3--20. Over 1--2 Gyr, the total mass, the 
mass in small particles ($r \lesssim$ 1~mm), and the luminosity decline 
as $t^{-n}$, with $n$ = 1.1--1.2. The ratio $\xi$ of the mass in small 
particles to the total mass grows by a factor of 2--3 from 1~Myr to 2~Gyr. 
After 1~Myr, the shape of the size distribution for $r \gtrsim$ 1~\mum\ is
fairly constant in time and approximately follows a power law, 
$N(r) \propto r^{-q}$. The slope $q$ depends on the form of the relation for 
the binding energy \qdstar: for constant \qdstar, $q$ = 3.5; for $\qdstar(r)$,
$q_s$ = 3.68 at small sizes and $q_l \approx$ 2.7--2.9 at large sizes.  Waves 
superimposed on the power law have modest amplitude and are most pronounced 
at the smallest and largest sizes in the cascade. 

Comparison of these results with the predictions of analytic models highlights
several robust outcomes. For the size distribution, analytic estimates for $q_s$, 
the overall level of waviness, and the approximately invariant shape from 
1~\mum\ to 6--30~km agree with results from the coagulation calculations. 
Analytic models underestimate the mass lost from destructive collisions.
Coagulation calculations have smaller total mass, dust mass, and 
dust luminosity than analytic models. These differences between analytic models and 
numerical calculations probably stem from the variation of the size of the largest 
object with time. In the coagulation calculations, large reductions in \rmax\ shorten
the collision time and speed up the evolution.

To place the collisional cascade calculations in context with planet formation models, 
we examine a suite of coagulation calculations where gravitational interactions
among the solids yield an internally consistent $e$ and $i$ as a function of 
particle size. For swarms with $\Sigma_0$ = 10~\gcms, \r0\ = $10^n$, and $n$ = 
0, 1, 2, $\ldots$, 8, it takes 0.1--1~Myr for the largest objects to reach sizes 
of 2000--4000~km. As these protoplanets grow, their gravity stirs up the $e$ and
$i$ for smaller particles, initiating a collisional cascade.  From 0.1--10~Myr, 
the sizes of the largest objects that suffer catastrophic collisions grows from
\rbrk\ $\approx$ 10--20~km to \rbrk\ $\approx$ 1000~km. 
When $\r0 \lesssim$ 100~km, the dust luminosity 
reaches a peak $L_d / \lstar \approx 2 - 3 \times 10^{-4}$ and then slowly 
declines. During this decline, giant impacts between 300--1000~km and larger
objects release 
large amounts of debris, producing pronounced spikes in $L_d$. Eventually, 
accretion by large objects and destructive collisions among smaller objects 
eliminate all of the smaller particles. The dust luminosity then drops to zero.

Compared to analytic and numerical models of pure collisional cascades, the planet
formation calculations reveal several stark differences. In planet formation 
calculations, the persistent growth in \rbrk\ involves larger and larger objects
in the cascade; \rbrk\ is constant in pure cascades.  Until the precipitous drop 
in $L_d$ when protoplanets achieve their final masses, the decline of $L_d$ in
planet formation models is slower in time. Throughout the evolution, these models 
also produce a higher frequency of spikes in $L_d$. Despite the similar maximum
$L_d$, planet formation models typically have a factor of 3--10 less mass in small
particles ($r \lesssim$ 1~mm) than pure cascade models.

Observations of known debris disks around solar-type stars provide interesting 
tests of these calculations.  Measurements of the frequency and relative fluxes 
as a function of stellar age suggest dust produced in collisional cascades and 
giant impacts contribute to the debris. The lack of dust from planet formation 
calculations with \r0\ $\gtrsim$ 1000~km suggests terrestrial planets grow out
of disks with 300--500~km or smaller planetesimals.  For older stars with ages 
exceeding 20--30~Myr, the predicted detection frequency agrees with observations.  
However, the observed detection frequency for 5--20~Myr old stars is much smaller 
than expected. Investigations into the accretion, collisional damping, and the 
strength of small particles might lead to calculations which produce somewhat 
less dust emission.  Better constraints on the frequency of (i) Earth-mass 
planets at 1--2~AU and (ii) debris disks with $L_d / L_\star \ll 10^{-4}$ would 
allow more stringent tests of the calculations.

\acknowledgements
We acknowledge a generous allotment of computer time on the NASA `discover' cluster. 
Comments from M. Geller, G. Kennedy, J. Najita, and an anonymous referee improved 
our presentation. 
Portions of this project were supported by the {\it NASA } {\it Astrophysics Theory}
and {\it Origins of Solar Systems} programs through grant NNX10AF35G and the
{\it NASA Outer Planets Program} through grant NNX11AM37G.

\appendix

\section{Appendix}
\label{sec: app}

To test the algorithms used in \orch, we compare numerical results with analytic 
solutions to the coagulation equation \citep{kl1998,kb2015a} and published 
results from other investigators \citep{kb2001,bk2006,kb2008,bk2011a,kb2015a}.
Here, we examine how \orch\ performs for collisional cascades.

\subsection{The Mass Spacing Parameter}
\label{sec: app-delta}

The accuracy of coagulation calculations depends on the mass spacing parameter
between adjacent mass bins, $\delta_k = m_{k+1} / m_k$ 
\citep[e.g.,][]{weth1990,kl1998,kb2015a}. At the start of each calculation, we 
set the typical mass $m_k$ and the boundaries $m_{k-1/2}$ and $m_{k+1/2}$ of 
each mass bin. These parameters remain fixed throughout a calculation. The 
initial average mass within each bin is $\bar{m}_k = M_k / N_k$; typically
$\bar{m}_k = m_k$. As the calculation proceeds, collisions add and remove 
material from all bins; the average mass in a bin $\bar{m}_k$ and the average
physical radius of particles in the bin 
$\bar{r}_k = (3 \bar{m}_k / 4 \pi \rho_p)^{1/3}$ then change with time. 

To illustrate how the evolution depends on $\delta$, we consider an annulus
centered at $a$ = 1~AU. The annulus has a width $\Delta a = 0.2 a$; thus the
inner and outer radii are $a_{in}$ = 0.9~AU and $a_{out}$ = 1.1~AU. We seed 
the annulus with particles having mass density $\rho_p$ = 3~\gcmc, $e$ = 0.1,
and $i$ = $e$/2. The annulus has surface density $\Sigma_0$ = 10.6~\gcms\ and 
total mass $M_0$ = 0.5~\mearth. 

\begin{figure} 
\includegraphics[width=6.5in]{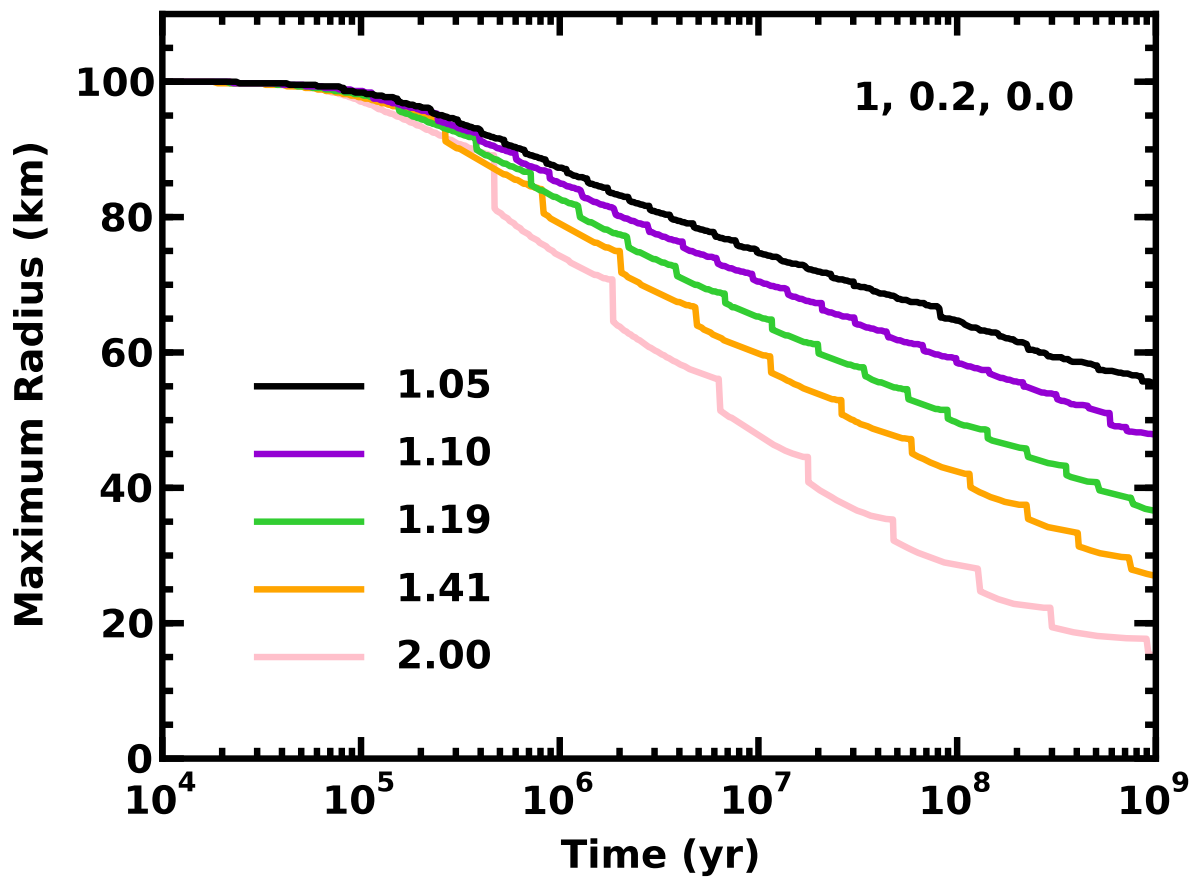}
\vskip 3ex
\caption{
Time evolution of \rmax\ in collisional cascades with a mono-disperse
ensemble of 100~km objects and no velocity evolution. The legend in
the lower left corner indicates the mass spacing factor $\delta$ for 
each calculation. In the upper right corner, the legend indicates
adopted values for $b_d$, $m_{l,0}$, and $b_l$.  Continuous cratering 
collisions produce a gradual decline in \rmax.  Downward jumps in 
\rmax\ occur when mass loss moves the largest objects into the next 
smallest mass bin.  Calculations with larger $\delta$ produce larger 
and less frequent jumps.
\label{fig: rmax0}
}
\end{figure}

Fig.~\ref{fig: rmax0} shows the time-variation of \rmax\ (the radius of the 
largest object) for calculations with a mono-disperse set of 100~km objects 
and five different values of $\delta$.  All curves have the same general shape: 
after a brief, $\sim 10^5$~yr period where \rmax\ is roughly constant, the size 
of the largest objects declines monotonically with time. Tracks with larger 
$\delta$ decline faster. 

Along each track, the evolution consists of a gradual reduction in \rmax\ interspersed 
with large jumps to smaller \rmax. Cratering collisions -- where somewhat smaller 
objects gradually chip away at the mass of larger objects -- produce continuous mass
loss from the largest objects. Thus, the average mass in the largest mass bin falls 
with time.  Eventually, this mass falls below the mass boundary between adjacent bins 
(e.g., $\bar{m}_k < m_{k-1/2}$). Objects in bin $k$ are then placed into bin $k-1$. 
Averaging the mass of the `old' objects in bin $k-1$ with the `new' objects from bin 
$k$ yields a new average mass $\bar{m}_{k-1}$ which is smaller than the average mass 
of bin $k$.  Thus, the radius (average mass) of the largest object jumps downward. 
Because the spacing of mass bins scales with $\delta$, calculations with larger 
$\delta$ have larger jumps than those with smaller $\delta$.

Although the mass loss rate from the grid is fairly insensitive to $\delta$ (see below),
the mass of the largest object declines faster in calculations with larger $\delta$. 
Cratering collisions are responsible for this difference. For all $\delta$, these
collisions are rare. Thus, only a few of the largest objects suffer substantial mass
loss from cratering collisions every time step.  When $\delta$ is small (1.05--1.10), 
these objects are placed into the next smallest mass bin; the average mass of the 
remaining objects in the mass bin is unchanged. When $\delta$ is large (1.41--2.00),
the amount of mass loss is not sufficient to place objects into the next smallest mass 
bin; the average mass of all objects in the bin then decreases. As a result, the average
mass of the largest objects declines faster when $\delta$ = 2 than when $\delta$ = 1.05.

\begin{figure} 
\includegraphics[width=6.5in]{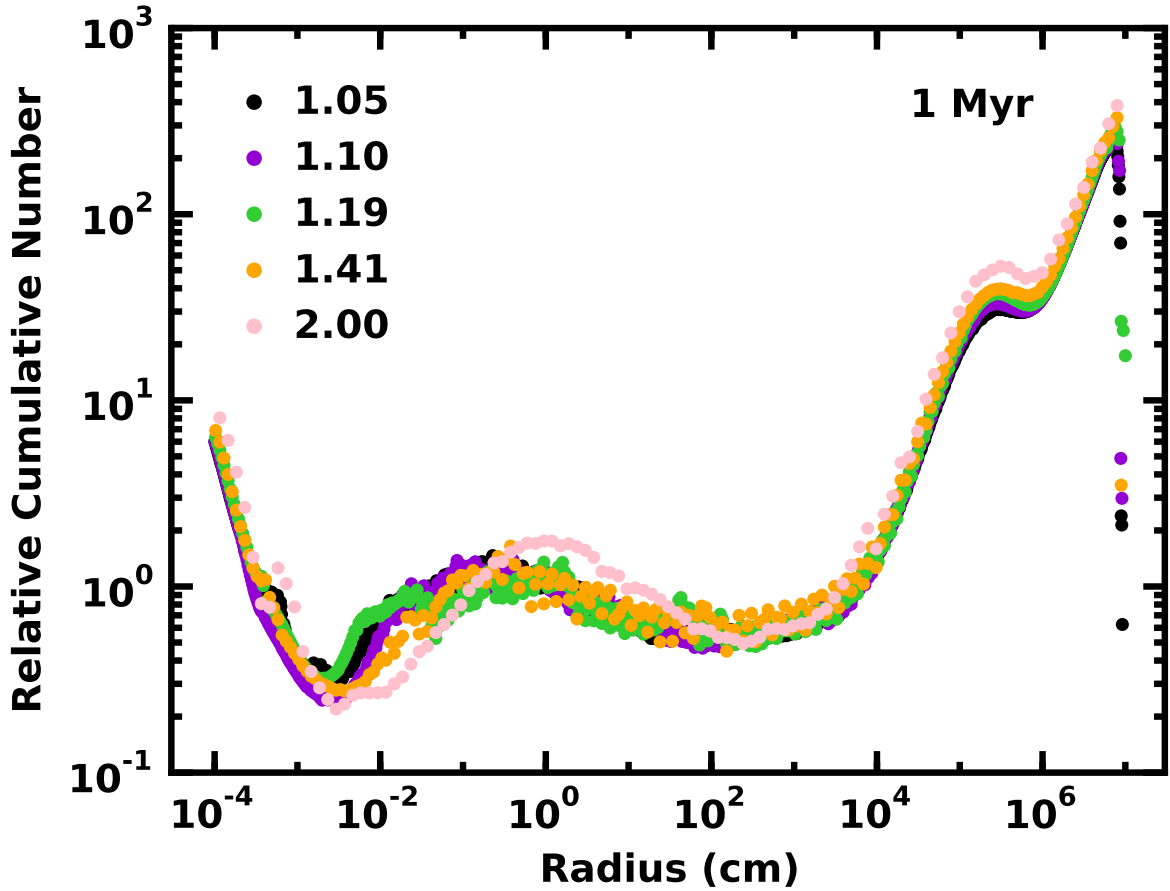}
\vskip 3ex
\caption{
Relative cumulative size distributions $N(>r) / r^{-2.5}$ for collisional 
cascades at 1~Myr.  All of the curves have a similar morphology:
(i) an excess for particle sizes $r \approx$ 1--5~\mum,
(ii) a deficit for $r \approx$ 5--50~\mum,
(iii) a constant part for $r \approx$ 0.01~cm to 0.1~km, and
(iv) a steep rise for $r \gtrsim$ 0.1~km with a small shoulder 
at $r \approx$ 1--3~km. Calculations with larger $\delta$
(as indicated in the legend) have larger fluctuations (waves) 
about a smooth curve.
\label{fig: sd0a}
}
\end{figure}

Despite this difference, other aspects of the evolution are fairly insensitive to $\delta$.
Fig.~\ref{fig: sd0a} shows the relative size distributions at 1~Myr. Each curve follows
a standard pattern, with an excess (deficit) of particles at 1--10~\mum\ (10--100~\mum),
a wavy pattern at 0.1~mm to 0.1~km, and then a rise at larger sizes. For $\delta$ = 1.05--1.41,
there is little difference between the curves. When $\delta$ = 2, the amplitude of the waves
is larger. 

\begin{figure} 
\includegraphics[width=6.5in]{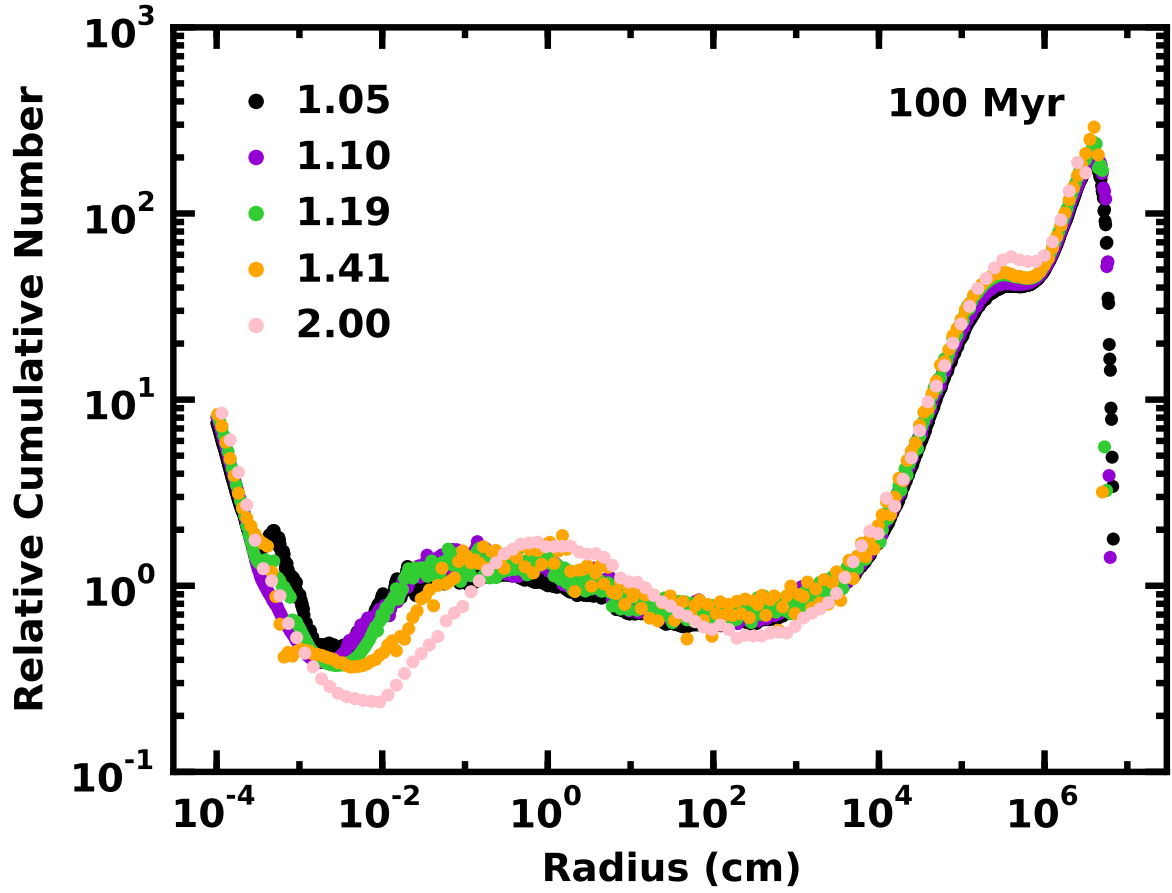}
\vskip 3ex
\caption{
As in Fig.~\ref{fig: sd0a} for 100~Myr. At later times, calculations
with smaller $\delta$ still have smaller fluctuations relative to 
calculations with larger $\delta$.
\label{fig: sd0b}
}
\end{figure}

Fig.~\ref{fig: sd0b} shows a similar pattern at 100~Myr. The three curves for $\delta$ =
1.05--1.19 are almost identical. Although the pattern for $\delta$ = 1.41 is very similar,
the waviness is more pronounced for particles with sizes of 100~\mum. This waviness is 
even larger for $\delta$ = 2.

\begin{figure} 
\includegraphics[width=6.5in]{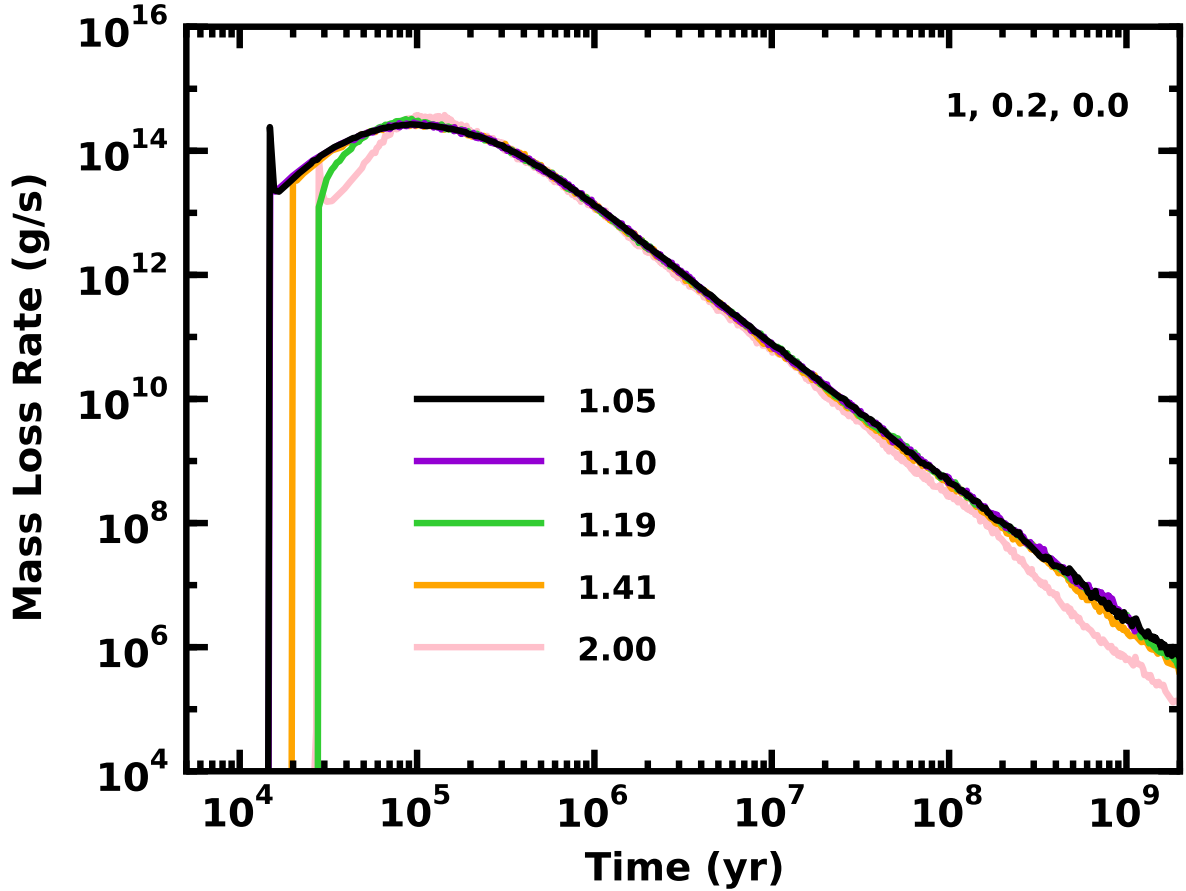}
\vskip 3ex
\caption{
Time evolution of the mass loss rate for collisional cascade
calculations with a mono-disperse swarm of 100~km particles, 
no velocity evolution, and different $\delta$ (as indicated 
in the legend in the lower left corner).  The legend in the 
upper right corner indicates adopted values for $b_d$, $m_{l,0}$, 
and $b_l$. 
The mass loss rate measures the production rate of objects with
$r <$~1~\mum\ from collisions between objects with typical radii
of 1--100~\mum. Starting from an ensemble of 100~km planetesimals,
the production rate is initially zero. Destructive collisions among 
large objects eventually produce sufficient numbers of small particles;
the mass loss rate then rises abruptly. Within $\sim$ 0.1~Myr, the
mass loss reaches a clear peak and then declines linearly with time.
\label{fig: mdot0a}
}
\end{figure}

To explore the evolution of the particle flux through the grid, Fig.~\ref{fig: mdot0a} shows
the production rate for particles with $r <$~1~\mum. At early times, collisions among
the mono-disperse set of 100~km objects produce little debris with sizes smaller than 1~mm.
The mass loss rate from the grid is then zero. As each calculation proceeds, destructive
collisions disperse more and more small particles. Collisions among these particles yield
particles smaller than 1~\mum. The mass loss rate then rises abruptly. The timing of this 
rise depends on stochastic variations in the collision rates among large particles. Thus,
the timing of the rise is not sensitive to $\delta$. The size distribution then takes awhile
to reach an equilibrium which depends on $\delta$ and on the timing of the rise in the mass
loss rate. As this equilibrium is established, the mass loss rate rises slowly. Afterwards,
the mass loss rate declines, $\dot{M} \propto t^{-\alpha}$ with $\alpha \approx$ 2.0--2.25.
For calculations with $\delta $ = 1.05--1.41, the long term time evolution of $\dot{M}$ is
independent of $\delta$.  When $\delta$ = 2.0, the decline of $\dot{M}$ with time is more
rapid at late times.

\begin{figure} 
\includegraphics[width=6.5in]{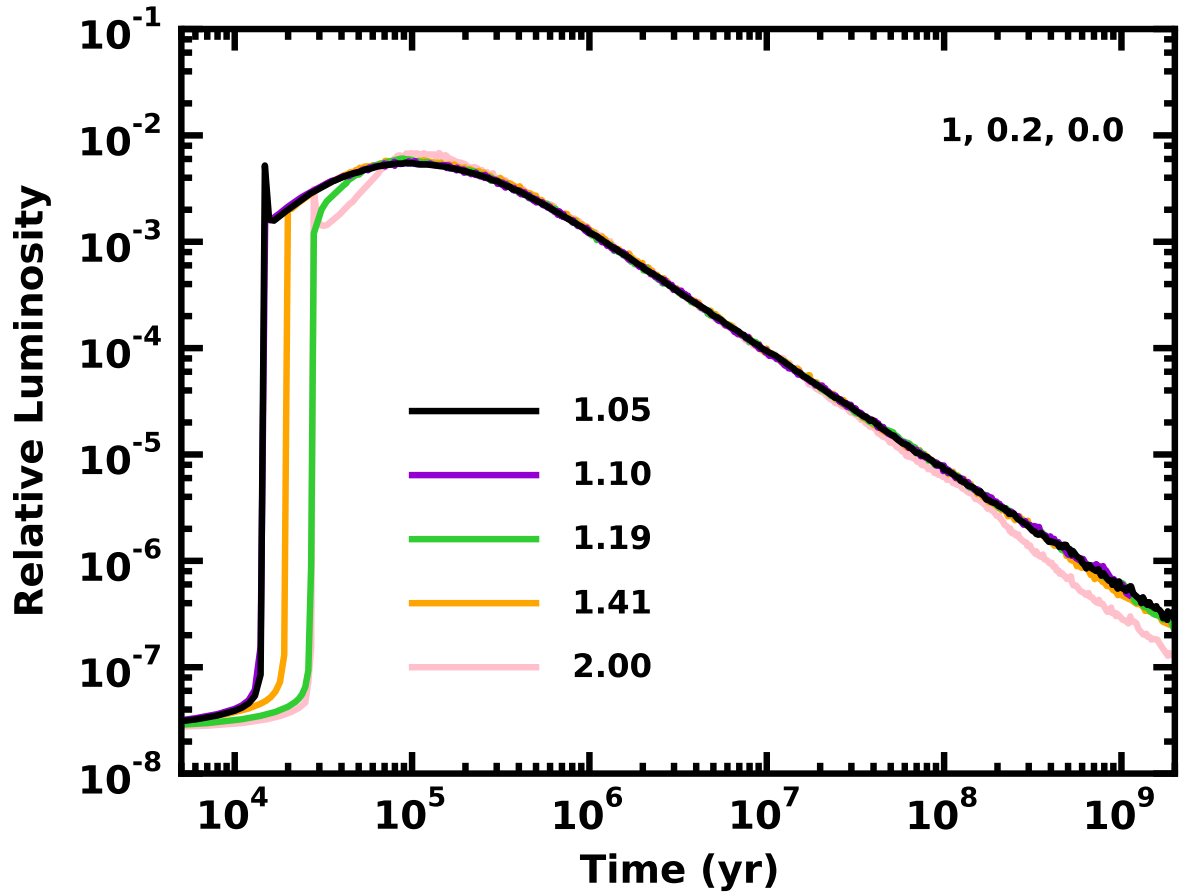}
\vskip 3ex
\caption{
As in Fig.~\ref{fig: mdot0a} for the dust luminosity $L_d / \lstar$.
Rare destructive collisions among 100~km objects 
gradually produce debris. When the debris contains sufficient 
numbers of 1--100~\mum\ objects, the dust luminosity rises.
For all calculations, the dust luminosity reaches a peak at 
0.1~Myr and then declines roughly linearly with time.
\label{fig: lum0}
}
\end{figure}

The evolution of the dust luminosity follows the evolution of $\dot{M}$ (Fig.~\ref{fig: lum0}).
Early on, it takes awhile for destructive collisions to produce a large ensemble of small
objects. Once these objects exist, the dust luminosity rises very rapidly. As the size
distribution of small particles adjusts to an equilibrium, the luminosity slowly rises to
a clear maximum. Once the size distribution reaches an equilibrium, the luminosity declines
roughly linearly with time. Calculations with $\delta$ = 1.05--1.41 yield nearly identical 
solutions for the time variation of the dust luminosity. At late times, the luminosity
declines more rapidly when $\delta$ = 2.00 than when $\delta$ = 1.05--1.41.

To conclude this discussion, we consider how the time evolution of $\dot{M}$ changes in
a `hybrid' model, where we adopt different values of $\delta$ for large and small particles.
We adopt $\delta$ = 2.0 for particles with $r \lesssim$ 0.2--0.4~km and $\delta$ = 1.05, 
1.10, 1.19, or 1.41 for particles with $r \gtrsim$ 3--5~km. In between these limits, 
$\delta$ varies linearly with particle size. This approach seeks to capture the better
accuracy of small $\delta$ models with a smaller expense of cpu time.

\begin{figure} 
\includegraphics[width=6.5in]{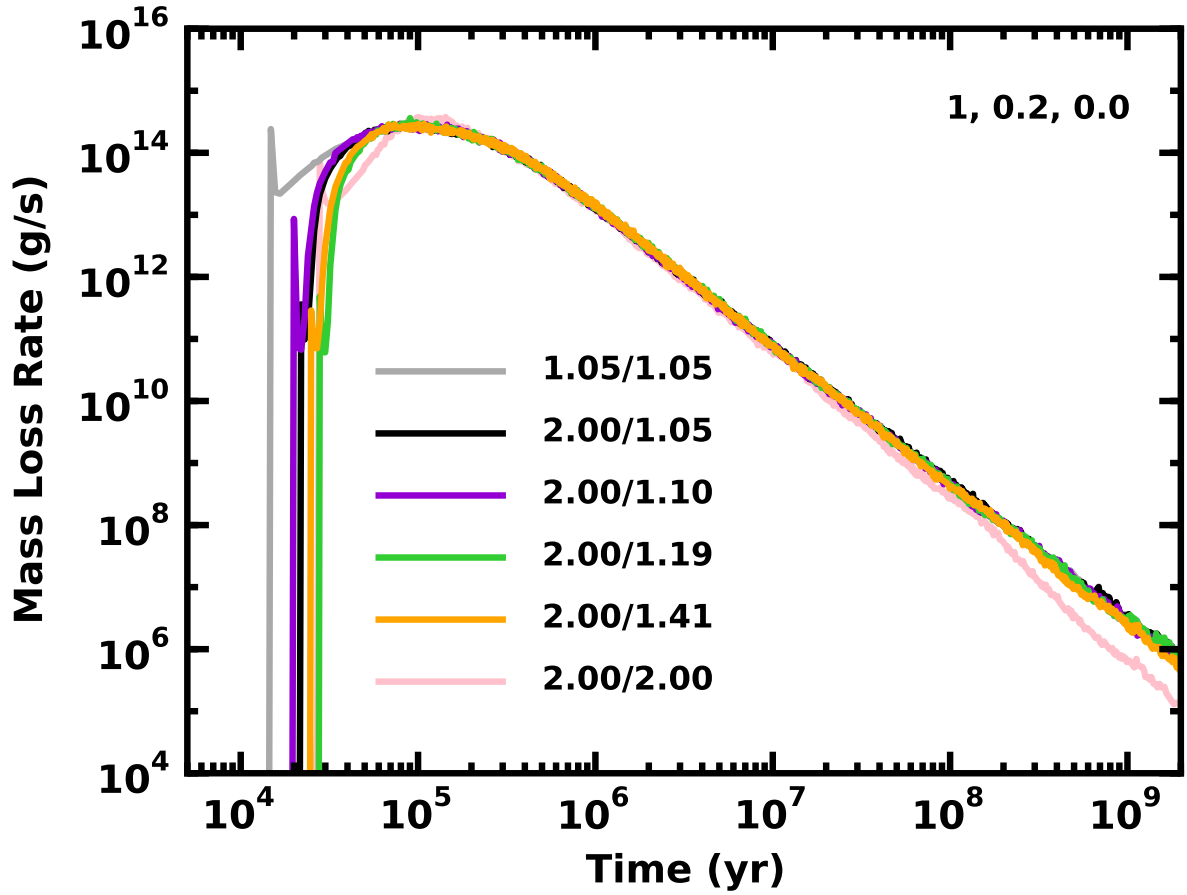}
\vskip 3ex
\caption{
As in Fig.~\ref{fig: mdot0a} for hybrid models where $\delta$ 
varies with particle size. As indicated in the legend, we
adopt one $\delta$ for $r \lesssim$ 0.2--0.4~km and a smaller
$\delta$ for $r \gtrsim$ 3--5~km. 
\label{fig: mdot0b}
}
\end{figure}

Fig.~\ref{fig: mdot0b} shows the results. Aside from the magenta curve (the pure 
$\delta$ = 2.00 from Fig.~\ref{fig: mdot0a}), all of the hybrid models follow the 
time variation of $\dot{M}$ established for models with $\delta$ = 1.05 for all
particles sizes. These results demonstrate that we can follow collisional cascades 
with $\delta$ = 2.0 for small particles and $\delta \lesssim$ 1.41 for large particles.

\subsection{Analytic Model for Wavy Size Distributions}
\label{sec: app-waves}

Wavy size distributions are characteristic of coagulation calculations with a low
mass cutoff \citep{campo1994,durda1998,obrien2003,kral2013}. To extend previous
analytic and numerical studies, \citet{wyatt2011} explore a new steady state
numerical model which shows how the amplitudes and positions of the waves depend 
on the collision velocity. Here, we derive an analytic solution to their model 
and briefly illustrate how the properties of the waves depend on \qdstar.

Following method I of the \citet{wyatt2011} framework (see their \S2.4.1), we 
establish a logarithmic grid of particle sizes extending from \rmin\ to \rmax\ with 
indices $k$ = 1 to $k$ = N.  Reversing the sense of the labeling from 
\citet{wyatt2011}, the smallest (largest) bin has $k$ = 1 ($k$ = N). If the mass
loss rate through the grid is independent of particle size, the mass $M_k$ 
contained in a mass bin is then a simple function of the rate of destructive 
collisions $R_k$ summed over all particles with index $i$ and $i < k$:
\begin{equation}
M_k = C_0 R_k^{-1} ~ ,
\label{eq: mks}
\end{equation}
where $C_0$ is an arbitrary constant \citep[see also eq. 15 of][]{wyatt2011}. 
The sum $R_k$ is
\begin{equation}
R_k = C_1 \sum_{i=1}^{i=k} \epsilon_{ik} M_i (r_i + r_k)^2 / r_i^3 ~ ,
\label{eq: rk}
\end{equation}
where $C_1$ is another constant. Recalling that $Q_c$ is the collision energy
per unit mass and \qdstar\ is the binding energy per unit mass,
\begin{equation}
\epsilon_{ik} = \left\{
\begin{array}{lll}
0 & & Q_c < \qdstar \\
\\
1 & & Q_c \ge \qdstar \\
\end{array}
\right.
\label{eq: eps}
\end{equation}
Thus, $R_k$ is the rate of collisions which disperse at least half the
mass of the colliding pair of particles.

The mass conservation equation in eq.~\ref{eq: mks} has a recursive 
solution. For $k$ = 1, $R_k$ has only one term: $R_k = 4 C_1 M_k / r_k$. 
Thus, $M_k^2 = C_0 / 4 C_1 r_k$.  For $k \ge$ 2, $R_k$ is a sum over terms 
with known $M_i$ and one term with $M_k$:
\begin{equation}
R_k = 4 C_1 M_k / r_k + C_1 \sum_{i=1}^{i=k-1} \epsilon_{ik} M_i (r_i + r_k)^2 / r_i^3 ~ .
\label{eq: rk2}
\end{equation}
Setting $R_{ki}$ equal to the second term in eq.~\ref{eq: rk2} leads to a
quadratic equation,
\begin{equation}
4 C_1 M_k^2 + R_{ki} M_k - C_0 = 0 ~ ,
\label{eq: mk2}
\end{equation}
which has only one possible solution with $M_k > 0$. 

To examine how steady-state mass distributions depend on \qdstar, 
we consider models with \qdstar\ = constant and 
$\qdstar = Q_b r^{\beta_b} + Q_g \rho_p r^{\beta_g}$ with
the standard fragmentation parameters (\S\ref{sec: calcs-casc}).
In these examples, the collision velocity $v$ = 3~\kms, the
mass density $\rho$ = 3~\gcmc, and $\delta_r = r_{i+1}/r_i$ = 
$10^{0.01}$ = 1.023.  The center-of-mass collision energy 
$Q_c = v_c^2 m_i m_k / 2 (m_i + m_k)^2$ then establishes the 
ratio $Q_c / \qdstar$ and $\epsilon_{ik}$ for each pair of particles. 
Recursive solution of eq.~\ref{eq: mk2} for $k$ = 1 (1~\mum)
to $k$ = 1101 (100~km) yields $M_k$ for each particle size.
Dividing $M_k$ by the expected $M_k \propto r_k^{\alpha}$
with $\alpha$ = 1/2 for \qdstar\ = constant and 
$\alpha$ = 0.32 for $\qdstar(r)$ yields the relative mass in
each bin.

\begin{figure} 
\includegraphics[width=6.5in]{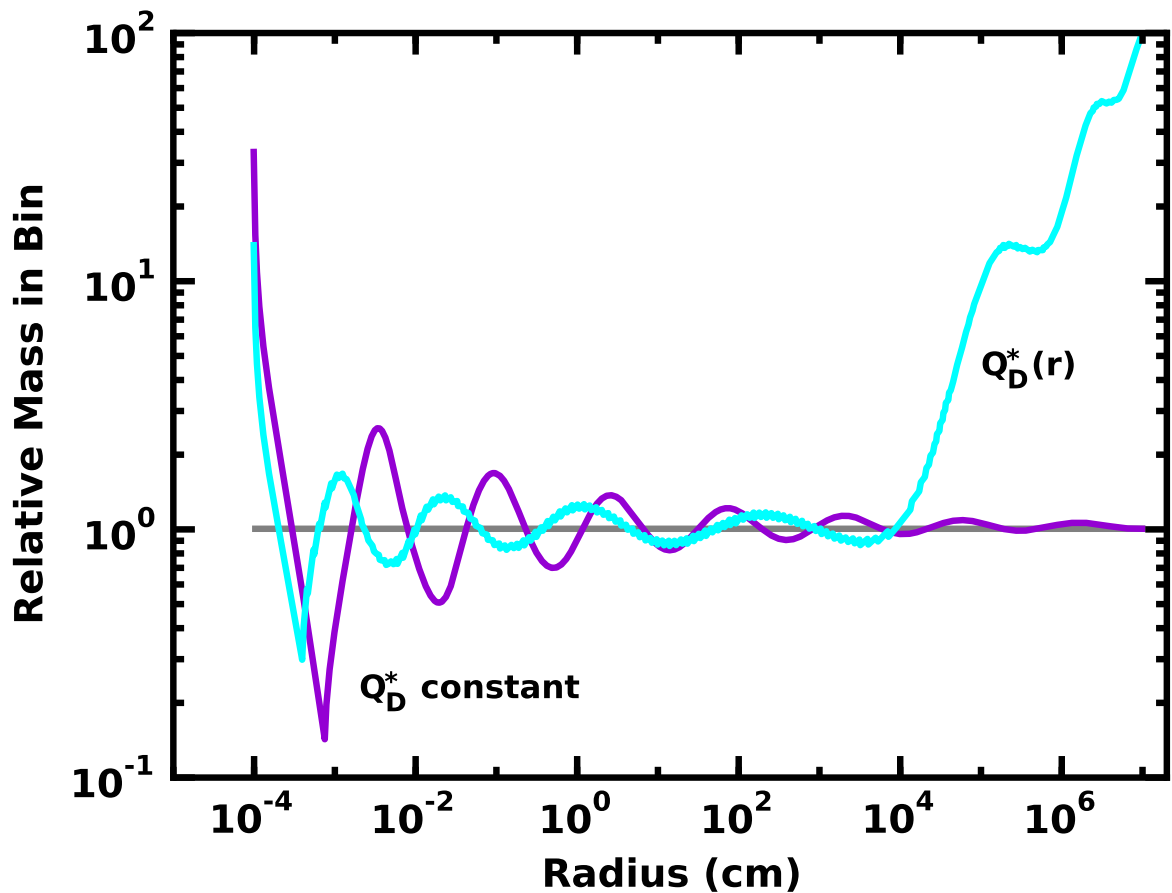}
\vskip 3ex
\caption{
Relative mass distributions derived from the analytic model
described by eqs.~\ref{eq: mks}--\ref{eq: mk2} with $v$ = 3~\kms.
The relative mass in each bin is $M_k / r^{1/2}$ (\qdstar\ =
constant, violet curve) or $M_k / r^{0.34}$ ($\qdstar(r)$,
cyan curve). Systems with \qdstar\ = constant have size 
distributions with larger amplitude and longer wavelength
waves than systems with $\qdstar(r)$. When \qdstar\ is a 
function of particle size, the change in slope at large 
sizes produces a corresponding change in the slope of the
size distribution.
\label{fig: mks}
}
\end{figure}

Fig.~\ref{fig: mks} compares results for the two models. When
\qdstar\ = constant (violet line), the relative mass generally
follows the power law. Waves with modest amplitude start at
the small-size cutoff and gradually diminish in amplitude to
large sizes. Relative to the power law, there is a clear excess
of particles at the smallest sizes: these bins require extra
mass to generate enough collisions to maintain a constant rate
of mass loss per bin.

When \qdstar\ is a function of particle size, the system follows
a power law from 1~\mum\ to 0.1~km. For larger sizes, the gravity
component of the binding energy dominates; the slope of the size
distribution changes from 3.68 to 3.0. Although systems where
\qdstar\ is a function of $r$ still display wavy size distributions,
the waves have smaller amplitude and shorter wavelength.

For both \qdstar\ relations, the shape of the size distribution 
derived from numerical simulations is remarkably close to the
analytic prediction. In particular, (i) the amplitudes of the 
waves at 1--10~\mum\ and (ii) the shape of the rise at 0.1--10~km
in systems with $\qdstar(r)$ are very similar in the analytic and 
numerical calculations. In the numerical simulations, cratering 
produces a longer wavelength compared to the analytic model. The
gradual reduction in the size of the largest object modifies the
shape of the size distribution at the largest sizes.

\bibliography{ms.bbl}

\clearpage
\begin{deluxetable}{ll}
\tablecolumns{7}
\tablewidth{0pc}
\tabletypesize{\small}
\tablenum{1}
\tablecaption{List of Variables }
\tablehead{
\colhead{Variable} &
\colhead{Definition}
}
\startdata
$a$ & semimajor axis, radial coordinate \\
$A_d$ & cross-sectional area of particles \\
$b_d$ & exponent in relation for debris from collisions \\
$b_l$ & exponent in relation for mass of largest particle in debris \\
$e$ & eccentricity \\
$e_H$ & Hill eccentricity \\
$f_c$ & fraction of solids converted into small particles ejected by radiation pressure \\
$f_d$ & detection frequency of dust emission \\
$f_g$ & gravitational focusing factor \\
$F$ & radiation flux emitted by a particle or a star at a specific wavelength \\
$h$ & horizontal velocity \\
$H$ & vertical scale height \\
$i$ & inclination \\
$L_d$ & luminosity of particles \\
\lstar\ & stellar luminosity \\
$m$ & particle mass \\
$\bar{n}$ & average mass of particle in a mass bin \\
$m_l$ & mass of largest particle in debris \\
$m_{l,0}$ & coefficient in relation for mass of largest particle in debris \\
$m_{max}$ & mass of largest particle \\
$m_{min}$ & mass of smallest particle \\
$M_d$ & total mass in particles \\
$\dot{M}$ & mass loss rate \\
\mstar\ & stellar mass \\
$n$ & exponent of power law or particle number density \\
$N$ & particle number \\
$N(>r)$ & cumulative number of particles larger than $r$ \\
$N_{max}$ & number of largest particles \\
$q$ & exponent of power law size distribution \\
$q_c$, $q_r$ & exponents for cumulative and relative cumulative size distribution \\
$q_d$ & exponent of power law size distribution for debris \\
$Q_b$, $Q_g$ & coefficients in \qdstar\ relation \\
$Q_c$ & center of mass collision energy \\
$\qdstar$ & collision energy required to eject 50\% of the mass \\
$r$ & particle radius \\
$\bar{r}$ & average radius of particle in a mass bin \\
$r_{brk}$ & radius of particles destroyed in equal mass collisions \\
$r_{max}$ & radius of largest particle \\
$r_{min}$ & radius of smallest particle \\
$r_Q $ & particle size with smallest \qdstar \\
$t$ & time \\
$t_c$ & collision time \\
$T$ & temperature \\
$v$ & relative collision velocity \\
$v_h$ & horizontal velocity \\
$v_K$ & orbital velocity \\
$v_z$ & vertical velocity \\
$V$ & volume \\
$\alpha$ & correction factor for collision time \\
$\beta$ & ratio of the radiation force to the gravitational force\\
$\beta_b$, $\beta_g$ & exponents in \qdstar\ relation \\
$\delta$ & mass spacing factor \\
$\delta a$ & width of annulus \\
$\xi$ & ratio of mass in small particles ($r <$ 1~mm) to total mass \\
$\rho$ & particle mass density \\
$\sigma$ & geometric cross section \\
$\Sigma$ & surface density \\
$\Omega$ & angular velocity \\
\enddata
\tablecomments{Variables with a subscript `0' refer to initial conditions;
e.g., $e_0$ is the initial eccentricity}
\label{tab: pars}
\end{deluxetable}
\clearpage

%
%

\clearpage

\clearpage

\clearpage

\clearpage

\clearpage

\clearpage

\clearpage

\clearpage

\clearpage

\clearpage

\clearpage
\clearpage

\clearpage

\clearpage

\clearpage

\clearpage

\clearpage

\clearpage

\clearpage

\clearpage

\clearpage

\clearpage

\clearpage

\clearpage

\clearpage

\clearpage

\clearpage

\clearpage

\clearpage

\clearpage

\end{document}